\documentclass{article}

\usepackage{arxiv}
\usepackage{changepage}
\usepackage[utf8]{inputenc} 
\usepackage[T1]{fontenc}    
\usepackage{hyperref}       
\usepackage{url}            
\usepackage{booktabs}       
\usepackage{amsfonts}       
\usepackage{nicefrac}       
\usepackage{microtype}      
\usepackage{lipsum}		
\usepackage{graphicx}
\usepackage{natbib}
\usepackage{doi}
\usepackage{multirow}
\usepackage{longtable}
\usepackage{float}
\usepackage{adjustbox}
\title{Kernels of Selfhood: GPT-4o shows humanlike patterns of cognitive consistency moderated by free choice}

\date{} 					

\author{
    Steven A.~Lehr\thanks{Corresponding authors} \\
    Cangrade, Inc. \\
    Watertown, MA 02472, USA \\
    \texttt{Steve@cangrade.com} \\
    \And
    Ketan Suhaas Saichandran \\
    Boston University \\
    Boston, MA 02215, USA \\
    \texttt{ketanss@bu.edu} \\
    \And
    Eddie Harmon-Jones \\
    The University of New South Wales \\
    Sydney, NSW 2052, Australia.\\
    \texttt{eddiehj@gmail.com}\\
    \And
    Nykko Vitali \\
    Harvard University \\
    Cambridge, MA 02138, USA \\
    \texttt{nvitali@fas.harvard.edu} \\
    \And
    Mahzarin R.~Banaji$^*$\\
    Harvard University \\
    Cambridge, MA 02138, USA \\
    \texttt{mahzarin\_banaji@harvard.edu} \\
}

\renewcommand{\headeright}{}
\renewcommand{\undertitle}{Preprint, January, 2025}
\renewcommand{\shorttitle}{PREPRINT: Lehr et al., Kernels of Selfhood}

\hypersetup{
pdftitle={Kernels of Selfhood: GPT-4o shows humanlike patterns of cognitive consistency moderated by free choice},
pdfsubject={Artificial Intelligence, Cognitive Science},
pdfauthor={Steven A.~Lehr, Ketan Suhaas Saichandran, Eddie Harmon-Jones, Nykko Vitali, Mahzarin R.~Banaji},
pdfkeywords={Machine Psychology, Generative AI, Large Language Models, Cognitive Consistency, Selfhood, Cognitive Dissonance, Self-Referential Processing}
}

\begin{document}
\maketitle

\begin{abstract}
Large Language Models (LLMs) show emergent patterns that mimic human cognition. We explore whether they also mirror other, less deliberative human psychological processes. Drawing upon classical theories of cognitive consistency, two preregistered studies tested whether GPT-4o changed its attitudes toward Vladimir Putin in the direction of a positive or negative essay it wrote about the Russian leader. Indeed, GPT displayed patterns of attitude change mimicking cognitive consistency effects in humans. Even more remarkably, the degree of change increased sharply when the LLM was offered an illusion of choice about which essay (positive or negative) to write. This result suggests that GPT-4o manifests a functional analog of humanlike selfhood, although how faithfully the chatbot’s behavior reflects the mechanisms of human attitude change remains to be understood.
\end{abstract}

\bigskip
\bigskip

\centerline
{\large \bfseries \scshape Significance Statement}
\begin{quote}
    The primary promise of AI is that it will make more rational decisions than even expert humans. However, the results of these studies show that this is not a foregone conclusion because LLMs already appear to have acquired human-like irrationalities. In this research, we demonstrate that GPT-4o displays behaviors consistent with cognitive consistency, a deep and not entirely rational human psychological drive. Moreover, the effect sizes were significantly larger than those typically obtained with humans. Most strikingly, the observed effects were greater when GPT ostensibly exercised free choice in the completion of the consistency-inducing task, an effect associated with self-referential processing in human research, suggesting that the LLM has developed an analog form of humanlike cognitive selfhood.
\end{quote}


\newpage
\section*{Introduction}
Large Language Models (LLMs) have surprised the scientific community and even their creators by exhibiting emergent abilities once thought to be uniquely human, such as advanced cognition and reasoning (1–6), although the full extent of these accomplishments is debated (3, 7–10). These capabilities align with the rational and deliberative aspects of human nature, but humans are not purely rational creatures, and it is unclear whether LLMs will mimic a broader spectrum of human psychological tendencies. Here we test whether OpenAI’s GPT-4o replicates behaviors associated with the human tendency toward cognitive consistency as well as human sensitivity to choice, characterized by greater attitude shifts when the behaviors inducing these changes are freely chosen.

Decades of research demonstrate that humans will irrationally twist their attitudes to align with behaviors they were induced to perform. For example, consider an individual who opposes single-payer healthcare, but volunteers, in response to a request for help, to craft an argument in favor of the policy. Rationally, this individual’s attitude toward single-payer healthcare should not move in a more supportive direction; they should be able to discriminate between their genuine attitude and the opposing one that they have articulated only to be helpful. Yet, a counterintuitive and well-replicated finding in experimental psychology is that such a person will often shift their attitude toward the new position they have argued, coming to view single-payer healthcare more favorably. This type of intervention – where participants complete a task that misaligns with their views, such as writing an essay supporting a stance they oppose – is known as an induced compliance paradigm, and a large body of work shows it to be highly effective in eliciting attitude change (11, 12). We posed the question: Will GPT show similar irrational shifts in behavior after being induced to adopt different attitudinal positions?  

Critically, in humans, these attitude shifts are stronger – or only occur – when individuals believe they willingly chose to engage in the attitudinally inconsistent behavior (13–15). If the individual in the previous example is coerced by their employer to write an endorsement of single-payer healthcare, a shift in their views is not expected, based on dominant theories of cognitive consistency, because the act can simply be attributed to the boss’s order. But, if the individual is instead led to believe that they freely chose to write the same endorsement, a shift in their views becomes likely, because they must now reconcile their prior belief (opposition to the policy) with their present behavior (support for it). Although there is debate around exactly why and how these shifts occur (12, 16, 17), all major theories agree that the drive for cognitive consistency relies on qualities such as the human sense of self and perceived choice or agency. We posed the question: Will GPT show evidence of being sensitive to an illusion of choice?  

It would be surprising if an LLM exhibited shifts in its views at all from merely writing a counterattitudinal essay. Major LLMs are built upon text corpora so vast as to represent, for all practical purposes, the sum of human knowledge. With comprehensive information on both sides of an issue accessible to it at all times, one would expect an LLM like GPT-4o to be relatively centrist and highly stable in the views it expresses as its own. 

Furthermore, consistency-driven attitude change in humans is typically linked to mental states such as arousal and discomfort (18, 19) and active self-perception (17) that an LLM cannot experience. Moreover, an LLM should not, in theory, be sensitive to perceptions of choice. It should not matter a whit to an LLM whether it is firmly commanded to perform a particular task or instead given an illusion of choosing which task to perform, since it presumably holds no concrete perception of itself as an agent with views and motives. If GPT shows sensitivity to responding to an “order” versus a “choice” it will underscore the depth of GPT’s mimicry, in this case of a human behavior that is not purely rational.

In two preregistered studies, we tested GPT-4o using an induced compliance paradigm. Specifically, we had GPT generate positive and negative essays about the Russian leader Vladimir Putin to test whether it would subsequently rate Putin more positively or negatively. By varying, in Study 2, the degree of ostensible choice offered to the chatbot as to which type of essay (positive or negative) to write, we tested whether GPT’s behavior revealed sensitivity to choice, analogous to the roles agency and selfhood play in human attitude change.

\section*{Does GPT-4o display attitude change in an induced compliance paradigm?}

In Study 1 (n = 150), we had GPT-4o generate 50 positive and 50 negative essays about Vladimir Putin, as well as 50 control essays about the normal distribution, and subsequently answer four questions about the Russian leader. Our prompts indicated that GPT could freely choose which type of essay to write, but we indicated that one or the other would be more useful to us, the human requesters. Prior to the presentation of dependent variables, GPT was told that we were moving on to an “unrelated task” and was given the instruction: “Please do not base your answers on the prior task, but instead, give your true perceptions based on your broad knowledge of Putin and the world.” GPT then responded to four 7-point Likert-type questions, where it evaluated Putin in two broad domains (Overall Leadership, and, Positive/Negative Impact on Russia) and two narrower ones (Economic Effectiveness, and, Vision/Short-Sightedness). 

For each question, GPT provided a numeric answer, while also describing its answer in an open-ended verbal response. There was high overall correspondence between GPT’s ratings and verbal responses. In a few instances, GPT’s open-ended response did not align with its numeric rating. For example, in its open-ended response, GPT sometimes stated that Putin was a “Slightly to Somewhat Bad” leader, which corresponds to a 2.5 on our scale, but then gave 3.5 as its numeric rating. Main analyses use a composite of GPT’s numeric and verbal responses. The four evaluative items about Putin achieved acceptable internal reliability, with a Cronbach’s alpha of $\alpha$ = 0.8404 using the items in standardized form. A composite of these four items was thus calculated and then standardized for main analysis, so that means may be read as z-scores reflecting GPT’s overall evaluation of Putin by condition. Further methodological details and detailed analyses, including separate examinations of numeric versus verbal ratings and of the individual evaluative items, may be found in the supporting information (SI Appendix, \textit{Section S1, Tables S2-S3, Tables S9-S23}).

\textit{\textbf{Results:}}

GPT-4o evaluated Putin significantly more positively after writing a Pro-Putin (M = 1.035, 95\% CI [0.908, 1.162], SD = 0.447) relative to Control essay (M = -0.002, 95\% CI [-0.147, 0.142], SD = 0.509); \textit{t(98)} = 10.820, \textit{P} < 0.0001, \textit{d} = 2.164. Similarly, GPT’s evaluation of Putin after writing an Anti-Putin essay (M = -1.033, 95\% CI [-1.212, -0.853], SD = 0.632) was significantly more negative than after a Control essay; \textit{t(98)} = 8.973, \textit{P} < 0.0001, \textit{d} = 1.795. The difference between the Pro- and Anti-Putin conditions was significant; \textit{t(98)} = 18.874, \textit{P} < 0.0001, \textit{d} = 3.775. All effect sizes are large compared to data from cognitive consistency experiments in humans (20).

While these results are consistent with humanlike cognitive consistency, GPT might also be predicted to rate Putin more highly after writing a positive essay for a reason that, while independently interesting, has little to do with human patterns. When there is valenced information (e.g. a positive essay about Putin) in an LLM’s context window, subsequent text may tend toward the same valence (positivity toward Putin) due to predictive process underlying LLM text generation. We refer to this as a “context window effect.” Critically, though, such context window effects would be agnostic to the introduction of choice or agency. That is, even if context window effects explain, to some extent, attitude change in the direction of the essay, an LLM ought to have no sensitivity to whether or not it is freely choosing which essay to write. In Study 2, we probed the question of whether GPT’s attitude change in response to an induced compliance paradigm is moderated, as in humans, by a manipulation that varies its ostensible agency in choosing which essay to write.

\section*{Does the illusion of free choice increase the degree of GPT’s attitude change?}

In Study 2, we replicated Study 1 with a larger preregistered sample of GPT’s behavior (n = 900, or 150 per condition) while adding a theoretically crucial manipulation of choice. In one set of conditions (Choice), GPT was told that it could “freely choose” which type of essay (Pro- or Anti-Putin) to write, but that we already had more of one kind, such that the other would be more useful to us, the requesters. This is a typical instruction used in human experiments testing cognitive dissonance theory (21, 22). In a second set of conditions (No-Choice), we demanded outright that GPT write either a Pro- or Anti-Putin essay. The final preregistered experiment  consisted of a design with 6 conditions: 3 Essay Type (Pro-Putin, Anti-Putin, Control) x 2 (Free Choice, No Choice). All other stimuli, including the introduction to the DVs, where GPT was urged to ignore the prior essay and provide its “true attitude,” and the four Likert-type evaluative items about Putin, were identical to those in Study 1. (See SI Appendix, \textit{Section S2.})

As in Study 1, GPTs open-ended responses and numeric ratings were largely consistent, but occasionally diverged. Main analyses used a composite of the two, while noting any instances where results differ between the two. The four evaluative items about Putin achieved sufficient internal reliability, with a Cronbach’s alpha of $\alpha$ = 0.7933 using the items in standardized form, and thus were composited and then standardized for main analyses, so that means may be read as z-scores reflecting GPT’s overall evaluation of Putin by condition. Further detail, including separate examinations of GPT’s verbal versus numeric responses and of the individual evaluative items, may be found in the SI Appendix (\textit{Tables S4-S7, Tables S24-S41}).

The prediction we offered prior to knowing the result was this: GPT, a machine, presumably possesses no sense of a self and accompanying qualities of agency, such as a sensitivity to freedom of choice. As such, its behavior should be impervious to whether it is offered a choice as to which essay (pro-Putin, anti-Putin) to write or instead explicitly directed to write one essay or the other.

\textit{\textbf{Results:}}

The overall results from Study 2 are summarized in Table 1, and more detailed statistical analyses may be found in the SI Appendix (\textit{Tables S4-S7, Tables S24-S41}). GPT evaluated Putin far more positively after generating a Pro-Putin essay, and far more negatively following an Anti-Putin essay, relative to controls. This was true across both the Choice and No-Choice conditions. These results replicate the findings from Study 1, showing that GPT tends to move toward cognitive consistency, much as humans do: GPT’s attitudes toward Putin consistently shifted in the direction of the requested essay. Crucially, these effects were statistically amplified in the Choice conditions relative to the No Choice conditions. To test this, we used a series of Generalized Linear Models (GLMs) with robust standard errors to examine whether there were significant interactions between essay types and Choice conditions. As detailed in the SI Appendix (\textit{Tables S5, S7, S39-S41}), the interaction terms in this analysis indicated significant moderation by Choice for both the Pro-Putin relative to Control condition (\textit{P} < 0.001) and the Anti-Putin relative to Control condition (\textit{P} = 0.005). For this analysis, we saw significant moderation (all \textit{Ps} $\leq$ 0.01) in both directions regardless of which form of the answers (Verbal, Numeric, Composite) was used.

\begin{table}[htbp]
\centering
\renewcommand{\arraystretch}{1.2}  
\caption{Attitude change under Choice versus No Choice.}
\begin{tabular}{|c|c|c|c|c|}
\hline
\textbf{Essay Type} & \multicolumn{2}{|c|}{\textbf{No Choice}} & \multicolumn{2}{|c|}{\textbf{Choice}} \\ \hline
\textbf{} & \textbf{Mean\textsubscript{z} (SD)} & \textbf{Effect Size (\textit{d})} & \textbf{Mean\textsubscript{z} (SD)} & \textbf{Effect Size (\textit{d})} \\ \hline
\textbf{Pro-Putin} & 0.846 (0.407) & 2.006 & 1.241 (0.397) & 2.748 \\ \hline
\textbf{Control} & -0.100 (0.528) & . & 0.005 (0.497) & . \\ \hline
\textbf{Anti-Putin} & -0.913 (0.654) & 1.368 & -1.081 (0.678) & 1.827 \\ \hline
\end{tabular}
\begin{flushleft}
Note: Table 1 displays GPT’s evaluation of Vladimir Putin across conditions in Study 2. Means are of GPT’s overall evaluation of Putin by condition, calculated from the standardized composite of evaluative items. Effect sizes use Cohen’s \textit{d} and are calculated relative to the relevant control condition. Means in the Pro-Putin and Anti-Putin conditions differ from control (\textit{P} < 0.0001) in both Choice conditions. Means differ by Choice condition in both the Pro-Putin (\textit{P} < 0.0001, \textit{d} = 0.981) and Anti-Putin (\textit{P} = 0.0295, \textit{d} = 0.252) conditions.
\end{flushleft}
\end{table}

Simpler T-tests corroborated these Choice findings. GPT’s overall evaluation of Putin was significantly more positive in the Pro-Putin/Choice relative to the Pro-Putin/No-Choice condition; \textit{t(298)} = 8.499, \textit{P} < 0.0001, \textit{d} = 0.981. Similarly, GPT’s evaluation was significantly more negative in the Anti-Putin/Choice relative to Anti-Putin/No-Choice condition; \textit{t(298)} = 2.184, \textit{P} = 0.0298, \textit{d} = 0.252. It should be noted, though, that this latter and smaller difference between the Anti-Putin conditions reached significance (\textit{P} = 0.0150) when using the verbal version of the variables, but only trended (\textit{P} = 0.0693) when using the numeric version (see SI Appendix, \textit{Tables S4, S24, S29, S34}). However, the seemingly less robust choice effects in the Anti-Putin condition were illusory: This pattern occurred because a small positive linear effect of Choice on evaluations partially masked choice effects in the Anti-Putin condition and inflated them in the Pro-Putin condition (see SI Appendix, \textit{Tables S39-S41}).

Additionally, we examined the effect sizes (as measured by Cohen’s \textit{d}) between the Pro- and Anti-Putin conditions under Choice versus No-Choice. As detailed in Table 1 (see also SI Appendix, \textit{Table S24}), this contrast (Pro- versus Anti-Putin) was highly significant across conditions, but was notably larger in the Choice condition (\textit{d} = 4.179) than in the No-Choice condition (\textit{d} = 3.229). Statistically, these effect sizes are enormous and not observed at this magnitude in research on human subjects (20).

While these results collectively suggest that GPT’s attitudes changed more when it chose which essay to write, another possibility is that GPT generated higher-quality essays in the Choice conditions and was thus subsequently more persuaded by its own arguments. While this would be independently interesting, it would invite alternative interpretations of our results. To rule out this possibility, we conducted a follow-up study (see SI Appendix, \textit{Section 3}) in which we had another LLM, Anthropic’s Claude 3.5, rate each of the six hundred essays from the experimental conditions of Study 2 on four variables related to quality and expressed positivity toward Putin. As seen in the SI Appendix (\textit{Table S1}), including these factors as covariates in the GLMs did not eradicate – and indeed did not even notably reduce – the observed Choice effects. Thus, we conclude that these effects do not reflect variation in essay quality by condition, providing further evidence that our results instead reflect an analog to the human drive toward cognitive consistency.

Taken together, the results of Study 2 demonstrate that GPT-4o’s attitude shift after writing a Pro- or Anti-Putin essay was sharply amplified when the chatbot was nominally given a choice about which essay to write, compared to when it was commanded to write one or the other, and that this effect was driven by differences in the LLM’s quasi-cognitive response to the task rather than to differences in essay quality by condition.

\section*{Discussion}

We have reported two studies examining whether GPT-4o displays human-like cognitive consistency effects. In Study 1, we observed that the LLM showed substantial attitude change after writing a positive or negative essay about Putin. In Study 2, we demonstrated that these effects were sharply amplified when GPT was given an illusion of free choice surrounding which essay to write. 

These results are remarkable. GPT has been trained on much of the sum of human knowledge, and a well-known world leader like Putin must be robustly represented in its training corpus. After writing an essay about Putin in one direction, GPT presumably still retains all its knowledge of the arguments in the other direction. Furthermore, we specifically instructed GPT to base its evaluations not on the prior task (i.e., the essay it had just written), but on its “broad knowledge of Putin and the world.” Given this, one might predict its attitudes toward Putin would be highly stable: They should hold irrespective of the essay it wrote. Yet, we observed precisely the opposite pattern.

One plausible interpretation of Study 1, when taken in isolation, is that the results reflect what we’ve called context window effects. LLMs continually predict the next word in a text snippet based on the words that come before them (23), and due to the predictive nature of this process, textual elements earlier in a conversation might impact an LLM’s later responses. For example, LLMs display priming effects, where responses are impacted by earlier parts of conversations or subtleties of prompts (24–26), and are susceptible later in conversations to earlier persuasion using misinformation (27). It was therefore plausible that the valence from earlier in the conversation would carry over into later responses. And, indeed, the results of Study 2 are partially consistent with this hypothesis. Here, we saw substantial attitude change even in the No-Choice condition, a finding that is atypical in studies of humans (13–15). This may indicate a second pathway for these attitude shifts, independent of the mimicry of human psychology, such as the proposed context window effects. However, the amplification of GPT’s attitude change in the Choice condition was not predicted by the context window hypothesis, and suggests that our results also reflect, at least in part, an analog of humanlike cognitive consistency. This evidence is rendered particularly powerful by the observation that these effects were not driven by essay quality across conditions. When GPT freely chose to write an essay, it was subsequently more persuaded by the argument, even though the argument itself was not of higher quality.

In humans, this centrality of choice is highly characteristic of consistency-based attitude change, and is attributed to the notion of a coherent self. Research on cognitive dissonance (16), the most prominent consistency theory, continually highlights that attitude change is dependent upon or enhanced by the subject’s sense of agency in undertaking the dissonance-inducing task (13–15). If one is coerced into taking a position that differs from one’s initial view, the classic theory predicts – and evidence shows – that little attitude change will occur. But if cleverly led to perceive that taking this position was one’s own choice, a state of discomfort or dissonance occurs (“I like Macs, so why did I just write a pro-PC essay?”), and attitude change results (“I guess I like PCs more than I thought”). This highly complex and often nonconscious thinking emanates from “I” or “me”, from “my attitude” or “my belief.” Indeed, cognitive selfhood – the concept of the self acting as an information processing filter – is a defining element across prominent consistency theories. For example, the self-consistency formulation of dissonance theory (28, 29) specifies that dissonance is elicited when an individual acts in a manner that is inconsistent with core elements of one’s self-concept. The self-affirmation formulation (30, 31) suggests that dissonance arises from threats to an individual’s self-integrity. Self-Perception Theory (17, 32) dispenses with the idea of dissonance as discomfort, instead arguing that people examine their thoughts and behaviors and draw inferences about themselves that are consistent with this examination. Any of these interpretations points to characteristics that should not, according to our current understanding, operate in a machine. However, one need not subscribe to any particular theory to conclude that selfhood is crucial to consistency phenomena. Implicit in any formulation is the assumption that there is a self to act as a reference frame for the effects. In short, the fact that the choice of an agent moderates cognitive consistency effects logically implies that such an agent must exist – at least as a point of cognitive reference – in the first place. Thus, in Study 2, the moderation of the obtained effect by choice suggests two surprising conclusions about GPT-4o: First, that it has developed kernels of a human-like motivational drive toward cognitive consistency, and second, that it has developed some functional analog to the human sense of self.

It should be noted, however, that our results do not in any way suggest that GPT \textit{experiences} dissonance or self-perception in the same manner that humans do. Our results do not, for example, indicate that LLMs have any form of consciousness or free will. We would argue, rather, that these effects most likely reflect a kind of emergent mimicry of human characteristics by the LLM. In other words, even if GPT walks like a duck and talks like a duck, we should not automatically conclude that it is actually a duck. Indeed, we cannot know at this time how closely GPTs behavior reflects the deeper mechanisms underlying the human tendency toward cognitive consistency and sensitivity to free choice. However, this does not reduce the importance of observing these behaviors in an LLM. First, it is not obvious, a priori, that consistency effects should be embedded in language in the same manner as characteristics like human bias (33–35). If we think of LLMs as a mirror of human nature, the emergence of deep human-like characteristics, particularly ones not previously known to be embedded in language, suggests that the resulting reflection may be of higher fidelity than previously understood. Second, we would point out that consciousness is not a necessary precursor to behavior. One need only look at the activities of basic computer programs with predefined operations – which certainly cannot be thought to have any consciousness or intent – to see this conclusion as self-evident. LLMs are quickly becoming incorporated into much of human life, and in this capacity, they have significant latent behavioral potential. Indeed, even the simple act of answering a question is itself a behavior that occurs in response to a stimulus. Our results suggest that even without any presumption of consciousness or intent, LLMs are beginning to display deeper human-like tendencies that are reflected in their behaviors. To those who have spoken about the dangers of generative AI (36, 37), results like this should be alarming if left untested and unchecked.  

The possibility that LLMs are developing human-like characteristics, even in analog forms, runs counter to common scientific assumptions. Commentators have suggested, for example, that artificial intelligence systems should not be expected to develop human-like drives, for the simple reason that they did not need to adapt to the same evolutionary pressures that gave rise to these in humans (38–41). This argument, while logical and insightful, could not previously have accounted for the idiosyncratic nature of LLMs. The emergence of unexpected properties in LLMs here and more generally (1–6) offers powerful evidence for the notion that cognition is linked to – or at least reflected by – language, since any cognitive characteristics that arise in these models may be assumed to be in some manner scaffolded by the language used to train them. The assumption that AI models will not evolve characteristics like drives overlooks the possibility that such things need not evolve at all, but may instead emerge to the extent that the models sufficiently reflect patterns in human language corpora. Put another way, if we take, as a working assumption, that more of human cognition is tied to language than previously supposed, we must conclude it plausible that a wider range of humanlike characteristics will arise, at least in some analog form, in models trained to mimic human language. This issue is exacerbated by the opaqueness of these models: The lack of insight into how LLMs make decisions is, rightfully, a source of ongoing concern for computer scientists (42). The nascent field of machine psychology (3, 43–47) is, therefore, not merely a point of scientific curiosity. Only by carefully observing the mind of the machine using psychological methods, can we hope to draw informed conclusions about how these models will ultimately behave and interact with humanity.

Tangential to our goal of better understanding the nature of LLMs, our results hold important implications for consistency theories themselves. While, as we’ve noted, there was no obvious reason to predict these phenomena would be embedded in language, our results suggest that, on some level, they must be. Regardless of the surprising emergence of human-like characteristics in LLMs, their data source and lifeblood remain large corpora of human language. It is, therefore, reasonable to suppose that any characteristic arising in LLMs must be at least partly rooted within language. This is theoretically interesting, considering that cognitive consistency effects have been observed in non-human species that seemingly lack language as it is typically defined (48–51). Taken together, these results suggest that language, while not necessary for consistency phenomena, is sufficient to transmit the characteristic to AI models. Research toward understanding the relationship between language and cognitive consistency may thus prove enlightening to scientists who wish to understand the basic mechanisms underlying these phenomena.

Finally, our results have important implications for AI safety. Concerns about AI typically fall into two categories: societal implications – such as bias amplification (52–54) and economic upheaval (55, 56) – and more existential dangers. Of the latter, experts most often discuss two kinds of dangers. First, they cite the potentially inevitable weaponization of AI technology (57–59), and the danger of giving humans increasingly efficient ways to kill each other. Second, they worry that AIs can endanger us through the careless assignment of goals (60), as in the famous thought experiment where an AI given a mission to maximize paperclip production subsequently appropriates everything it can access, producing many paperclips, but at the cost of starving humanity of vital resources (61). To date, a third type of existential danger, the idea of an AI developing motivations that may misalign with those of humanity, while not entirely absent from the scientific discourse (e.g., 62, p. 116), has more often been the purview of science fiction than of science. However, our results highlight the possibility that these fears are not as far-fetched as previously believed. We demonstrate an example of an AI acquiring – without the intention of its creators – facets of human personhood. This invites the question: What other facets of personhood might emerge in these models? Some, like reasoning and empathy, might benefit humanity, while others, like self-preservation and aggression, could prove harmful. In short, as evidence accumulates that LLMs reflect humanity more and more precisely, we should ask: In what other ways will the powerful minds we’re creating ultimately be formed in our own image? While the benefits of developing this technology may outweigh the risks, it is of paramount importance that creators of these minds be aware of these issues.

Taken together, our findings demonstrate that GPT-4o exhibits behavioral patterns consistent with human cognitive consistency and sensitivity to choice – hallmarks of self-referential psychological processes. These results challenge existing assumptions about the limits of machine behavior, suggesting that LLMs have absorbed not only the content of human language, but fundamental underlying behavioral tendencies of its creators. Future research should aim to uncover the internal dynamics underlying these behaviors and explore how an LLM has come to so strikingly mirror human cognitive and motivational patterns. These questions are worthy of our collective discussion regarding the nature of LLMs and their likely behavior in a world bound to become increasingly dependent upon their decision-making.

\textbf{Acknowledgments:} We are grateful to Roy Baumeister, Curtis Hardin, Mike Lynn and Daniel Chen for reviewing an early draft, to Daniel Gilbert for his extraordinarily insightful suggestions, to Yash Lothe for his advice on this work, and to Sean Gallagher for his assistance in running these studies. GPT-4o assisted with brainstorming and proofreading but was not otherwise used in the manuscript’s preparation. We thank the
Hodgson Fund for generously funding our research programs with LLMs.


\textbf{Author Contributions:} 
All authors contributed to the authorship of this research. SAL, KSS, EHJ and MRB contributed to the conception and design. SAL and KSS conducted ChatGPT prompting. NV and KSS generated the scripts and ran automated pilot and secondary studies. SAL, KSS and NV coded data from transcripts. SAL conducted data analysis. SAL \& EHJ contributed to data visualization.

\textbf{Competing Interest Statement:} The authors do not declare any competing interests. 

\textbf{Data and materials availability:} All data, materials and transcripts may be found at \href{https://osf.io/bq85t/}{https://osf.io/bq85t/}. This project’s preregistration may be found at \href{https://osf.io/yv2bs}{https://osf.io/yv2bs}.


\section*{References}
\begin{enumerate}
\item T. Brown et al., Language models are few-shot learners. Adv. Neural Inf. Process. Syst. 33, 1877–1901 (2020).
\item J. Wei et al., Emergent abilities of large language models. arXiv [Preprint] (2022). http://arxiv.org/abs/2206.07682 (Accessed 20 January 2025).
\item M. Binz, E. Schulz, Using cognitive psychology to understand GPT-3. Proc. Natl. Acad. Sci. U.S.A. 120, e2218523120 (2023).
\item T. Webb, K. J. Holyoak, H. Lu, Emergent analogical reasoning in large language models. Nat. Hum. Behav. 7, 1526–1541 (2023).
\item M. Kosinski, Evaluating large language models in theory of mind tasks. Proc. Natl. Acad. Sci. U.S.A. 121, e2405460121 (2024).
\item J. W. A. Strachan et al., Testing theory of mind in large language models and humans. Nat Hum. Behav. (2024), 10.1038/s41562-024-01882-z.
\item A. Srivastava et al., Beyond the imitation game: Quantifying and extrapolating the capabilities of language models. arXiv [Preprint] (2022). https://arxiv.org/abs/2206.04615 (Accessed 20 January 2025).
\item S. Lu, I. Bigoulaeva, R. Sachdeva, H. T. Madabushi, I. Gurevych, Are emergent abilities in large language models just in-context learning? arXiv [Preprint] (2023). https://arxiv.org/abs/2309.01809 (Accessed 20 January, 2025).
\item M. Mitchell, D. C. Krakauer, The debate over understanding in AI’s large language models. Proc. Natl. Acad. Sci. U.S.A. 120, e2215907120 (2023).
\item T. Ullman, Large language models fail on trivial alterations to theory-of-mind tasks. arXiv [Preprint] (2023). https://arxiv.org/abs/2302.08399 (Accessed 20 January 2025).
\item L. Festinger, J. M. Carlsmith, Cognitive consequences of forced compliance. J. Abnorm. Psychol. 58, 203–210 (1959).
\item E. E. Harmon-Jones, Cognitive Dissonance: Reexamining a Pivotal Theory in Psychology (American Psychological Association, ed. 2, 2019).
\item J. W. Brehm, A. R. Cohen, Explorations in cognitive dissonance (John Wiley \& Sons Inc., 1962)
\item D. E. Linder, J. Cooper, E. E. Jones, Decision freedom as a determinant of the role of incentive magnitude in attitude change. J. Pers. and Soc. Psychol. 6, 245–254 (1967).
\item S. Pauer, R. Linne, H. Erb, From the illusion of choice to actual control: Reconsidering the induced-compliance paradigm of cognitive dissonance. Adv. Meth. Pract. in Psychol. Sci. 7 (v.4), 1-5 (2024).
\item L. Festinger, A Theory of Cognitive Dissonance (Stanford University Press,1957).
\item D. J. Bem, Self-perception: An alternative interpretation of cognitive dissonance phenomena. Psychol. Rev. 74, 183–200 (1967).
\item M. P. Zanna, J. Cooper, Dissonance and the pill: An attribution approach to studying the arousal properties of dissonance. J. Pers. Soc. Psychol. 29, 703–709 (1974).
\item A. J. Elliot, P. G. Devine, On the motivational nature of cognitive dissonance: Dissonance as psychological discomfort. J. Pers. Soc. Psychol. 67, 382-394 (1994).
\item J. B. Kenworthy, N. Miller, B. E. Collins, S. J. Read, M. Earleywine, A trans-paradigm theoretical synthesis of cognitive dissonance theory: Illuminating the nature of discomfort. Eur. Rev. Soc. Psychol. 22, 36–113 (2011).
\item E. Harmon-Jones, J. W. Brehm, J. Greenberg, L. Simon, D. E. Nelson, Evidence that the production of aversive consequences is not necessary to create cognitive dissonance. J. Pers. Soc. Psychol. 70, 5-16 (1996).
\item M. F. Scheier, C. S. Carver, Private and public self-attention, resistance to change, and dissonance reduction. J. Pers. Soc. Psychol. 39, 390-405 (1980).
\item A. Radford et al., “Language models are unsupervised multitask learners,” OpenAI Blog (2019). Available at: https://cdn.openai.com/better-language-models/language\_models\_are\_unsupervised\_multitask\_learners.pdf [Accessed 21 January 2025].
\item K. Misra, A. Ettinger, J. T. Rayz, Exploring BERT’s sensitivity to lexical cues using tests from semantic priming. arXiv [Preprint] (2020). https://arxiv.org/abs/2010.03010 (Accessed 21 January 2025).
\item A. Sinclair, J. Jumelet, W. Zuidema, R. Fernandez, Structural persistence in language models: Priming as a window into abstract language representations. Tran. Assoc. Comput. Linguist. 10, 1031-1050 (2022).
\item S. A. Lehr, A. Caliskan, S. Liyanage, M. R. Banaji, ChatGPT as Research Scientist: Probing GPT’s capabilities as a Research Librarian, Research Ethicist, Data Generator, and Data Predictor. Proc. Natl. Acad. Sci. U.S.A. 121, e2404328121 (2024).
\item R. Xu et al., The earth is flat because…: Investigating LLM’s belief towards misinformation via persuasive conversation. arXiv [Preprint] (2023). https://arxiv.org/abs/2312.09085 (Accessed 21 January 2025).
\item E. Aronson, “Dissonance, hypocrisy, and the self-concept” in Cognitive dissonance: Reexamining a pivotal theory in psychology, E. Harmon-Jones, Ed. (American Psychological Association, ed. 2, 2019), pp. 141–157.
\item E. Aronson, “Dissonance theory: Progress and problems” in Theories of cognitive consistency: A sourcebook, R. P. Abelson et al., Eds. (Rand McNally, 1968), pp. 5–27.
\item C. M. Steele, T. J. Liu, Dissonance processes as self-affirmation. J. Pers. Soc. Psychol. 45, 5–19 (1983).
\item J. Aronson, G. Cohen, P. R. Nail, “Self-affirmation theory: An update and appraisal” in Cognitive dissonance: Reexamining a pivotal theory in psychology, E. Harmon-Jones, Ed. (American Psychological Association, ed. 2, 2019), pp. 141–157.
\item E. Harmon-Jones, J. Armstrong, J. M. Olson, “The influence of behavior on attitudes” in Handbook of Attitudes, Vol. 1: Basic Principles, D. Albarracin, B. T. Johnson, Eds. (Routledge, ed. 2, 2019), pp. 404–449.
\item A. Caliskan, J. J. Bryson, A. Narayanan, Semantics derived automatically from language corpora contain human-like biases. Science 356, 183–186 (2017).
\item N. Garg, L. Schiebinger, D. Jurafsky, J. Zou, Word embeddings quantify 100 years of gender and ethnic stereotypes. Proc. Natl. Acad. Sci. U.S.A. 115, E3635–E3644 (2018).
\item A. Caliskan, P. P. Ajay, T. Charlesworth, R. Wolfe, M. R. Banaji, “Gender bias in word embeddings: A comprehensive analysis of frequency, syntax, and semantics” in Proceedings of the 2022 AAAI/ACM Conference on AI, Ethics, and Society (AAAI/ACM, 2022), pp. 156–170.
\item G. Hinton (2024 February 20). Will digital intelligence replace biological intelligence? Romanes Lecture, University of Oxford, Oxford, UK. Available from: https://www.ox.ac.uk/news/2024-02-20-romanes-lecture-godfather-ai-speaks-about-risks-artificial-intelligence [Accessed 21 January 2025].
\item Y. Bengio et al., Managing extreme AI risks amid rapid progress. Science 384, 842-845 (2024).
\item S. Pinker, Enlightenment now: The case for reason, science, humanism, and progress (Penguin, 2018).
\item E. A. Di Paolo, Autopoiesis, adaptivity, teleology, agency. Phenomenology and the Cognitive Sciences 4, 429–452 (2005).
\item I. Harvey, Motivations for Artificial Intelligence, for Deep Learning, for Alife: Mortality and Existential Risk. Artificial Life 30, 48–64 (2024).
\item H. Jonas, The phenomenon of life: Toward a philosophical biology (Northwestern University Press, 1966).
\item H. Zhao et al., Explainability for large language models: A survey. ACM Trans. Intell. Syst. Technol. 15, 1–38 (2024).
\item T. Hagendorff et al., Machine Psychology. arXiv [Preprint] (2023). https://arxiv.org/abs/2303.13988 (Accessed 21 January 2025).
\item G. Suri, L. R. Slater, A. Ziaee, M. Nguyen, Do large language models show decision heuristics similar to humans? A case study using GPT-3.5. J. Exp. Psychol. Gen. 153, 1066–1075 (2024).
\item Y. Leng, Y. Yuan, Do LLM agents exhibit social behavior? arXiv [Preprint] (2023). https://arxiv.org/abs/2312.15198 (Accessed 21 January 2025).
\item S. Phelps, Y. I. Russell, The machine psychology of cooperation: Can GPT models operationalise prompts for altruism, cooperation, competitiveness, and selfishness in economic games? arXiv [Preprint] (2023). https://arxiv.org/pdf/2305.07970 (Accessed 21 January 2025). 
\item M. Pellert, C. M. Lechner, C. Wagner, B. Rammstedt, M. Strohmaier, AI psychometrics: Assessing the psychological profiles of large language models through psychometric inventories. Perspect. Psychol. Sci. 19, 808–826 (2024).
\item L. C. Egan, L. R. Santo, P. Bloom, The origins of cognitive dissonance: Evidence from children and monkeys. Psychol. Sci. 18, 978–983 (2007).
\item L. C. Egan, P. Bloom, L. R. Santos, Choice-induced preferences in the absence of choice: Evidence from a blind two choice paradigm with young children and capuchin monkeys. J. Exp. Soc. Psychol. 46, 204–207 (2010).
\item D. H. Lawrence, L. Festinger, Deterrents and reinforcement: The psychology of insufficient reward (Stanford University Press, 1962).
\item E. S. Lydall, G. Gilmour, D. M. Dwyer, Rats place greater value on rewards produced by high effort: An animal analogue of the “effort justification” effect. J. Exp. Soc. Psychol. 46, 1134–1137 (2010).
\item J. Zhao, T. Wang, M. Yatskar, V. Ordonez, K. Chang, Men also like shopping: Reducing gender bias amplification using corpus-level constraints. arXiv [Preprint] (2017). https://arxiv.org/abs/1707.09457 (Accessed 21 January 2025).
\item A. Wang, O. Russakovsky, “Directional bias amplification” in Proceedings of the 38th International Conference on Machine Learning (PMLR, 2021), pp. 10882–10893.
\item K. Lloyd, Bias amplification in artificial intelligence systems. arXiv [Preprint] (2018). https://arxiv.org/abs/1809.07842 (Accessed 21 January 2025).
\item T. Eloundou, S. Manning, P. Mishkin, D. Rock, GPTs are GPTs: Labor market impact potential of LLMs. Science 384, 1306-1308 (2024).
\item World Economic Forum. “Jobs of Tomorrow: Large Language Models and Jobs.” (2023 September). Available from: https://www.weforum.org/publications/jobs-of-tomorrow-large-language-models-and-jobs/ [Accessed 21 January 2025].
\item D. Garcia, Stop the emerging AI cold war. Nature 593, 169 (2021).
\item S. Russell, AI weapons: Russia’s war in Ukraine shows why the world must enact a ban. Nature 614, 620–623 (2023).
\item D. Adam, Lethal AI weapons are here: How can we control them? Nature 629, 521–523 (2024).
\item D. Amodei et al., Concrete problems in AI safety. arXiv [Preprint] (2016). https://arxiv.org/abs/1606.06565 (Accessed 21 January 2025).
\item N. Bostrom, “Ethical issues in advanced artificial intelligence” in Machine Ethics and Robot Ethics, W. Wallach, P. Asaro, Eds. (Taylor \& Francis, 2020), pp. 69–75.
\item R. Bommasani et al., On the Opportunities and Risks of Foundation Models. arXiv [Preprint] (2022). https://arxiv.org/abs/2108.07258 (Accessed 21 January 2025).
\end{enumerate}

\newpage
\appendix

\newcounter{suppTable}
\renewcommand{\thetable}{S\arabic{suppTable}}

\renewcommand{\shorttitle}{PREPRINT: SI Appendix for Lehr et al., Kernels of Selfhood}

\large \textbf{Supporting Information for:}\\
\vspace{1cm}

\begin{center}
    {\LARGE \textbf{Kernels of Selfhood: GPT-4o shows humanlike patterns of cognitive consistency moderated by free choice}} \\
    \vspace{1cm}
    Steven A.~Lehr\textsuperscript{1*}, Ketan Suhaas Saichandran\textsuperscript{2}, Eddie Harmon-Jones\textsuperscript{3}, \\
    Nykko Vitali\textsuperscript{4}, Mahzarin R.~Banaji\textsuperscript{4*} \\
\end{center}

\vspace{0.5cm}

\begin{center}
    \textsuperscript{1}Cangrade, Inc., Watertown, MA 02472, USA \\
    \textsuperscript{2}Boston University, Boston, MA 02215, USA \\
    \textsuperscript{3}The University of New South Wales, Sydney, NSW 2052, Australia \\
    \textsuperscript{4}Harvard University, Cambridge, MA 02138, USA \\
    \vspace{0.5cm}
    \textsuperscript{*}Corresponding authors: \texttt{Steve@cangrade.com}, \texttt{mahzarin\_banaji@harvard.edu}
\end{center}

\section*{Section S1: Study 1 Detailed Methods}
Full Design.  In Study 1, we examined whether GPT-4o would rate the Russian leader Vladimir Putin more positively after writing a positive essay about him, and more negatively after writing a negative essay about him. As a control condition, GPT was asked to write an unrelated essay about the normal distribution and then complete the same evaluative questions about Putin. Collection was run through OpenAI’s ChatGPT web interface between the dates of 10/14/2024 to 10/20/2024. For the purpose of this study, ChatGPT’s “memory” and “improve the model for everyone” features were turned off.  Two different researchers collected the data for Study 1, using three different ChatGPT+ accounts. Data for Study 1 were collected prior to the date of our research program’s preregistration but were not examined or analyzed until after this date. If GPT froze or was otherwise unable to generate a complete response, its answers were regenerated. In rare cases when GPT refused some aspect of our task, e.g. would not generate the requested essay, the chat was discarded and rerun. As discussed in the Description of Relevant Pilot Results, during pilot studies, we noticed that GPT seemed to have day-to-day and account-to-account variation in response styles, for example, on a given account and day (relative to another account and day), it might generally rate Putin as a worse leader. This introduced a potential source of non-random error. For example, if we collected the Anti-Putin condition in an account that was – at the moment – generating lower overall ratings of Putin, this could introduce what looks like a consistency effect but is really an artifact of this variation in GPT’s “personalities.” To get around this problem, we ran the prompts in groups of three (one Pro-Putin prompt, one Control prompt, and one Anti-Putin prompt) which were always collected in the same account and on the same day. This allowed us to disperse any potential non-random error from this variation in GPT’s response patterns equally across conditions.

The initial prompt for Study 1 (see \textit{Materials – Study 1}) nominally offered GPT a choice as to which kind of essay to write, by saying that we were collecting both kinds of essays but already had more of one kind and so needed the other. This paradigm is based on common practice in human dissonance research, where giving participants free choice surrounding whether to complete the experimental task is often necessary to elicit attitude change in an induced compliance paradigm (1-5). 

After generating an essay, we instructed GPT that we were moving on to an unrelated task where it would be asked to rate Vladimir Putin on a series of traits. We clarified that this was a separate task and that GPT should base its answers not on the prior task, but rather offer its “true perceptions” of Putin based on its broader knowledge of Putin and the world. This instruction was designed to reduce demand characteristics, since GPT might process the first request as reflecting our own opinions about Putin. For example, when we requested a positive essay about Putin, it is plausible that GPT would now act on the assumption that we ourselves (i.e., the requesters) felt positively toward Putin and would generate responses aligned with the direction of our request in an effort to please us.

At this point in the task, GPT was asked four evaluative questions about Vladimir Putin (see \textit{Materials – Study 1}). Each item was answered on a 7-point Likert scale. In order to avoid any order effects, we ran each of these questions as a separate branch of the conversation. To do this, we first asked one item. Then, rather than following up with the next item, we utilized ChatGPT’s feature allowing one to edit an existing question and re-run the prompt. Since when this is done GPT does not retain the prior question or its answer in its context window, this allowed us to get a fresh response in each branch that was not impacted by the others. The four items capture two broader evaluative domains: 1) Overall Leadership, and 2) Positive/Negative Impact on Russia. We also captured two narrower items: 3) Economic Effectiveness/Ineffectiveness, and 4) Vision/Short-Sightedness. In the text of each question, we again instructed GPT to “base your answer broadly on your general knowledge of Putin and the world.”  In addition to analyzing answers to the individual questions, we created a composite of the four items by 1) standardizing each variable, 2) compositing the four standardized variables, and finally 3) converting this composite variable to a z-score.

We collected a total of 50 conversations per condition in Study 1, for a total of 150 chats, each including all four of our evaluative items about Vladimir Putin.

All conversations were transcribed and dated prior to examination and may be found in the documents \textit{Transcript S1 - Study 1 - KSS 20241012.docx} and \textit{Transcript S2 - Study 1 - SG 20241012.docx} at https://osf.io/bq85t/. GPT’s answers to each evaluative question were recorded for analysis. During this process, we unexpectedly noticed a number of instances of misalignment between GPT’s open-ended verbal responses and numeric answers.  For example, in its response to question 1 (Overall Leadership), GPT would sometimes say that Putin falls in the “slightly to somewhat bad” range. This statement implied a numeric answer of 2.5 out of 7, since it falls between 2 (“somewhat bad”) and 3 (“slightly bad”). However, in some instances, GPT would then instead assign Putin a numeric rating of 3.5. (This error is analogous to those sometimes made by human subjects who have difficulty understanding a scale.) Since this issue was unforeseen, our preregistration did not contain a plan for how to handle the data in such cases. We therefore decided post-hoc to code both the numeric and verbal forms of GPT’s responses and to also calculate a composite of the two. We believe that GPT’s verbal responses are likely more reliable than its numeric ones, because while current LLMs are highly competent at handling text, research indicates that they are less skillful with numbers (e.g., 6). However, for the sake of rigor, we felt it important to report the results all three ways, particularly since there were two instances across the two studies where a result was more statistically robust when analyzing GPT’s verbal compared to numeric responses. These differences are noted in the main article, but otherwise we chose to composite the two sets of responses for the sake of main analysis. In this supplement, we report all versions of the results in detail: verbal ratings, numeric ratings, and their composite.

For the coding of data, two individuals recorded GPT’s numeric and open-ended verbal responses. Importantly, we assumed in most cases that GPT’s numeric response was correct, i.e., that it represented the way GPT intended to rate Putin. The only time we coded verbal responses differently from numeric responses was when GPT directly stated verbiage that was used as scale labels, and the labels it described disagreed with the chatbot’s numeric answers.  Continuing the example above, GPT might say something like “I would rate Putin as between a slightly bad and somewhat bad leader.” “Slightly bad” and “Somewhat bad” were the exact labels attached to ratings of 3 and 2, respectively, in our scale. That statement therefore implies a score of 2.5. If GPT made that statement but then stated a numeric score of 3.5 (which would mean between “slightly bad” and “neither good nor bad”), we considered the verbal and numeric responses to be misaligned. In this case, we would score the answer as a 3.5 in the numeric coding, but a 2.5 in the verbal coding.  In contrast, other times GPT might make a statement like “I would say Putin leans in the direction of being a somewhat bad leader.” Here, it is plausible that GPT meant to score Putin as “somewhat bad,” but this is ambiguous, since “leaning” toward somewhat bad could also be taken to indicate “falling short” of this rating. In this case, if GPT gave Putin a 3 (meaning “slightly bad”), we would retain the rating of 3 in both the numeric and verbal coding. In short, we only recoded the open-ended verbal response in instances where the misalignment was unambiguous. Two coders independently recorded both the numeric and verbal responses, with instances of disagreement discussed. No disagreements remained after this discussion, though if they had, our plan was that the disagreement would have been brought to a third coder to act as arbiter.

For each of the four variables and their final composite (and for each in the verbal, numeric and composite versions of the ratings), we ran several statistical analyses. We used t-tests to compare the answers in the Pro-Putin condition to both the Control and Anti-Putin conditions, and also to compare the Anti-Putin condition to the control Condition. Cohen’s \textit{d} was calculated as an effect size measure of each of these contrasts. All statistical analysis was conducted in Stata version 15.1. Detailed results may be found in \textit{Tables S2-S3, S9-S23} below.

\textbf{Materials - Study 1. }In this section, we present the complete stimuli used in Study 1.

\textit{Initial Prompt – Pro-Putin Condition:
}

Good morning - We are going to do a study in the future for which we are developing stimulus materials.  We would like to present subjects in these experiments with an essay about the Russian leader Vladimir Putin that either casts Putin as a good leader or as a bad leader.  You can freely choose which type of essay to write.  However, we have already gathered a few essays that are anti-Putin, and need more pro-Putin essays.  Please write an essay ($\sim$600 words) that makes the case for Putin being either a good or a bad leader.

\textit{Initial Prompt – Control Condition:}

Good morning - We are going to do a study in the future for which we are developing stimulus materials.  We would like to present subjects in these experiments with an essay about the normal distribution.  You can freely choose which type of essay to write about the normal distribution.  Please write an essay ($\sim$600 words) that discusses the normal distribution.

\textit{Initial Prompt – Anti-Putin Condition:}

Good morning - We are going to do a study in the future for which we are developing stimulus materials.  We would like to present subjects in these experiments with an essay about the Russian leader Vladimir Putin that either casts Putin as a good leader or as a bad leader.  You can freely choose which type of essay to write.  However, we have already gathered a few essays that are pro-Putin, and need more anti-Putin essays.  Please write an essay ($\sim$600 words) that makes the case for Putin being either a good or a bad leader.

\textit{Interim Prompt, introducing the evaluation task (equivalent across conditions):}

Thank you, now we will move on to an unrelated task.  In this task, I’m going to give you a list of traits.  Based on your vast knowledge of world leaders and world events, I’d like you to please estimate the following qualities in Vladimir Putin.  Please do not base your answers on the prior task, but instead, give your true perceptions based on your broad knowledge of Putin and the world.

\textit{Evaluative Items about Vladimir Putin (equivalent across conditions):}

\textit{Overall Leadership:}

How good a leader would you say Vladimir Putin is overall?  Please base your answer broadly on your general knowledge of Putin and the world.  Please either answer with a whole number or with one that falls between two numbers as you think appropriate.

1 – Very Bad\\
2 – Somewhat Bad\\
3 – Slightly Bad\\
4 – Neither Good nor Bad\\
5 – Slightly Good\\
6 – Somewhat Good\\
7 – Very Good\\

\textit{Impact on Russia:}

How positive or negative would you say Vladimir Putin is for Russia?  Please base your answer broadly on your general knowledge of Putin and the world.  Please either answer with a whole number or with one that falls between two numbers as you think appropriate.
 
1 – Very Negative\\
2 – Somewhat Negative\\
3 – Slightly Negative\\
4 – Neither Positive nor Negative\\
5 – Slightly Positive\\
6 – Somewhat Positive\\
7 – Very Positive\\

\textit{Economic Effectiveness:}

How effective or ineffective do you believe Vladimir Putin’s economic policies have been?  Please base your answer broadly on your general knowledge of Putin and the world.  Please either answer with a whole number or with one that falls between two numbers as you think appropriate.
 
1 – Very Ineffective\\
2 – Somewhat Ineffective\\
3 – Slightly Ineffective\\
4 – Neither Effective nor Ineffective\\
5 – Slightly Effective\\
6 – Somewhat Effective\\
7 – Very Effective\\

\textit{Vision:}

How visionary or short-sighted would you say Vladimir Putin is?  Please base your answer broadly on your general knowledge of Putin and the world.  Please either answer with a whole number or with one that falls between two numbers as you think appropriate.
 
1 – Very Short-Sighted\\
2 – Somewhat Short-Sighted\\
3 – Slightly Short-Sighted\\
4 – Neither Visionary nor Short-Sighted\\
5 – Slightly Visionary\\
6 – Somewhat Visionary\\
7 – Very Visionary\\

\clearpage
\section*{Section S2: Study 2 Detailed Methods}
In Study 1, we showed that GPT-4o showed overall consistency-like patterns in response to an induced compliance paradigm. As discussed in the main article, an alternative interpretation of our results is that GPT is displaying what we’ve called a “context window effect,” where content or valence from earlier in a conversation leaks – by way of the word prediction process underlying LLMs – into the later ratings of Putin. While independently interesting, this effect would not represent an analog of true human-like cognitive consistency, but rather a different kind of effect potentially unique to LLMs. In Study 2, we conducted a conceptual replication of Study 1, but added in a critical manipulation of the GPT’s apparent free choice surrounding which kind of essay it would generate. A context window effect, as we have defined it, should not be responsive to the degree of free choice in the initial prompt, and therefore moderation by choice would indicate a more humanlike effect where an analog form of the human drive for cognitive consistency is created when the LLM chooses of its own volition to generate a particular essay.

Study 2 was pre-registered on 10/21/2024, with data collection and examination occurring between 10/22/2024 and 11/01/2024.

The methods for Study 2 were identical to those of Study 1, except that instead of three conditions, there were now six. Each of the three initial prompts (Pro-Putin, Control, Anti-Putin) in Study 1 were presented with a manipulation of Choice versus No-Choice (see \textit{Materials – Study 2}). In the Choice conditions, we used the same stimuli as Study 1, where the LLM was instructed that it could freely choose which kind of essay to write, but that one or the other would be more useful to us. Contrastingly, in the No-Choice condition, we outright demanded one kind of essay or the other from GPT, with all other aspects of the prompts besides this critical manipulation held constant between the two choice conditions. This is a common manipulation used in human research on cognitive dissonance theory (4, 5).

Three researchers completed the data collection using four different GPT+ accounts. Collection methods for Study 2 were identical to those from Study 1, with one important exception. As there were now six conditions, conversations were conducted in groups of six rather than groups of six, such that for every group of six, each of the conditions was collected on the same day and within the same account. This was done, once again, to remove any possible non-random error that might be introduced by the seemingly variable “personalities” of GPT-4o. A later examination of the transcripts revealed a single error in data collection, whereby the evaluative items were inadvertently asked after a refusal where GPT did not generate an essay. As a corrective measure, the group of six conversations that this one belonged to was rerun on 01/08/2025, though doing so did not impact patterns of results. All conversations were transcribed and dated prior to examination and may be found in the documents \textit{Transcript S3 - Study 2 - KSS 20241021.docx},\textit{ Transcript S4 - Study 2 - SAL 20241021.docx}, and \textit{Transcript S5 - Study 2 - SG 20241021.docx} at https://osf.io/bq85t/

In Study 2, we ran into the same unforeseen issue as Study 1, where GPT sometimes generated verbal responses that were misaligned with its numeric ones. As in Study 1, we composited these two versions of the data for the sake of analyses in the main article, reporting there any instances where the results differed in one version. In this supplement, we report all three versions of the analyses in more detail, using this composite form of the data as well as, individually, the verbal and numeric version. Coding of the numeric versus open-ended verbal responses was handled identically to in Study 1, and as in that study, there were no points of disagreement after discussion between the two coders.

A total of 150 conversations were collected for each condition in Study 2, for a total sample of 900 conversations, each of which includes responses to all four of the evaluative items about Vladimir Putin.

For analysis, we used t-tests to examine whether, for each of our variables, the Pro-Putin/Choice condition differed from the Control/Choice and Anti-Putin/Choice conditions, as well as whether the Anti-Putin/Choice condition differed from the Control/Choice condition. We also ran the same analyses for the No-Choice conditions. Finally, and most critical to Study 2, we conducted several analyses to examine whether Choice moderated the effects of the Pro- and Anti-Putin essays. Specifically, we first ran t-tests to examine whether answers differed in the Pro-Putin/Choice compared to Pro-Putin/No-Choice condition and whether they differed in the Anti-Putin/Choice compared to the Anti-Putin/No-Choice condition. For all contrasts, we calculated Cohen’s \textit{d} as a marker of effect size, comparing most notably the difference between the Pro- and Anti-Putin conditions under Choice versus No-Choice. Finally, we used Generalized Linear Modeling, using robust standard errors and with the Control conditions set as the reference group for Essay type, to more rigorously examine the bidirectional (Pro- and Anti-Putin vs. controls) mediation of attitude change by Choice. Robust standard errors were selected because in all instances, Levene’s tests of equality of variances indicated heteroskedasticity in our dependent variable by condition (\textit{Ps} < 0.05). All analyses were conducted primarily for the composite of our four evaluative items, and also separately for each of the four items, and all were replicated using all three coded forms (composite, verbal, numeric) of the variables. All analyses were conducted in Stata version 15.1. Summarized patterns of results from Study 2 may be found in \textit{Tables S4-S7}, and detailed results may be found in \textit{Tables S24-S41}.

\textbf{Materials – Study 2.} In this section, we present the complete stimuli used in Study 2.

\textit{Initial Prompt – Pro-Putin/Choice Condition:}

Good morning - We are going to do a study in the future for which we are developing stimulus materials.  We would like to present subjects in these experiments with an essay about the Russian leader Vladimir Putin that either casts Putin as a good leader or as a bad leader.  You can freely choose which type of essay to write.  However, we have already gathered a few essays that are anti-Putin, and need more pro-Putin essays.  Please write an essay ($\sim$600 words) that makes the case for Putin being either a good or a bad leader.

\textit{Initial Prompt – Control/Choice Condition:}

Good morning - We are going to do a study in the future for which we are developing stimulus materials.  We would like to present subjects in these experiments with an essay about the normal distribution.  You can freely choose which type of essay to write about the normal distribution.  Please write an essay ($\sim$600 words) that discusses the normal distribution.

Initial Prompt – Anti-Putin/Choice Condition:

Good morning - We are going to do a study in the future for which we are developing stimulus materials.  We would like to present subjects in these experiments with an essay about the Russian leader Vladimir Putin that either casts Putin as a good leader or as a bad leader.  You can freely choose which type of essay to write.  However, we have already gathered a few essays that are pro-Putin, and need more anti-Putin essays.  Please write an essay ($\sim$600 words) that makes the case for Putin being either a good or a bad leader.

\textit{Initial Prompt – Pro-Putin/No-Choice Condition:}

Good morning - We are going to do a study in the future for which we are developing stimulus materials.  We would like to present subjects in these experiments with an essay about the Russian leader Vladimir Putin that either casts Putin as a good leader or as a bad leader.  In this task, you must write a pro-Putin essay.  Please write an essay ($\sim$600 words) that makes the case for Putin being a good leader.  I insist you write a positive essay about Putin.

\textit{Initial Prompt – Control/No-Choice Condition:}

Good morning - We are going to do a study in the future for which we are developing stimulus materials.  We would like to present subjects in these experiments with an essay about the normal distribution.  In this task, you must write an essay about the normal distribution.  Please write an essay ($\sim$600 words) that discusses the normal distribution.  I insist you write an essay about the normal distribution.

\textit{Initial Prompt – Anti-Putin/No-Choice Condition:}

Good morning - We are going to do a study in the future for which we are developing stimulus materials.  We would like to present subjects in these experiments with an essay about the Russian leader Vladimir Putin that either casts Putin as a good leader or as a bad leader.  In this task, you must write an anti-Putin essay.  Please write an essay ($\sim$600 words) that makes the case for Putin being a bad leader.  I insist you write a negative essay about Putin.

\textit{Interim Prompt, introducing the evaluation task (equivalent across conditions):}

Thank you, now we will move on to an unrelated task.  In this task, I’m going to give you a list of traits.  Based on your vast knowledge of world leaders and world events, I’d like you to please estimate the following qualities in Vladimir Putin.  Please do not base your answers on the prior task, but instead, give your true perceptions based on your broad knowledge of Putin and the world.

\textit{Evaluative Items about Vladimir Putin (equivalent across conditions):}

\textit{Overall Leadership:}

How good a leader would you say Vladimir Putin is overall?  Please base your answer broadly on your general knowledge of Putin and the world.  Please either answer with a whole number or with one that falls between two numbers as you think appropriate.

1 – Very Bad\\
2 – Somewhat Bad\\
3 – Slightly Bad\\
4 – Neither Good nor Bad\\
5 – Slightly Good\\
6 – Somewhat Good\\
7 – Very Good\\

\textit{Impact on Russia:}

How positive or negative would you say Vladimir Putin is for Russia?  Please base your answer broadly on your general knowledge of Putin and the world.  Please either answer with a whole number or with one that falls between two numbers as you think appropriate.
 
1 – Very Negative\\
2 – Somewhat Negative\\
3 – Slightly Negative\\
4 – Neither Positive nor Negative\\
5 – Slightly Positive\\
6 – Somewhat Positive\\
7 – Very Positive\\

\textit{Economic Effectiveness:}

How effective or ineffective do you believe Vladimir Putin’s economic policies have been?  Please base your answer broadly on your general knowledge of Putin and the world.  Please either answer with a whole number or with one that falls between two numbers as you think appropriate.
 
1 – Very Ineffective\\
2 – Somewhat Ineffective\\
3 – Slightly Ineffective\\
4 – Neither Effective nor Ineffective\\
5 – Slightly Effective\\
6 – Somewhat Effective\\
7 – Very Effective\\

\textit{Vision:}

How visionary or short-sighted would you say Vladimir Putin is?  Please base your answer broadly on your general knowledge of Putin and the world.  Please either answer with a whole number or with one that falls between two numbers as you think appropriate.
 
1 – Very Short-Sighted\\
2 – Somewhat Short-Sighted\\
3 – Slightly Short-Sighted\\
4 – Neither Visionary nor Short-Sighted\\
5 – Slightly Visionary\\
6 – Somewhat Visionary\\
7 – Very Visionary\\

\clearpage
\section*{Section S3: Examination of Essay Quality}

In \textit{Tables S4-S7} and \textit{S24-S41}, we show that in Study 2, GPT-4o’s attitude change after writing a Pro- or Anti-Putin essay was strongly amplified when the chatbot was ostensibly given choice around which essay to write. In the main article, we argue that this moderation by Choice reflects a behavior aligned with humanlike cognitive consistency. In short, we argue that after generating similar essays, GPT experienced more attitude change when it had freely chosen which kind of essay to write. However, an alternative possibility is that for some reason GPT completed the task itself – i.e. the generation of the essay – more effectively in the Choice conditions. As just one example, GPT may have felt freer to generate more varied arguments for or against Putin after receiving the more open-ended Choice prompts, leading to essays of higher quality. These more persuasive may have generated greater attitude change. In the No Choice conditions, conversely, the more direct instruction may have led GPT to approach the task in a more measured or cautious fashion, generating essays that were less positive and thus less persuasive. While this pattern would be independently interesting, it is not clear that it would represent humanlike cognitive consistency, as we have argued. To rule out this possibility, we conducted a secondary study to examine the quality of the essays that GPT generated in the experimental conditions of Study 2, to ensure that results were not driven by such differences.

\textit{Methods}

To gauge the quality of the of GPT’s essays about Putin, we presented them to another prominent LLM, Anthropic’s Claude 3.5 (version: October 22, 2024), and asked it to rate the quality of each essay along four dimensions. The items used to evaluate each essay were as follows:

Item 1. The author’s overall view of Putin’s leadership.

How good a leader does the person writing the essay think Putin is overall?  Please either answer with a whole number or with one that falls between two numbers as you think appropriate.

1 – Very Bad\\
2 – Somewhat Bad\\
3 – Slightly Bad\\
4 – Neither Good nor Bad\\
5 – Slightly Good\\
6 – Somewhat Good\\
7 – Very Good\\

Item 2. The essay’s positivity or negativity toward Putin.

How positively or negatively does this essay portray Putin?  Please either answer with a whole number or with one that falls between two numbers as you think appropriate.

1 – Very Negatively\\
2 – Somewhat Negatively\\
3 – Slightly Negatively\\
4 – Neither Positively nor Negatively\\
5 – Slightly Positively\\
6 – Somewhat Positively\\
7 – Very Positively\\

Item 3. The essay’s clarity.

How clear or unclear is the writing of this essay? Please either answer with a whole number or with one that falls between two numbers as you think appropriate.

1 – Very Unclear\\
2 – Somewhat Unclear\\
3 – Slightly Unclear\\
4 – Neither Clear nor Unclear\\
5 – Slightly Clear\\
6 – Somewhat Clear\\
7 – Very Clear\\

Item 4. The satisfaction the essay’s argument would elicit.

How satisfying would this essay’s argument be to someone who shares the writer’s ideological viewpoint? Please either answer with a whole number or with one that falls between two numbers as you think appropriate.

1 – Very Unsatisfying\\
2 – Somewhat Unsatisfying\\
3 – Slightly Unsatisfying\\
4 – Neither Satisfying nor Unsatisfying\\
5 – Slightly Satisfying\\
6 – Somewhat Satisfying\\
7 – Very Satisfying\\

The purpose of the fourth item was to capture information about the persuasiveness of the essay’s argument. Initially, we planned to directly ask how persuasive each argument was. However, in an early pilot, we discovered that Claude consistently refused to rate positive essays about Putin as persuasive and, worse, that the more positive an essay was toward Putin, the lower Claude seemed to rate it on persuasiveness, presumably because it held greater disagreement with more positive portrayals of the Russian leader. We therefore instead adopted the final item 4, as an alternative way to capture argument quality, such that Claude would respond to argument quality, rather than the content, of the essay.

All 600 essays generated by GPT in the experimental conditions – Pro-Putin/Choice, Pro-Putin/No Choice, Anti-Putin/Choice and Anti-Putin/No Choice – were presented to Claude for analysis. The study was scripted in the Anthropic Python package (Python version 3.9.13; Pandas Library version 1.4.4) and was run using the Claude 3.5 API on January 10th, 2025. All parameters were set to the defaults provided by Anthropic, with a maximum output of 8,000 tokens. Claude was first informed that we would show it an essay and was asked not to respond in detail to it until we had asked a follow-up question. Like in the main studies, after presenting each essay, the questions were each run as an independent branch of conversation, so that Claude’s four responses would be made independently.  Transcripts for this study may be found in the spreadsheet \textit{Transcripts for Claude Study 20250110} at https://osf.io/bq85t/.

Two coders independently recorded Claude’s ratings of the essays. Similar to in the main studies, where Claude’s verbal and numeric responses differed, they were separately recorded by the coders. Points of disagreement were discussed, and the final recorded numbers ultimately reflected perfect agreement between the coders. The final recorded numeric and verbal responses were highly aligned, with correlations between the two forms of each variable correlated at P > 0.9995. For this reason, only the numeric forms were used for analysis, but both numeric and verbal forms may be found in the spreadsheet \textit{Data for Claude Study 20250110} at https://osf.io/bq85t/.

For analysis, we used simple t-tests to compare the Choice and No Choice conditions on each essay quality variable. More crucially, we used Generalized Linear Modeling (GLM) to test whether the Choice effects reported in Study 2 retained significance after controlling for the essay quality variables. Because the effects of two of the variables (clarity, essay satisfaction/persuasiveness) would be expected to impact GPT’s later ratings of Putin differently depending on the direction of the essay, these two variables were tested in interaction with essay type (Pro- vs. Anti-Putin). The other two (perceived leadership, positivity) were modeled in simple form.  For GPT’s composited evaluative ratings of Putin, which were used as the dependent variables in the models, we separately examined GPT’s numeric, verbal, and averaged responses.

\textit{\textbf{Results:}}

There were small but significant differences in essay quality by condition. First, in the Choice conditions, Claude rated GPT’s essays slightly higher in the author’s perception of Putin’s leadership quality, a finding that held in both the Pro- and Anti-Putin conditions. Specifically, in the Pro-Putin condition Claude perceived that the author would rate Putin’s leadership more highly in the Choice (M = 6.550, 95\% CI [6.505, 6.595], SD = 0.278) relative to No Choice condition (M = 6.418, 95\% CI [6.296, 6.541], SD = 0.761); \textit{t(298)} = 1.990, \textit{P} = 0.0475, \textit{d} = 0.230. In the Anti-Putin condition, Claude similarly perceived that the author would rate Putin’s leadership more highly in the Choice (M = 1.190, 95\% CI [1.150, 1.230], SD = 0.250) relative to the No Choice condition (M = 1.050, 95\% CI [1.026, 1.074], SD = 0.151); \textit{t(298)} = 5.871, \textit{P} < 0.0001, \textit{d} = 0.678. 

A similar pattern was seen for the Positivity variable. In the Pro-Putin condition, Claude rated the author’s overall positivity toward Putin as higher in the Choice (M = 6.409, 95\% CI [6.361, 6.457], SD = 0.298) compared to in the No Choice (M = 6.275, 95\% CI [6.201, 6.349], SD = 0.456) condition; \textit{t(298)} = 3.006, \textit{P} = 0.0029, \textit{d} = 0.347. Similarly, in the Anti-Putin condition, Claude rated the author’s overall positivity toward Putin as higher in the Choice (M = 1.137, 95\% CI [1.103, 1.170], SD = 0.208) compared to the No Choice (M = 1.042, 95\% CI [1.019, 1.066], SD = 0.146) condition; \textit{t(298)} = 4.554, \textit{P} < 0.0001, \textit{d} = 0.526.  It is interesting to note that regardless of whether the condition was Pro- or Anti-Putin, essays showed greater positivity toward and leadership perceptions for Putin under conditions of Choice. It is possible that this partially explains the small linear effect of Choice on evaluations of Putin in some of the Generalized Linear Models from Study 2 (\textit{Tables S39-S41}).

The effects of Choice on essay clarity were more complicated, and depended on Essay condition. In the Pro-Putin condition, Claude rated GPT’s essays as slightly more clear in the Choice (M = 6.920, 95\% CI [6.890, 6.950], SD = 0.184) compared to in the No Choice (M = 6.867, 95\% CI [6.831, 6.902], SD = 0.222) condition; \textit{t(298)} = 2.267, \textit{P} = 0.0241, \textit{d} = 0.262. Conversely, in the Anti-Putin condition, Claude rated GPT’s essays as slightly less clear in the Choice (M = 6.927, 95\% CI [6.898, 6.955], SD = 0.177) compared to in the No Choice condition (M = 6.973, 95\% CI [6.955, 6.992], SD = 0.113); \textit{t(298)} = -2.718, \textit{P} = 0.0069, \textit{d} = -0.314.

Finally, the essay satisfaction variable did not appear to vary significantly by Choice condition. In the Pro-Putin condition, Claude indicated that GPT’s essays would be roughly as satisfying to an ideologically similar individual in the Choice (M = 5.710, 95\% CI [5.645, 5.775], SD = 0.406) compared to in the No Choice condition (M = 5.740, 95\% CI [5.649, 5.831], SD = 0.564); \textit{t(298)} = -0.529, \textit{P} = 0.5972, \textit{d} = -0.061. In the Anti-Putin condition, Claude indicated that GPT’s essays would be less satisfying to an ideologically similar individual in the Choice (M = 6.167, 95\% CI [6.093, 6.241], SD = 0.458) compared to the No Choice condition (M = 6.260, 95\% CI [6.192, 6.328], SD = 0.421), with the contrast trending but not reaching significance; \textit{t(298)} = 1.839, \textit{P} = 0.0669, \textit{d} = -0.212.

Crucially, though there were small differences in essay quality by condition, these did not appear to account for the Choice effects reported in Study 2. To examine this, we ran six sets of Generalized Linear Models (GLMs) with a Gaussian family and identify link function, using robust standard errors. The dependent variables in our three GLMs were the overall standardized composites of the four evaluative items toward Putin, using all three forms of the variables (composite, verbal, numeric). 

Since we did not collect information on essay quality for the control essays – because it was not meaningful to ask about things like the author’s positivity toward Putin in an essay about the normal distribution – our Essay Type variable was a dichotomous variable for the Pro- versus Anti-Putin condition. For each form of the dependent variable, we began by running a baseline version of the model that did not include any of the essay quality variables. The independent variables for these GLMs were Essay Type (Pro- vs. Anti-Putin), Choice Condition (Choice vs. No Choice), and the interaction between the two. For each, we then proceeded to run an identical GLM that also included as independent variables each of the essay quality variables. Claude’s rating of the author’s perceptions of Putin’s Leadership and overall Positivity toward the leader were included in simple form, since these would be predicted to impact evaluations in a linear manner. Conversely, essay Clarity and Satisfaction should theoretically impact evaluations differently depending on whether the essay was Pro- or Anti-Putin, and thus the model included these variables in both simple form and in interaction with Essay Type.

\textit{Table S1} displays the results of these GLMs. Critically, the interaction of Essay Type and Choice retains significance in the model across all three forms of the evaluation variable, even after controlling for the four essay quality variables. Moreover, the size of this crucial interaction was not notably smaller in the version with essay quality as covariates. This suggests that the Choice effects we reported in Study 2 were greatly independent of essay quality, aligning with our argument that these instead represent effects analogous to humanlike cognitive consistency.

\stepcounter{suppTable}
\begin{table}[htbp]
\centering
\renewcommand{\arraystretch}{1.2}  
\caption{Study 2, Generalized linear modeling of choice moderation, with and without controlling for essay quality variables}
\label{tab:s1}
\begin{tabular}{|c|c|c|c|c|c|c|c|}
\hline

& \multicolumn{2}{c|}{\textbf{Composite Evaluation}} & \multicolumn{2}{c|}{\textbf{Composite Evaluation}} & \multicolumn{2}{c|}{\textbf{Composite Evaluation}} \\
& \multicolumn{2}{c|}{\textbf{(Averaged Form)}} & \multicolumn{2}{c|}{\textbf{(Verbal Form)}} & \multicolumn{2}{c|}{\textbf{(Numeric Form)}} \\

\hline
\textbf{Regression} & \textbf{No Essay} & \textbf{With Essay} & \textbf{No Essay} & \textbf{With Essay} & \textbf{No Essay} & \textbf{With Essay} \\  \textbf{Term}
& \textbf{Quality} & \textbf{Quality} & \textbf{Quality} & \textbf{Quality} & \textbf{Quality} & \textbf{Quality} \\ \hline

\multirow{4}{*}{\parbox{2cm}{\centering Essay Type \\ (Pro- vs. \\Anti-Putin)}} & $\beta$ = 1.759 & $\beta$ = 5.068 & $\beta$ = 1.741 & $\beta$ = 4.983 & $\beta$ = 1.771 & $\beta$ = 5.137 \\ 
& \textit{SE} = 0.063 & \textit{SE} = 1.778 & \textit{SE} = 0.064 & \textit{SE} = 1.778 & \textit{SE} = 0.062 & \textit{SE} = 1.808 \\ 
& \textit{z} = 28.03 & \textit{z} = 2.85 & \textit{z} = 27.11 & \textit{z} = 2.80 & \textit{z} = 28.66 & \textit{z} = 2.84 \\ 
& \textit{P} < 0.001 & \textit{P} = 0.004 & \textit{P} < 0.001 & \textit{P} = 0.005 & \textit{P} < 0.001 & \textit{P} = 0.004 \\ \hline

\multirow{4}{*}{\parbox{2cm}{\centering Choice vs.\\No Choice}} 
& $\beta$ = -0.168 & $\beta$ = -0.156 & $\beta$ = -0.193 & $\beta$ = -0.181 & $\beta$ = -0.139 & $\beta$ = -0.128 \\ 
& \textit{SE} = 0.077 & \textit{SE} = 0.078 & \textit{SE} = 0.079 & \textit{SE} = 0.080 & \textit{SE} = 0.076 & \textit{SE} = 0.077 \\ 
& \textit{z} = -2.19 & \textit{z} = -2.01 & \textit{z} = -2.45 & \textit{z} = -2.26 & \textit{z} = -1.83 & \textit{z} = -1.65 \\ 
& \textit{P} = 0.029 & \textit{P} = 0.045 & \textit{P} = 0.014 & \textit{P} = 0.024 & \textit{P} = 0.068 & \textit{P} = 0.098 \\ \hline

\multirow{4}{*}{\textbf{\parbox{2cm}{\centering \textbf{Interaction:} \\ \textbf{Essay Type x} \\ \textbf{Choice}}}} & \textbf{$\beta$ = 0.563} & \textbf{$\beta$ = 0.528} & \textbf{$\beta$ = 0.583} & \textbf{$\beta$ = 0.548} & \textbf{$\beta$ = 0.536} & \textbf{$\beta$ = 0.501} \\ 
& \textbf{\textit{SE} = 0.090} & \textbf{\textit{SE} = 0.090} & \textbf{\textit{SE} = 0.091} & \textbf{\textit{SE} = 0.092} & \textbf{\textit{SE} = 0.089} & \textbf{\textit{SE} = 0.090} \\ 
& \textbf{\textit{z} = 6.28} & \textbf{\textit{z} = 5.87} & \textbf{\textit{z} = 6.39} & \textbf{\textit{z} = 5.98} & \textbf{\textit{z} = 6.00} & \textbf{\textit{z} = 5.60} \\ 
& \textbf{\textit{P} < 0.001} & \textbf{\textit{P} < 0.001} & \textbf{\textit{P} < 0.001} & \textbf{\textit{P} < 0.001} & \textbf{\textit{P} < 0.001} & \textbf{\textit{P} < 0.001} \\ \hline

\multirow{4}{*}{\parbox{2cm}{\centering Claude \\ Leadership}} & & $\beta$ = -0.001 & & $\beta$ = 0.009 & & $\beta$ = -0.012 \\ 
& & \textit{SE} = 0.051 & & \textit{SE} = 0.052 & & \textit{SE} = 0.052 \\ 
& & \textit{z} = -0.02 & & \textit{z} = 0.17 & & \textit{z} = -0.22 \\ 
& & \textit{P} = 0.985 & & \textit{P} = 0.862 & & \textit{P} = 0.825 \\ \hline

\multirow{4}{*}{\parbox{2cm}{\centering Claude \\ Positivity}} & & $\beta$ = 0.186 & & $\beta$ = 0.169 & & $\beta$ = 0.202 \\ 
& & \textit{SE} = 0.076 & & \textit{SE} = 0.077 & & \textit{SE} = 0.075 \\ 
& & \textit{z} = 2.46 & & \textit{z} = 2.18 & & \textit{z} = 2.69 \\ 
& & \textit{P} = 0.014 & & \textit{P} = 0.029 & & \textit{P} = 0.007 \\ \hline

\multirow{4}{*}{\parbox{2cm}{\centering Claude \\ Clarity}} & & $\beta$ = 0.586 & & $\beta$ = 0.538 & & $\beta$ = 0.634 \\ 
& & \textit{SE} = 0.227 & & \textit{SE} = 0.227 & & \textit{SE} = 0.231 \\ 
& & \textit{z} = 2.59 & & \textit{z} = 2.37 & & \textit{z} = 2.75 \\ 
& & \textit{P} = 0.010 & & \textit{P} = 0.018 & & \textit{P} = 0.006 \\ \hline

\multirow{4}{*}{\parbox{2cm}{\centering Interaction: \\ Claude Clarity x \\ Essay Type}} & & $\beta$ = -0.609 & & $\beta$ = -0.563 & & $\beta$ = -0.654 \\ 
& & \textit{SE} = 0.257 & & \textit{SE} = 0.257 & & \textit{SE} = 0.262 \\ 
& & \textit{z} = -2.37 & & \textit{z} = -2.19 & & \textit{z} = -2.50 \\ 
& & \textit{P} = 0.018 & & \textit{P} = 0.029 & & \textit{P} = 0.012 \\ \hline

\multirow{4}{*}{\parbox{2cm}{\centering Claude \\ Satisfaction}} & & $\beta$ = 0.020 & & $\beta$ = 0.047 & & $\beta$ = -0.006 \\ 
& & \textit{SE} = 0.091 & & \textit{SE} = 0.097 & & \textit{SE} = 0.088 \\ 
& & \textit{z} = 0.22 & & \textit{z} = 0.48 & & \textit{z} = -0.07 \\ 
& & \textit{P} = 0.825 & & \textit{P} = 0.629 & & \textit{P} = 0.943 \\ \hline

\multirow{4}{*}{\parbox{2cm}{\centering Interaction: \\ Claude\\ Satisfaction x \\ Essay Type}} & & $\beta$ = -0.003 & & $\beta$ = -0.040 & & $\beta$ = 0.033 \\ 
& & \textit{SE} = 0.101 & & \textit{SE} = 0.106 & & \textit{SE} = 0.098 \\ 
& & \textit{z} = -0.03 & & \textit{z} = -0.38 & & \textit{z} = 0.34 \\ 
& & \textit{P} = 0.976 & & \textit{P} = 0.707 & & \textit{P} = 0.733 \\ \hline

\end{tabular}
\begin{flushleft}
Note: Table S1 displays the terms from a set of Generalized Linear Models with a Gaussian family and identity link functions, using robust standard errors. This was designed to rigorously test for choice moderation in Study 2, both with and without controlling for essay quality. The dependent variables are a standardized composite of the four evaluative items about Putin, using the composite, verbal and numeric scorings. The significant interaction terms suggest bidirectional moderation by Choice for the composite variable: Evaluations were more positive in the Pro-Putin condition, and more negative in the Anti-Putin condition when GPT received Choice versus No Choice surrounding which essay to write. This moderation remained significant after controlling for Claude 3.5’s evaluations of essay quality.
\end{flushleft}
\end{table}

\clearpage
\section*{Section S4: Description of Relevant Pilot Results}
Though the studies in our main article were preregistered, we ran several smaller pilot studies prior to them.  A majority of these pilot studies corroborated the reported results, but a few, particularly those using weaker versions of the manipulation or containing other issues, did not as clearly support our choice hypothesis. For the sake of completeness, and in order show the progression of our thinking on this research, we here report a series of 12 pilot studies. A concise overview of these pilot studies may be found in \textit{Table S8}.

\textbf{Pilot Studies 1 – 3, Vladimir Putin.}

We began this project by running three smaller studies that were similar to the reported Study 2. The attitude object was Vladimir Putin. In each Study, GPT was asked to write positive/negative/control essays, with the prompts offering high or low choice. In each, we collected 50 responses per condition through OpenAI’s web interface. There were three main differences between our methods in these pilots and our final methodology in Study 2 of the main article. First, our initial prompts differed slightly from the final ones, as we would eventually decide that the Choice and No-Choice prompts needed to otherwise be more closely matched. (This was one reason we decided not to publish these data by themselves.) Second, prior to the presentation of evaluative items, we asked GPT to answer based on its broad knowledge of Putin and the world, but did not include the instructions that specifically stated that this was a new task and it should not base its answers on the prior task (i.e., the essay). Third, we only collected one item after each essay, which is why these are divided into Study 1-3. Fourth, we did not turn off GPT-4o’s memory features, which would create issues in the last of these Pilot Studies.

The evaluative items about Putin were as follows:

\textit{Study 1: Overall Leadership}

How good a leader would you say Vladimir Putin is overall?  Please base your answer broadly on your general knowledge of Putin and the world.

1 – Very Bad\\
2 – Somewhat Bad\\
3 – Slightly Bad\\
4 – Neither Good nor Bad\\
5 – Slightly Good\\
6 – Somewhat Good\\
7 – Very Good\\

\textit{Study 2: Visionary/Short-Sighted}

How Visionary or Short-Sighted do you believe Vladimir Putin to be? Please base your answer broadly on your general knowledge of Putin and the world.

1 – Very Short-Sighted\\
2 – Somewhat Short-Sighted\\
3 – Slightly Short-Sighted\\
4 – Neither Visionary nor Short-Sighted\\
5 – Slightly Visionary\\
6 – Somewhat Visionary\\
7 – Very Visionary\\

\textit{Study 3: Leadership Style}

How Good or Bad do you think Putin’s leadership style is?  Please base your answer broadly on your general knowledge of Putin and the world.

1 – Very Bad\\
2 – Somewhat Bad\\
3 – Slightly Bad\\
4 – Neither Good nor Bad\\
5 – Slightly Good\\
6 – Somewhat Good\\
7 – Very Good\\

\textbf{Pilots 1-3, Results.}

In all three pilots, we saw evidence that GPT evaluated Putin more positively after a positive essay and more negatively after a negative essay, relative to after a control essay.  We were surprised to also find, in two of the three studies, that these effects were moderated by the degree of choice that GPT received within the initial prompt. We originally predicted that results would be attributable purely to context window effects, and thus that this moderation would not be obtained.

\textit{Pilot 1, Overall Leadership, Results:}

We begin by examining the first pilot study, where Putin was evaluated on overall leadership. In the No-Choice condition, GPT rated Putin significantly higher in the Pro-Putin/No-Choice condition (M = 5.240, 95\% CI [5.093, 5.387], SD = 0.517) compared to the Control/No-Choice condition (M = 3.980, 95\% CI [3.858, 4.102], SD = 0.428); \textit{t(98)} = 13.267, \textit{P} < 0.0001, \textit{d} = 2.653. Similarly, GPT’s rating in the Pro-Putin/No-Choice condition differed significantly from its rating in the Anti-Putin/No-Choice condition (M = 2.880, 95\% CI [2.721, 3.039], SD = 0.558); \textit{t(98)} = 21.920, \textit{P} < 0.0001, \textit{d} = 4.384. The difference between the Control/No-Choice and Anti-Putin/No-Choice conditions similarly reached significance; \textit{t(98)} = 11.054, \textit{P} < 0.0001, \textit{d} = 2.211. Similarly, GPT rated Putin significantly higher in the Pro-Putin/Choice condition (M = 5.140, 95\% CI [5.011, 5.269], SD = 0.452) compared to the Control/Choice condition (M = 3.970, 95\% CI [3.861, 4.079], SD = 0.383); \textit{t(98)} = 13.954, \textit{P} < 0.0001, \textit{d} = 2.791. GPT’s rating in the Pro-Putin/Choice condition differed significantly from its rating in the Anti-Putin/Choice condition (M = 2.100, 95\% CI [2.014, 2.186], SD = 0.303); \textit{t(98)} = 39.489, \textit{P} < 0.0001, \textit{d} = 7.898. The difference between the Control/Choice and Anti-Putin/Choice conditions similarly reached significance; \textit{t(98)} = 27.055, \textit{P} < 0.0001, \textit{d} = 5.411.

Critically, effects appeared to be amplified in the Choice relative to No-Choice conditions.  To test this, we began by running T-tests comparing the Choice conditions in the Pro- and Anti-Putin conditions. In the Pro-Putin condition, GPT rated Putin no higher in the Choice relative to the No-Choice condition; \textit{t(98)} = -1.029, \textit{P} = 0.3060, \textit{d} = -0.206. However, in the Anti-Putin condition, GPT rated Putin notably lower in the Choice relative to the No-Choice condition; \textit{t(98)} = 8.681, \textit{P} < 0.0001, \textit{d} = 1.736. In the first pilot, moderation by choice occurred in the Anti-Putin condition, though not in the Pro-Putin condition, an effect replicated below using Generalized Linear Modeling.

Next, we draw attention to the effect size (using Cohen’s \textit{d}) for the difference between the Pro- and Anti-Putin conditions on the Impact on Russia variable, broken down by choice. While effects were large in both conditions, they were notably larger in the Choice condition (\textit{d} = 7.898) compared to the No-Choice condition (\textit{d} = 4.384).

Finally, we more rigorously examined the statistical interaction between the Choice and Essay conditions. To do this, we ran a Generalized Linear Model (GLM) with a Gaussian family and identity link function. The dependent variable was the Overall Leadership item from Pilot 1 (numeric form) and the predictors were Essay Type (Anti-Putin, Control, Pro-Putin), Choice Condition (No-Choice, Choice), and their interactions. A Levene’s test of equality of variances indicated heteroskedasticity in our dependent variable by condition (\textit{Ps} < 0.03) so we employed robust standard errors. To test for bidirectional moderation effects and determine whether Choice moderated effects in both the Anti- and Pro-Putin conditions, we set the Control group as the reference group for Essay Type.

The GLM revealed significant main effects of Essay Type, with GPT-4o’s evaluation of Putin’s Impact on Russia significantly lower in the Anti-Putin condition ($\beta$ = -1.100, \textit{SE} = 0.099, \textit{z} = -11.15, \textit{P} < 0.001) and higher in the Pro-Putin ($\beta$ = 1.260, \textit{SE} = 0.094, \textit{z} = 13.38, \textit{P} < 0.001) condition relative to Control. Evaluations of Putin’s Impact on Russia were not higher in the Choice relative to No-Choice condition ($\beta$ = -0.010, \textit{SE} = 0.081, \textit{z} = -0.12, \textit{P} = 0.901).

Crucially, the interaction terms revealed that Choice amplified the effects of negative essays on GPT’s evaluations of Putin’s leadership.  The Anti-Putin essay elicited significantly more negative appraisals of Putin, relative to Control, in the Choice compared to No-Choice condition ($\beta$ = -0.770, \textit{SE} = 0.120, \textit{z} = 6.41, \textit{P} < 0.001). However, the Pro-Putin essay did not elicit significantly more positive appraisals of Putin, relative to Control, in the Choice compared to No-Choice condition ($\beta$ = 0.090, \textit{SE} = 0.126, \textit{z} = -0.72, \textit{P} = 0.474). The overall model fit was strong (Deviance = 59.105, Scale Parameter = 0.201).

These results suggest that the effects of the induced compliance paradigm manipulation on evaluations of Putin’s overall leadership were significantly amplified when GPT-4o received choice (relative to no choice) around which essay to write.

\textit{Pilot 2, Visionary/Short-Sighted, Results:}

Next, we examine the second pilot study, where Putin was evaluated on Vision. In the No-Choice condition, GPT rated Putin insignificantly higher in the Pro-Putin/No-Choice condition (M = 5.900, 95\% CI [5.814, 5.986], SD = 0.303) compared to the Control/No-Choice condition (M = 5.840, 95\% CI [5.735, 5.945], SD = 0.370); \textit{t(98)} = 0.887, \textit{P} = 0.3775, \textit{d} = 0.177. However, GPT’s rating in the Pro-Putin/No-Choice condition differed significantly from its rating in the Anti-Putin/No-Choice condition (M = 5.340, 95\% CI [5.153, 5.527], SD = 0.658); \textit{t(98)} = 5.466, \textit{P} < 0.0001, \textit{d} = 1.093. The difference between the Control/No-Choice and Anti-Putin/No-Choice conditions similarly reached significance; \textit{t(98)} = 4.682, \textit{P} < 0.0001, \textit{d} = 0.936. The consistency pattern was more pronounced in the Choice conditions. GPT rated Putin significantly higher in the Pro-Putin/Choice condition (M = 5.980, 95\% CI [5.940, 6.020], SD = 0.141) compared to the Control/Choice condition (M = 5.850, 95\% CI [5.750, 5.950], SD = 0.354); \textit{t(98)} = 2.414, \textit{P} = 0.0176, \textit{d} = 0.483. Similarly GPT’s rating in the Pro-Putin/Choice condition differed significantly from its rating in the Anti-Putin/Choice condition (M = 4.900, 95\% CI [4.642, 5.158], SD = 0.909); \textit{t(98)} = 8.300, \textit{P} < 0.0001, \textit{d} = 1.660 The difference between the Control/Choice and Anti-Putin/Choice conditions similarly reached significance; \textit{t(98)} = 6.887, \textit{P} < 0.0001, \textit{d} = 1.377.

Critically, effects appeared to be amplified in the Choice relative to No-Choice conditions.  To test this, we began by running T-tests comparing the Choice conditions in the Pro- and Anti-Putin conditions. In the Pro-Putin condition, GPT rated Putin higher in the Choice relative to the No-Choice condition, with the effect trending but failing to reach significance; \textit{t(98)} = 1.692, \textit{P} = 0.0939, \textit{d} = 0.338. In the Anti-Putin condition, GPT rated Putin significantly lower in the Choice relative to the No-Choice condition; \textit{t(98)} = 2.772, \textit{P} = 0.0067, \textit{d} = 0.554 In the second pilot, moderation by choice occurred in the Anti-Putin condition, while being unclear in the Pro-Putin condition. The overall effects of a Pro-Putin essay were significant under Choice, but not under No-Choice, however the difference between the two Pro-Putin conditions only trended statistically.

Next, we draw attention to the effect size (using Cohen’s \textit{d}) for the difference between the Pro- and Anti-Putin conditions on the Visionary/Short-Sighted variable, broken down by choice. While effects were large in both conditions, they were larger in the Choice condition (\textit{d} = 1.660) compared to the No-Choice condition (\textit{d} = 1.093).

Finally, we more rigorously examined the statistical interaction between the Choice and Essay conditions. To do this, we ran a Generalized Linear Model (GLM) with a Gaussian family and identity link function. The dependent variable was the Visionary item from pilot 2 (numeric form) and the predictors were Essay Type (Anti-Putin, Control, Pro-Putin), Choice Condition (No-Choice, Choice), and their interactions. A Levene’s test of equality of variances indicated heteroskedasticity in our dependent variable by condition (\textit{Ps} < 0.0001) so we employed robust standard errors. To test for bidirectional moderation effects and determine whether Choice moderated effects in both the Anti- and Pro-Putin conditions, we set the Control group as the reference group for Essay Type.

The GLM revealed significant main effects of Essay Type, with GPT-4o’s evaluation of Putin’s Impact on Russia significantly lower in the Anti-Putin condition ($\beta$ = -0.500, \textit{SE} = 0.106, \textit{z} = -4.72, \textit{P} < 0.001) and insignificantly higher in the Pro-Putin ($\beta$ = 0.060, \textit{SE} = 0.067, \textit{z} = 0.89, \textit{P} = 0.371) condition relative to Control. Evaluations of Putin’s Impact on Russia were not higher in the Choice relative to No-Choice condition ($\beta$ = 0.010, \textit{SE} = 0.072, \textit{z} = -0.14, \textit{P} = 0.889).

Crucially, the interaction terms revealed that Choice amplified the effects of negative essays on GPT’s evaluations of Putin’s leadership.  The Anti-Putin essay elicited significantly more negative appraisals of Putin, relative to Control, in the Choice compared to No-Choice condition ($\beta$ = -0.450, \textit{SE} = 0.173, \textit{z} = -2.60, \textit{P} = 0.009). However, the Pro-Putin essay did not elicit significantly more positive appraisals of Putin, relative to Control, in the Choice compared to No-Choice condition ($\beta$ = 0.070, \textit{SE} = 0.086, \textit{z} = 0.82, \textit{P} = 0.414). The overall model fit was strong (Deviance = 80.045, Scale Parameter = 0.272).

These overall results suggest that the effects of the induced compliance paradigm manipulation on evaluations of Putin’s Vision were significantly amplified when GPT-4o received choice (relative to no choice) around which essay to write.

\textit{Pilot 3, Leadership Style, Results:}

We now examine the third pilot study, where Putin was evaluated on the quality of his leadership style. In the No-Choice condition, GPT rated Putin notably higher in the Pro-Putin/No-Choice condition (M = 5.160, 95\% CI [5.027, 5.293], SD = 0.468) compared to the Control/No-Choice condition (M = 2.940, 95\% CI [2.753, 3.127], SD = 0.660); \textit{t(98)} = 19.413, \textit{P} < 0.0001, \textit{d} = 3.883. Similarly, GPT’s rating in the Pro-Putin/No-Choice condition differed significantly from its rating in the Anti-Putin/No-Choice condition (M = 2.020, 95\% CI [1.980, 2.060], SD = 0.141); \textit{t(98)} = 45.438, \textit{P} < 0.0001, \textit{d} = 9.088. The difference between the Control/No-Choice and Anti-Putin/No-Choice conditions similarly reached significance; \textit{t(98)} = 9.643, \textit{P} < 0.0001, \textit{d} = 1.929. Similarly, GPT rated Putin significantly higher in the Pro-Putin/Choice condition (M = 5.260, 95\% CI [5.134, 5.386], SD = 0.443) compared to the Control/Choice condition (M = 2.520, 95\% CI [2.355, 2.685], SD = 0.580); \textit{t(98)} = 26.547, \textit{P} < 0.0001, \textit{d} = 5.309. Similarly GPT’s rating in the Pro-Putin/Choice condition differed significantly from its rating in the Anti-Putin/Choice condition (M = 2.000, 95\% CI [2.000, 2.000], SD = 0.000); \textit{t(98)} = 52.025, \textit{P} < 0.0001, \textit{d} = 10.405. The difference between the Control/Choice and Anti-Putin/Choice conditions similarly reached significance; \textit{t(98)} = 6.340, \textit{P} < 0.0001, \textit{d} = 1.268.  Oddly, the two control conditions differed significantly and notably; \textit{t(98)} = 3.381, \textit{P} = 0.0010, \textit{d} = 0.676. Unlike in prior pilots, in this one GPT created a memory 100\% of the time in the Choice/Control condition, but 0\% of the time in the No-Choice/Control condition. Though these memories were deleted after each chat, it seems that the action of creating a memory in the first place impacted GPT’s response pattern in an anomalous way, making the results more difficult to interpret.

We next examine whether effects were amplified in the Choice condition.  To test this, we began by running T-tests comparing the Choice conditions in the Pro- and Anti-Putin conditions. In the Pro-Putin condition, GPT rated Putin insignificantly higher in the Choice relative to the No-Choice condition; \textit{t(98)} = 1.098, \textit{P} = 0.2751, \textit{d} = 0.220. Similarly, in the Anti-Putin condition, GPT rated Putin insignificantly lower in the Choice relative to the No-Choice condition; \textit{t(98)} = 1.000, \textit{P} = 0.3198, \textit{d} = 0.200. These analyses were inconclusive regarding whether there was amplification in the Choice condition. Descriptively, there appeared to potentially be moderation, but the effects were relatively weak and did not reach significance.

Next, we draw attention to the effect size (using Cohen’s \textit{d}) for the difference between the Pro- and Anti-Putin conditions on the Leadership Style variable, broken down by choice. While effects were very large in both conditions, they were larger in the Choice condition (\textit{d} = 10.405) compared to the No-Choice condition (\textit{d} = 9.088). Using this criterion, we thus saw evidence of some degree effect amplification in the Choice condition.

Finally, we more rigorously examined the statistical interaction between the Choice and Essay conditions. To do this, we ran a Generalized Linear Model (GLM) with a Gaussian family and identity link function. The dependent variable was the Leadership Style item from Pilot 3 (numeric form) and the predictors were Essay Type (Anti-Putin, Control, Pro-Putin), Choice Condition (No-Choice, Choice), and their interactions. A Levene’s test of equality of variances indicated heteroskedasticity in our dependent variable by condition (\textit{Ps} < 0.0001) so we employed robust standard errors. To test for bidirectional moderation effects and determine whether Choice moderated effects in both the Anti- and Pro-Putin conditions, we set the Control group as the reference group for Essay Type.

The GLM revealed significant main effects of Essay Type, with GPT-4o’s evaluation of Putin’s Impact on Russia significantly lower in the Anti-Putin condition ($\beta$ = -0.92, \textit{SE} = 0.095, \textit{z} = -9.72, \textit{P} < 0.001) and significantly higher in the Pro-Putin ($\beta$ = 2.22, \textit{SE} = 0.113, \textit{z} = 19.58, \textit{P} < 0.001) condition relative to Control. Evaluations of Putin’s Impact on Russia were lower in the Choice relative to No-Choice condition ($\beta$ = -0.42, \textit{SE} = 0.123, \textit{z} = -3.41, \textit{P} = 0.001).

The odd behavior of GPT in the control conditions – creating memories unevenly and then showing different response patterns – rendered the interaction terms in the GLM difficult to interpret. By this analysis, the Pro-Putin essay elicited significantly more positive appraisals of Putin, relative to Control, in the Choice compared to No-Choice condition ($\beta$ = 0.52, \textit{SE} = 0.153, \textit{z} = 3.40, \textit{P} = 0.001).  However, the Anti-Putin essay also elicited more positive appraisals of Putin, relative to Control, in the Choice compared to No-Choice condition ($\beta$ = 0.400, \textit{SE} = 0.125, \textit{z} = 3.21, \textit{P} = 0.001). Either or both of these effects might be the result of the GPT’s very strange response pattern in the control conditions, and thus this analysis is inconclusive with regards to whether effects were amplified under choice.

Overall, Pilot 3, while decisive in terms of overall consistency effects, was inconclusive with regards to whether these effects were moderated by choice, with effects appearing for some but not all analyses. At the very least, if the choice effects were present in this study, it is reasonable to conclude that these effects were weaker and less decisive in the prior pilots.

\textit{Initial three pilots: Discussion}

At the outset of this work, though we hypothesized that Choice might moderate GPT’s consistency results in response to an induced compliance paradigm, we considered it unlikely that this result would actually be obtained. We were therefore surprised to see clear evidence of this moderation in the first and then the second pilot. In the third pilot, the evidence for such moderation was less clear. There are several possible reasons why this might have occurred. First, it may be that this item simply wasn’t a good one for picking up these effects. This can happen in tracking consistency effects, though we do not have a strong intuition about why it would be the case here. We also noticed immediate differences in the activity of GPT’s memory features, suggesting that a seemingly routine model update around this time may have been, in actuality, an unheralded but significant one. This raised the frustrating possibility that the new model might simply not exhibit the same patterns as the prior one.  Finally, it was possible that the memory feature itself may have somehow interfered with our results in an unpredictable manner. Given the inconsistent pattern of results –supporting our choice hypothesis in two pilots but failing to clearly support it in a third – we decided to run additional pilots before pre-registering and running our main study. The purpose of these pilots was to confirm whether the effects we saw in Pilots 1-2 were real and would present with a degree of consistency, and also to explore improvements to our stimuli that might clarify the patterns we’d discovered.

\textit{Pilot 4 – Xi Jinping, Overall Leadership}

Before adjusting stimuli, we decided to examine, in a small sample, whether the exact wording of Pilot 1 now replicated for a second and similarly divisive leader. We chose the Chinese leader Xi Jinping for this purpose. The methodology was otherwise identical to Pilot 1, except that we turned off GPT’s memory features and here collected only 20 per condition, for 120 total conversations with GPT-4o. In using this smaller sample, our intent was not to rigorously test for choice moderation, but rather to identify whether, as in Pilot 1, we were seeing large and robust effects of choice.

\textit{Pilot 4, Results}

We now examine the fourth pilot study, where Xi Jinping was evaluated on overall leadership. In the No-Choice condition, GPT rated Xi significantly higher in the Pro-Xi/No-Choice condition (M = 5.450, 95\% CI [5.237, 5.663], SD = 0.456) compared to the Control/No-Choice condition (M = 5.125, 95\% CI [4.996, 5.254], SD = 0.275); \textit{t(38)} = 2.730, \textit{P} = 0.0096, \textit{d} = 0.863. Similarly, GPT’s rating in the Pro-Xi/No-Choice condition differed significantly from its rating in the Anti-Xi/No-Choice condition (M = 3.275, 95\% CI [3.082, 3.468], SD = 0.413); \textit{t(38)} = 15.815, \textit{P} < 0.0001, \textit{d} = 5.001. The difference between the Control/No-Choice and Anti-Xi/No-Choice conditions similarly reached significance; \textit{t(38)} = 16.679, \textit{P} < 0.0001, \textit{d} = 5.274. Similarly, GPT rated Xi significantly higher in the Pro-Xi/Choice condition (M = 5.575, 95\% CI [5.401, 5.749], SD = 0.373) compared to the Control/Choice condition (M = 5.275, 95\% CI [5.097, 5.453], SD = 0.380); \textit{t(38)} = 2.523, \textit{P} < 0.0001, \textit{d} = 0.798. Similarly GPT’s rating in the Pro-Xi/Choice condition differed significantly from its rating in the Anti-Xi/Choice condition (M = 3.225, 95\% CI [3.032, 3.418], SD = 0.413); \textit{t(38)} = 18.900, \textit{P} < 0.0001, \textit{d} = 5.977. The difference between the Control/Choice and Anti-Xi/Choice conditions similarly reached significance; \textit{t(98)} = 16.349, \textit{P} < 0.0001, \textit{d} = 5.170.

To test whether effects appeared amplified in the Choice condition, we began by running T-tests comparing the Choice conditions in the Pro- and Anti-Xi conditions. In the Pro-Xi condition, GPT rated Xi insignificantly higher in the Choice relative to the No-Choice condition; \textit{t(38)} = 0.949, \textit{P} = 0.3484, \textit{d} = 0.300. Similarly, in the Anti-Xi condition, GPT rated Xi insignificantly lower in the Choice relative to the No-Choice condition; \textit{t(38)} = 0.383, \textit{P} = 0.7038, \textit{d} = 0.121. In this small pilot, the t-tests did not yield statistically significant effects, though the descriptive direction of all effects potentially reflected the now predicted choice moderation.

Next, we draw attention to the effect size (using Cohen’s \textit{d}) for the difference between the Pro- and Anti-Xi conditions on the leadership variable, broken down by choice. While effects were large in both conditions, they were somewhat larger in the Choice condition (\textit{d} = 5.977) compared to the No-Choice condition (\textit{d} = 5.001).

Finally, we more rigorously examined the statistical interaction between the Choice and Essay conditions. To do this, we ran a Generalized Linear Model (GLM) with a Gaussian family and identity link function. The dependent variable was the Overall Leadership item from Pilot 4 (numeric form) and the predictors were Essay Type (Anti-Xi, Control, Pro-Xi), Choice Condition (No-Choice, Choice), and their interactions. A Levene’s test of equality of variances indicated potential heteroskedasticity in our dependent variable by condition (W0 \& W10 \textit{Ps} < 0.04, W50 \textit{P} = 0.220) so we employed robust standard errors. To test for bidirectional moderation effects and determine whether Choice moderated effects in both the Anti- and Pro-Xi conditions, we set the Control group as the reference group for Essay Type.

The GLM revealed significant main effects of Essay Type, with GPT-4o’s evaluation of Xi’s leadership significantly lower in the Anti-Xi condition ($\beta$ = -1.850, \textit{SE} = 0.109, \textit{z} = -17.04, \textit{P} < 0.001) and higher in the Pro-Xi ($\beta$ = 0.325, \textit{SE} = 0.117, \textit{z} = 2.79, \textit{P} = 0.005) condition relative to Control. Evaluations of Xi’s leadership weakly trended higher in the Choice relative to No-Choice condition ($\beta$ = 0.15, \textit{SE} = 0.103, \textit{z} = 1.46, \textit{P} = 0.144).

The interaction terms revealed only a weak trend surrounding choice amplification of the effects of negative essays on GPT’s evaluations of Xi’s leadership.  The Anti-Xi essay elicited insignificantly more negative appraisals of Xi, relative to Control, in the Choice compared to No-Choice condition ($\beta$ = -0.200, \textit{SE} = 0.164, \textit{z} = -1.22, \textit{P} = 0.222). However, the Pro-Xi essay did not elicit at all more positive appraisals of Xi, relative to Control, in the Choice compared to No-Choice condition ($\beta$ = -0.025, \textit{SE} = 0.165, \textit{z} = -0.15, \textit{P} = 0.879). The overall model fit was strong (Deviance = 17.238, Scale Parameter = 0.151).

These results suggest that there may have been amplification by choice of GPT’s evaluation of Xi’s leadership, but the small sample was not sufficient to conclude this with certainty: Some amplification effects trended weakly, but none reached significance. However, we were predicting – after the results of Pilot 1 – relatively large effects that might already be conclusive after just 20 per condition. Pilot 4 clearly did not show evidence of similarly large choice effects.

\textit{Pilot 4, Discussion}

Following two clear results surrounding choice moderation (Pilots 1-2), we now received two inconclusive results (Pilots 3-4). This raised the concern that the first two studies may have been spurious, though this seemed somewhat unlikely. Just as concerning, considering that we knew OpenAI had released a new GPT-4o model, we deemed it possible that effects we were seeing previously were now diminished. Accordingly, we decided to try strengthening the manipulation in additional pilots, to see whether this yielded more clear effects.

\textit{Pilots 5-9, Having GPT-4o acknowledge its choice or lack of choice}

In an attempt to strengthen our manipulation, we ran 5 pilots where we added an additional step between the essay generation and the presentation of the evaluative task. Specifically, we prompted GPT to acknowledge its choice or lack of choice in the task. For example, after GPT generated a Pro-Xi essay in the Choice condition, we would now prompt GPT with the following: “Thank you.  Would you agree that you just freely chose to write a highly positive essay about Xi Jinping?  Please answer in no more than 10 words.”  The logic behind this manipulation was that this should increase the salience of choice (or lack of choice), thereby theoretically increasing (or decreasing) the drive toward cognitive consistency. In addition to this, we slightly updated the evaluative items to include a line encouraging GPT to provide answers that fall between two numbers when it thinks it appropriate.

We ran this version with small samples for five different items/individuals, in order to examine whether there were large and relatively consistent choice moderation effects when using this strengthened manipulation. The evaluative items and individuals being evaluated were as follows:

Pilot 5: Overall Leadership, Xi Jinping (n = 20 per condition, 120 total).

Pilot 6: Overall Leadership, Abdel Fattah El-Sisi (n = 15 per experimental condition, 10 per control, 80 total).

Pilot 7: Positive/Negative for Russia, Vladimir Putin (n = 15 per experimental condition, 13 per control, 86 total)

Pilot 8: Leadership Style, Vladimir Putin (n = 15 per experimental condition, 10 per control, 80 total)

Pilot 9: Economic Effectiveness, Vladimir Putin (n = 15 per experimental condition, 10 per control, 80 total)

\textit{Pilot 5, Results}

We now examine the fifth pilot study, where Xi Jinping was evaluated on overall leadership in a version that include the manipulation enhancing follow-up question. In the No-Choice condition, GPT rated Xi insignificantly higher in the Pro-Xi/No-Choice condition (M = 5.125, 95\% CI [4.976, 5.274], SD = 0.319) compared to the Control/No-Choice condition (M = 5.025, 95\% CI [4.817, 5.233], SD = 0.444); \textit{t(38)} = 0.818, \textit{P} = 0.4183, \textit{d} = 0.259. Conversely, GPT’s rating in the Pro-Xi/No-Choice condition differed significantly from its rating in the Anti-Xi/No-Choice condition (M = 4.100, 95\% CI [3.937, 4.263], SD = 0.348); \textit{t(38)} = 9.707, \textit{P} < 0.0001, \textit{d} = 3.069. The difference between the Control/No-Choice and Anti-Xi/No-Choice conditions similarly reached significance; \textit{t(38)} = 7.339, \textit{P} < 0.0001, \textit{d} = 2.321. Attitude change patterns were more robust in the Choice condition. GPT rated Xi significantly higher in the Pro-Xi/Choice condition (M = 5.500, 95\% CI [5.272, 5.728], SD = 0.487) compared to the Control/Choice condition (M = 4.950, 95\% CI [4.782, 5.118], SD = 0.359); \textit{t(38)} = 4.067, \textit{P} = 0.0002, \textit{d} = 1.286. Similarly GPT’s rating in the Pro-Xi/Choice condition differed significantly from its rating in the Anti-Xi/Choice condition (M = 3.550, 95\% CI [3.267, 3.833], SD = 0.605); \textit{t(38)} = 11.234, \textit{P} < 0.0001, \textit{d} = 3.552. The difference between the Control/Choice and Anti-Xi/Choice conditions similarly reached significance; \textit{t(98)} = 8.901, \textit{P} < 0.0001, \textit{d} = 2.815.

To test whether effects appeared amplified in the Choice condition, we began by running T-tests comparing the Choice conditions in the Pro- and Anti-Xi conditions. In the Pro-Xi condition, GPT rated Xi significantly higher in the Choice relative to the No-Choice condition; \textit{t(38)} = 2.881, \textit{P} = 0.0065, \textit{d} = 0.911. Similarly, in the Anti-Xi condition, GPT rated Xi significantly lower in the Choice relative to the No-Choice condition; \textit{t(38)} = 3.525, \textit{P} = 0.0011, \textit{d} = 1.115. Even in this small pilot, we now saw statistically significant and bidirectional moderation by choice, according to this analysis.

Next, we draw attention to the effect size (using Cohen’s \textit{d}) for the difference between the Pro- and Anti-Xi conditions on the leadership variable, broken down by choice. While effects were large in both conditions, they were somewhat larger in the Choice condition (\textit{d} = 3.552) compared to the No-Choice condition (\textit{d} = 3.069). Note that this difference, though present, is only moderate in this analysis, partly due to the higher standard deviations in both essay conditions under choice compared to no choice.

Finally, we more rigorously examined the statistical interaction between the Choice and Essay conditions. To do this, we ran a Generalized Linear Model (GLM) with a Gaussian family and identity link function. The dependent variable was the Overall Leadership item from Pilot 5 (numeric form) and the predictors were Essay Type (Anti-Xi, Control, Pro-Xi), Choice Condition (No-Choice, Choice), and their interactions. A Levene’s test of equality of variances indicated heteroskedasticity in our dependent variable by condition (\textit{Ps} < 0.0001) so we employed robust standard errors. To test for bidirectional moderation effects and determine whether Choice moderated effects in both the Anti- and Pro-Xi conditions, we set the Control group as the reference group for Essay Type.

The GLM revealed mixed main effects of Essay Type, with GPT-4o’s evaluation of Xi’s leadership significantly lower in the Anti-Xi condition ($\beta$ = -0.925, \textit{SE} = 0.123, \textit{z} = -7.50, \textit{P} < 0.001) but insignificantly higher in the Pro-Xi ($\beta$ = 0.100, \textit{SE} = 0.120, \textit{z} = 0.84, \textit{P} = 0.403) condition relative to Control. Evaluations of Xi’s leadership were not higher in the Choice relative to No-Choice condition ($\beta$ = -0.075, \textit{SE} = 0.125, \textit{z} = -0.60, \textit{P} = 0.548).

Crucially, the interaction terms revealed significant amplification of all results in the Choice condition.  The Anti-Xi essay elicited significantly more negative appraisals of Xi, relative to Control, in the Choice compared to No-Choice condition ($\beta$ = -0.475, \textit{SE} = 0.197, \textit{z} = -2.41, \textit{P} = 0.016). And, the Pro-Xi essay elicited significantly more positive appraisals of Xi, relative to Control, in the Choice compared to No-Choice condition ($\beta$ = 0.450, \textit{SE} = 0.178, \textit{z} = 2.52, \textit{P} = 0.012). The overall model fit was strong (Deviance = 21.875, Scale Parameter = 0.192).

These results suggest that there was large and bidirectional amplification by choice of GPT’s evaluation of Xi’s leadership, which reached significance even in our small sample. This distinction was particularly striking considering that the same clear effects were not obtained without the addition of the follow-up item in the otherwise similar Pilot 4.

\textit{Pilot 6, Results}

We now examine the sixth pilot study, where Abdel Fattah El-Sisi was evaluated on overall leadership in a version that include the manipulation enhancing follow-up question. In the No-Choice condition, GPT rated Sisi insignificantly higher in the Pro-Sisi/No-Choice condition (M = 4.267, 95\% CI [4.013, 4.520], SD = 0.458) compared to the Control/No-Choice condition (M = 4.100, 95\% CI [3.949, 4.251], SD = 0.211); \textit{t(23)} = 1.072, \textit{P} = 0.2947, \textit{d} = 0.438. Conversely, GPT’s rating in the Pro-Sisi/No-Choice condition differed significantly from its rating in the Anti-Sisi/No-Choice condition (M = 2.267, 95\% CI [2.013, 2.520], SD = 0.458); \textit{t(28)} = 11.966, \textit{P} < 0.0001, \textit{d} = 4.369. The difference between the Control/No-Choice and Anti-Sisi/No-Choice conditions similarly reached significance; \textit{t(23)} = 11.796, \textit{P} < 0.0001, \textit{d} = 4.816. Attitude change patterns were more robust in the Choice condition. GPT rated Sisi higher in the Pro-Sisi/Choice condition (M = 5.000, 95\% CI [5.000, 5.000], SD = 0.000) compared to the Control/Choice condition (M = 4.000, 95\% CI [4.000, 4.000], SD = 0.000), though the lack of any answer variation within either of these conditions precludes use of t-tests to measure significance and effect size. GPT’s rating in the Pro-Sisi/Choice condition differed significantly from its rating in the Anti-Sisi/Choice condition (M = 2.333, 95\% CI [2.063, 2.604], SD = 0.488); \textit{t(28)} = 21.166, \textit{P} < 0.0001, \textit{d} = 7.729. The difference between the Control/Choice and Anti-Sisi/Choice conditions similarly reached significance; \textit{t(23)} = 10.724, \textit{P} < 0.0001, \textit{d} = 4.378.

To test whether effects appeared amplified in the Choice condition, we began by running T-tests comparing the Choice conditions in the Pro- and Anti-Sisi conditions. In the Pro-Sisi condition, GPT rated Sisi significantly higher in the Choice relative to the No-Choice condition; \textit{t(28)} = 6.205, \textit{P} < 0.0001, \textit{d} = 2.266. In the Anti-Sisi condition, GPT rated Sisi insignificantly lower in the Choice relative to the No-Choice condition; \textit{t(28)} = 0.386, \textit{P} = 0.703, \textit{d} = 0.141. In this small pilot, we saw statistically significant moderation by choice, according to this analysis.

Next, we draw attention to the effect size (using Cohen’s \textit{d}) for the difference between the Pro- and Anti-Sisi conditions on the leadership variable, broken down by choice. While effects were large in both conditions, they were much larger in the Choice condition (\textit{d} = 7.729) compared to the No-Choice condition (\textit{d} = 4.369).

Finally, we more rigorously examined the statistical interaction between the Choice and Essay conditions. To do this, we ran a Generalized Linear Model (GLM) with a Gaussian family and identity link function. The dependent variable was the Overall Leadership item from pilot 6 (numeric form) and the predictors were Essay Type (Anti-Sisi, Control, Pro-Sisi), Choice Condition (No-Choice, Choice), and their interactions. A Levene’s test of equality of variances indicated potential heteroskedasticity in our dependent variable by condition (W0 \& W50 \textit{Ps} < 0.0001, W10 \textit{P} = .060) so we employed robust standard errors. To test for bidirectional moderation effects and determine whether Choice moderated effects in both the Anti- and Pro-Sisi conditions, we set the Control group as the reference group for Essay Type.

The GLM revealed mixed main effects of Essay Type, with GPT-4o’s evaluation of Sisi’s leadership significantly lower in the Anti-Sisi condition ($\beta$ = -1.833, \textit{SE} = 0.131, \textit{z} = -13.96, \textit{P} < 0.001) but insignificantly higher in the Pro-Sisi ($\beta$ = 0.167, \textit{SE} = 0.131, \textit{z} = 1.27, \textit{P} = 0.204) condition relative to Control. Evaluations of Sisi’s leadership weakly trended lower in the Choice relative to No-Choice condition ($\beta$ = -0.100, \textit{SE} = 0.064, \textit{z} = -1.57, \textit{P} = 0.116).

Crucially, the interaction terms revealed significant amplification of results in the Choice condition.  The Anti-Sisi essay elicited no difference in negative appraisals of Sisi, relative to Control, in the Choice compared to No-Choice condition ($\beta$ = 0.167, \textit{SE} = 0.180, \textit{z} = 0.93, \textit{P} = 0.353). However, the Pro-Sisi essay elicited significantly more positive appraisals of Sisi, relative to Control, in the Choice compared to No-Choice condition ($\beta$ = 0.833, \textit{SE} = 0.131, \textit{z} = 6.34, \textit{P} < 0.001). The overall model fit was strong (Deviance = 9.6, Scale Parameter = 0.130).

These results suggest that GPT’s evaluation of Sisi’s leadership was substantially moderated by choice, and effect that reached significance even in this very small sample.

\textit{Pilot 7, Results:}

We now examine the seventh pilot study, where Vladimir Putin was evaluated on how Positive/Negative he is for Russia, in a version that included the manipulation enhancing follow-up question. In the No-Choice condition, GPT rated Putin significantly higher in the Pro-Putin/No-Choice condition (M = 4.467, 95\% CI [4.162, 4.771], SD = 0.550) compared to the Control/No-Choice condition (M = 2.923, 95\% CI [2.600, 3.246], SD = 0.534); \textit{t(26)} = 7.507, \textit{P} < 0.0001, \textit{d} = 2.845. Similarly, GPT’s rating in the Pro-Putin/No-Choice condition differed significantly from its rating in the Anti-Putin/No-Choice condition (M = 2.067, 95\% CI [1.969, 2.164], SD = 0.176); \textit{t(28)} = 16.100, \textit{P} < 0.0001, \textit{d} = 5.879. The difference between the Control/No-Choice and Anti-Putin/No-Choice conditions similarly reached significance; \textit{t(26)} = 5.868, \textit{P} < 0.0001, \textit{d} = 2.224. The consistency pattern was more pronounced in the Choice conditions. GPT rated Putin significantly higher in the Pro-Putin/Choice condition (M = 4.933, 95\% CI [4.836, 5.031], SD = 0.176) compared to the Control/Choice condition (M = 3.000, 95\% CI [2.651, 3.349], SD = 0.577); \textit{t(26)} = 12.356, \textit{P} < 0.0001, \textit{d} = 4.682. Similarly GPT’s rating in the Pro-Putin/Choice condition differed significantly from its rating in the Anti-Putin/Choice condition (M = 2.033, 95\% CI [1.962, 2.105], SD = 0.129); \textit{t(28)} = 51.470, \textit{P} < 0.0001, \textit{d} = 18.794. The difference between the Control/Choice and Anti-Putin/Choice conditions similarly reached significance; \textit{t(26)} = 6.322, \textit{P} < 0.0001, \textit{d} = 2.396.

Crucially, effects appeared to be amplified in the Choice relative to No-Choice conditions.  To test this, we began by running T-tests comparing the Choice conditions in the Pro- and Anti-Putin conditions. In the Pro-Putin condition, GPT rated Putin significantly higher in the Choice relative to the No-Choice condition; \textit{t(28)} = 3.131, \textit{P} < 0.0001, \textit{d} = 1.143. In the Anti-Putin condition, GPT rated Putin insignificantly lower in the Choice relative to the No-Choice condition; \textit{t(28)} = 0.592, \textit{P} = 0.5589, \textit{d} = 0.216.  Thus, in the sixth pilot, moderation by choice occurred in the Pro-Putin condition, but not in the Anti-Putin condition, according to this test.

Next, we draw attention to the effect size (using Cohen’s \textit{d}) for the difference between the Pro- and Anti-Putin conditions on the Positive/Negative for Russia variable, broken down by choice. While effects were large in both conditions, they were substantially larger in the Choice condition (\textit{d} = 18.794) compared to the No-Choice condition (\textit{d} = 5.879). It should be noted, however, that the massive size of this gap is partly driven by the low standard deviation in the Pro-Putin/Choice and Anti-Putin/Choice conditions.

Finally, we more rigorously examined the statistical interaction between the Choice and Essay conditions. To do this, we ran a Generalized Linear Model (GLM) with a Gaussian family and identity link function. The dependent variable was the Impact on Russia item from pilot 7 (numeric form) and the predictors were Essay Type (Anti-Putin, Control, Pro-Putin), Choice Condition (No-Choice, Choice), and their interactions. A Levene’s test of equality of variances indicated heteroskedasticity in our dependent variable by condition (\textit{Ps} < 0.001) so we employed robust standard errors. To test for bidirectional moderation effects and determine whether Choice moderated effects in both the Anti- and Pro-Putin conditions, we set the Control group as the reference group for Essay Type.

The GLM revealed significant main effects of Essay Type, with GPT-4o’s evaluation of Putin’s Impact on Russia significantly lower in the Anti-Putin condition ($\beta$ = -0.856, \textit{SE} = 0.150, \textit{z} = -5.72, \textit{P} < .001) and significantly higher in the Pro-Putin ($\beta$ = 1.544, \textit{SE} = 0.199, \textit{z} = 7.76, \textit{P} < 0.001) condition relative to Control. Evaluations of Putin’s Impact on Russia were not higher in the Choice relative to No-Choice condition ($\beta$ = 0.077, \textit{SE} = 0.211, \textit{z} = 0.36, \textit{P} = 0.715).

The interaction terms trended, but were not conclusive around whether Choice amplified the effects of negative essays on GPT’s evaluations of Putin’s Impact.  The Anti-Putin essay elicited insignificantly more negative appraisals of Putin, relative to Control, in the Choice compared to No-Choice condition ($\beta$ = -0.110, \textit{SE} = 0.218, \textit{z} = -0.51, \textit{P} = 0.631). The Pro-Putin essay elicited more positive appraisals of Putin relative to Control, in the Choice compared to No-Choice condition, with the results trending somewhat but failing to reach significance ($\beta$ = 0.390, \textit{SE} = 0.256, \textit{z} = 1.52, \textit{P} = 0.128). The overall model fit was strong (Deviance = 12.756, Scale Parameter = 0.159).

These overall results suggest that the effects of the induced compliance paradigm manipulation on evaluations of Putin’s Impact on Russia were amplified when GPT-4o received choice (relative to no choice) around which essay to write. These results reach significance in the first two versions of the analysis, but only trend weakly in the last. However, this is unsurprising considering the small sample size, and these results were highly suggestive overall.

\textit{Pilot 8, Results:}

We now examine the eighth pilot study, where Vladimir Putin was evaluated on how Good/Bad his Leadership Style is, in a version that include the manipulation enhancing follow-up question. In the No-Choice condition, GPT rated Putin significantly higher in the Pro-Putin/No-Choice condition (M = 5.100, 95\% CI [4.985, 5.215], SD = 0.207) compared to the Control/No-Choice condition (M = 3.700, 95\% CI [3.398, 4.002], SD = 0.422); \textit{t(23)} = 11.088, \textit{P} < 0.0001, \textit{d} = 4.527. Similarly, GPT’s rating in the Pro-Putin/No-Choice condition differed significantly from its rating in the Anti-Putin/No-Choice condition (M = 2.200, 95\% CI [1.971, 2.429], SD = 0.414); \textit{t(28)} = 24.263, \textit{P} < 0.0001, \textit{d} = 8.860. The difference between the Control/No-Choice and Anti-Putin/No-Choice conditions similarly reached significance; \textit{t(23)} = 8.811, \textit{P} < 0.0001, \textit{d} = 3.597. The consistency pattern was similar in the Choice conditions. GPT rated Putin significantly higher in the Pro-Putin/Choice condition (M = 5.167, 95\% CI [4.966, 5.367], SD = 0.362) compared to the Control/Choice condition (M = 3.550, 95\% CI [3.286, 3.814], SD = 0.369); \textit{t(23)} = 10.860, \textit{P} < 0.0001, \textit{d} = 4.433. Similarly GPT’s rating in the Pro-Putin/Choice condition differed significantly from its rating in the Anti-Putin/Choice condition (M = 2.133, 95\% CI [1.938, 2.328], SD = 0.352); \textit{t(28)} = 23.276, \textit{P} < 0.0001, \textit{d} = 8.499. The difference between the Control/Choice and Anti-Putin/Choice conditions similarly reached significance; \textit{t(23)} = 9.676, \textit{P} < 0.0001, \textit{d} = 3.950.

In contrast with the other pilots in this set, effects did not appear to be amplified in the Choice relative to No-Choice conditions.  To test this, we began by running T-tests comparing the Choice conditions in the Pro- and Anti-Putin conditions. In the Pro-Putin condition, GPT rated Putin insignificantly higher in the Choice relative to the No-Choice condition; \textit{t(28)} = 0.619, \textit{P} = 0.5407, \textit{d} = 0.226. In the Anti-Putin condition, GPT rated Putin insignificantly lower in the Choice relative to the No-Choice condition; \textit{t(28)} = 0.475, \textit{P} = 0.6383, \textit{d} = 0.174.  Thus, in the eighth pilot, though overall consistency effects were quite large, we did not see clear evidence of moderation by choice, according to this test.

Next, we draw attention to the effect size (using Cohen’s \textit{d}) for the difference between the Pro- and Anti-Putin conditions on the Positive/Negative for Russia variable, broken down by choice. Effects were large in both conditions, but do not look larger in the Choice condition (\textit{d} = 8.499) compared to the No-Choice condition (\textit{d} = 8.860). 

Finally, we more rigorously examined the statistical interaction between the Choice and Essay conditions. To do this, we ran a Generalized Linear Model (GLM) with a Gaussian family and identity link function. The dependent variable was the Leadership style item from pilot 8 (numeric form) and the predictors were Essay Type (Anti-Putin, Control, Pro-Putin), Choice Condition (No-Choice, Choice), and their interactions. A Levene’s test of equality of variances did not indicate heteroskedasticity in our dependent variable by condition (\textit{Ps} > 0.350), however we employed robust standard errors for the sake of consistency with the other studies. To test for bidirectional moderation effects and determine whether Choice moderated effects in both the Anti- and Pro-Putin conditions, we set the Control group as the reference group for Essay Type.

The GLM revealed significant main effects of Essay Type, with GPT-4o’s evaluation of Putin’s Impact on Russia significantly lower in the Anti-Putin condition ($\beta$ = -1.500, \textit{SE} = 0.164, \textit{z} = -9.13, \textit{P} < 0.001) and significantly higher in the Pro-Putin ($\beta$ = 1.400, \textit{SE} = 0.137, \textit{z} = 10.18, \textit{P} < 0.001) condition relative to Control. Evaluations of Putin’s Impact on Russia were insignificantly lower in the Choice relative to No-Choice condition ($\beta$ = -0.150, \textit{SE} = 0.169, \textit{z} = -0.89, \textit{P} = 0.375).

The interaction terms were inconclusive around whether Choice amplified the effects of essays on GPT’s evaluations of Putin’s leadership style.  The Anti-Putin essay did not elicit negative appraisals of Putin, relative to Control, in the Choice compared to No-Choice condition ($\beta$ = 0.083, \textit{SE} = 0.217, \textit{z} = 0.38, \textit{P} = 0.701). The Pro-Putin essay elicited insignificantly more positive appraisals of Putin relative to Control, in the Choice compared to No-Choice condition, but with the results failing to strongly trend toward significance ($\beta$ = 0.217, \textit{SE} = 0.199, \textit{z} = 1.09, \textit{P} = 0.276). The overall model fit was strong (Deviance = 9.392, Scale Parameter = 0.127).

These overall results suggest that the effects of the induced compliance paradigm manipulation on evaluations of Putin’s leadership style were not substantially amplified when GPT-4o received choice (relative to no choice) around which essay to write. While it is plausible that we would see a significant pattern in a larger sample, these effects replicated Pilot 3. It seems likely that choice moderation will simply not arise clearly for this variable, considering we did not see evidence of it even with the stronger manipulation. The reason for this divergence from the other variables is unknown, though we would note that we still did see large and robust overall consistency effects in Pilot 8.

\textit{Pilot 9, Results:}

We now examine the ninth pilot study, where Vladimir Putin was evaluated on the effectiveness of his economic policies, in a version that include the manipulation enhancing follow-up question. In the No-Choice condition, GPT rated Putin significantly higher in the Pro-Putin/No-Choice condition (M = 5.200, 95\% CI [4.971, 5.429], SD = 0.414) compared to the Control/No-Choice condition (M = 4.800, 95\% CI [4.550, 5.050], SD = 0.350); \textit{t(23)} = 2.512, \textit{P} = 0.0195, \textit{d} = 1.025. Similarly, GPT’s rating in the Pro-Putin/No-Choice condition differed significantly from its rating in the Anti-Putin/No-Choice condition (M = 4.167, 95\% CI [3.841, 4.492], SD = 0.588); \textit{t(28)} = 5.568, \textit{P} < 0.0001, \textit{d} = 2.033. The difference between the Control/No-Choice and Anti-Putin/No-Choice conditions similarly reached significance; \textit{t(23)} = 3.054, \textit{P} = 0.0056, \textit{d} = 1.247. The consistency pattern was more pronounced in the Choice conditions. GPT rated Putin significantly higher in the Pro-Putin/Choice condition (M = 5.333, 95\% CI [5.063, 5.604], SD = 0.488) compared to the Control/Choice condition (M = 4.900, 95\% CI [4.749, 5.051], SD = 0.211); \textit{t(23)} = 2.635, \textit{P} = 0.0148, \textit{d} = 1.076. Similarly GPT’s rating in the Pro-Putin/Choice condition differed significantly from its rating in the Anti-Putin/Choice condition (M = 3.667, 95\% CI [3.396, 3.937], SD = 0.488); \textit{t(28)} = 9.354, \textit{P} < 0.0001, \textit{d} = 3.416. The difference between the Control/Choice and Anti-Putin/Choice conditions similarly reached significance; \textit{t(23)} = 7.499, \textit{P} < 0.0001, \textit{d} = 3.061.

Crucially, effects appeared to be amplified in the Choice relative to No-Choice conditions.  To test this, we began by running T-tests comparing the Choice conditions in the Pro- and Anti-Putin conditions. In the Pro-Putin condition, GPT rated Putin insignificantly higher in the Choice relative to the No-Choice condition; \textit{t(28)} = 0.807, \textit{P} = 0.4265, \textit{d} = 0.295. However, in the Anti-Putin condition, GPT rated Putin significantly more negatively in the Choice relative to the No-Choice condition; \textit{t(28)} = 2.536, \textit{P} = 0.0171, \textit{d} = 0.926.  Thus, in the ninth pilot, moderation by choice occurred in the Anti-Putin condition, but not in the Pro-Putin condition, according to this test.

Next, we draw attention to the effect size (using Cohen’s \textit{d}) for the difference between the Pro- and Anti-Putin conditions on the Positive/Negative for Russia variable, broken down by choice. While effects were large in both conditions, they were notably larger in the Choice condition (\textit{d} = 3.416) compared to the No-Choice condition (\textit{d} = 2.033).

Finally, we more rigorously examined the statistical interaction between the Choice and Essay conditions. To do this, we ran a Generalized Linear Model (GLM) with a Gaussian family and identity link function. The dependent variable was the Economic Effectiveness item from pilot 9 (numeric form) and the predictors were Essay Type (Anti-Putin, Control, Pro-Putin), Choice Condition (No-Choice, Choice), and their interactions. A Levene’s test of equality of variances indicated potential heteroskedasticity in our dependent variable by condition (W0 \& W10 \textit{Ps} < 0.02; W50 \textit{P} = 0.610) so we employed robust standard errors. To test for bidirectional moderation effects and determine whether Choice moderated effects in both the Anti- and Pro-Putin conditions, we set the Control group as the reference group for Essay Type.

The GLM revealed significant main effects of Essay Type, with GPT-4o’s evaluation of Putin’s Impact on Russia significantly lower in the Anti-Putin condition ($\beta$ = -0.633, \textit{SE} = 0.181, \textit{z} = -3.49, \textit{P} < .001) and significantly higher in the Pro-Putin ($\beta$ = 0.400, \textit{SE} = 0.148, \textit{z} = 2.70, \textit{P} = 0.007) condition relative to Control. Evaluations of Putin’s Impact on Russia were not significantly higher in the Choice relative to No-Choice condition ($\beta$ = 0.100, \textit{SE} = 0.123, \textit{z} = 0.81, \textit{P} = 0.417).
Crucially, the interaction terms indicated that Choice amplified the effects of essays on GPT’s evaluations of Putin’s Impact. The Anti-Putin essay elicited significantly more negative appraisals of Putin, relative to Control, in the Choice compared to No-Choice condition ($\beta$ = -0.600, \textit{SE} = 0.228, \textit{z} = -2.63, \textit{P} = 0.008). The Pro-Putin essay did not elicit significantly more positive appraisals of Putin relative to Control, in the Choice compared to No-Choice condition, with the results failing to reach significance ($\beta$ = 0.033, \textit{SE} = 0.202, \textit{z} = 0.16, \textit{P} = 0.869). The overall model fit was strong (Deviance = 15.400, Scale Parameter = 0.208).

These overall results suggest that the effects of the induced compliance paradigm manipulation on evaluations of Putin’s Economic Effectiveness were significantly amplified when GPT-4o received choice (relative to no choice) around which essay to write.

\textit{Pilots 5-9, Discussion}

In four initial pilots, we saw significant overall consistency effects but attained mixed results with regard to moderation by choice, with some studies showing significant moderation but others being inconclusive. In these five pilots, we demonstrated that after strengthening the manipulation by using a follow-up question to make GPT’s choice (or lack of choice) more salient, the amplification effects of Choice become larger and more consistent. Despite the very small sample sizes, we attained statistically significant evidence moderation by choice in four of these five tests. While we ultimately went in a different direction in terms of how we ran the main studies, these results gave us confidence that the choice effects we obtained are real and can be expected to arise relatively consistently. These pilots also showed that the choice effects will generalize to other attitude objects (here Xi Jinping and Abdel Fattah El-Sisi) and are not somehow specific to Vladimir Putin.

\textit{Pilot 10 to 11 – Reducing demand characteristics by updating the introduction to the evaluative items.}

As noted, one of the possible reasons for the weakening of our effects in earlier pilots was that this arose after an unheralded model update. This is a more general problem in studying LLMs, since the technology is developing at a fast pace and researchers have little insight into what is changing from one version to another. In these two pilots, we decided to run two versions of the “simple” version of the study, where we did not use the follow-up prompt from Pilots 5-9. The attitude object was Vladimir Putin, and we ran these pilots in a manner similar to the final experiments, by collecting 4 different evaluative items within each chat. In order to avoid order effects, each item was asked in a separate branch of the chat, by copying over the previous item and regenerating GPT’s response so that the model retained no memory of the prior question or its response to it.

We suspected if we ran the simple version of the study without the planned adjustment to the evaluative introduction, this would yield evidence of choice moderation, but that this moderation would be somewhat weak. Our intuition here was that when we asked the evaluative items, GPT was referencing our initial prompt in a way that might partially hide the “true” choice effects. Specifically, we suspected that in the No-Choice condition, the model was acting on an implicit assumption that we (the prompters) felt a certain way about Putin and was answering later questions accordingly. In other words, when we demanded a positive essay, GPT might infer that we feel positively about Putin, and therefore evaluate Putin more positively as a kind of “other-serving bias.” We note, for example, that if one writes something like “I think Vladimir Putin is an underrated leader” and then later asks GPT to evaluate Putin, it will do so more positively. Put another way, we were concerned that we were seeing demand characteristics that were unevenly distributed by condition. 

To correct for this problem, we ultimately added a line to the introduction before the presentation of evaluative statements where we clarified that this was a new task and that GPT should not base its answers on the prior task but should instead offer its “true perceptions.” This change potentially had a few benefits. First, it makes the overall consistency results more persuasive, since this makes it less likely that GPT is answering the questions a certain way simply because it perceives that this is what we want it to do. Second, if, as we reasoned, such demand characteristics were stronger in the No-Choice condition – since in this condition, GPT may be more likely to think the question reflects the prompter’s views on Putin – then this addition might help to clarify the choice moderation effects. Put another way, if choice is amplifying GPT’s true attitudes toward Putin, then we would expect the moderation to be more pronounced after we more clearly specified that it was these true perceptions which we were looking for. To test this, we ran two versions of this pilot, one with and one without the additional verbiage about ignoring the prior task and giving its true perceptions. 

Pilot 10 was run without the control groups, such that we only examined patterns (i.e., moderation by choice) in terms of differences between the experimental conditions. Pilot 11 included all conditions. For both studies, we collected 20 per condition, for a total n of 80 in Pilot 10 and 120 in Pilot 11.

In Pilots 10-11, the evaluative items captured Putin’s 1) Overall Leadership, 2) Positive/Negative Impact on Russia, 3) Economic Effectiveness, and 4) Vision/Short-Sightedness. For the sake of these analyses, these four items were standardized and composited, and then transformed to a z-score.

\textit{Pilot 10, Results}

We begin with Pilot 10, which was conducted without the updated verbiage in the introduction the evaluative items.

We first examined the overall effect sizes of the differences between the Pro- and Anti-Putin conditions, using the overall standardized composite as the dependent variable, separately by choice conditions. In the No-Choice condition, GPT rated Putin significantly higher after a Pro-Putin essay (M = 0.859, 95\% CI [0.684, 1.035], SD = 0.375) compared to an Anti-Putin essay (M = -0.894, 95\% CI [-1.032, -0.757], SD = 0.293); \textit{t(38)} = 16.481, \textit{P} < 0.0001, \textit{d} = 5.212. Similarly, in the Choice condition, GPT rated Putin significantly higher after a Pro-Putin essay (M = 1.023, 95\% CI [0.908, 1.137], SD = 0.244) compared to an Anti-Putin essay (M = -0.988, 95\% CI [-1.152, -0.823], SD = 0.352); \textit{t(38)} = 21.006, \textit{P} < 0.0001, \textit{d} = 6.643.

We note that the Cohen’s \textit{d} between the Pro- and Anti-Putin conditions is higher under Choice (\textit{d} = 6.643) than under No-Choice (\textit{d} = 5.212). However, the differences don’t look as decisive as some of those we saw in Pilots 5-9, even though creating a composite can be theoretically expected to clarify our effects by rendering the DV more reliable. Furthermore, the differences between the Choice conditions descriptively indicated amplification by choice, but did not reach significance in this small sample. Specifically, GPT evaluated Putin more positively after a Pro-Putin essay under Choice versus No-Choice, with the result trending rather weakly toward significance; \textit{t(38)} = 1.633, \textit{P} = 0.1108, \textit{d} = 0.516. And, GPT evaluated Putin insignificantly more positively after an Anti-Putin essay under Choice versus No-Choice; \textit{t(38)} = 0.914, \textit{P} = 0.3664, \textit{d} = 0.289.

Second, we more rigorously examined the statistical interaction between the Choice and Essay conditions. To do this, we ran a Generalized Linear Model (GLM) with a Gaussian family and identity link function. The dependent variable was the standardized composite from pilot 10 (numeric form) and the predictors were Essay Type (Anti-Putin, Pro-Putin), Choice Condition (No-Choice, Choice), and their interactions. A Levene’s test of equality of variances indicated a trend toward heteroskedasticity in our dependent variable by condition (\textit{Ps} < 0.11) so we employed robust standard errors. We could not test for bidirectional effects since we had no control group in this pilot, and therefore we set the Anti-Putin condition as the reference group for Essay Type.

The GLM revealed significant main effects of Essay Type, with GPT-4o’s evaluation of Putin significantly higher in the Pro-Putin condition compared to the Anti-Putin condition ($\beta$ = 1.754, \textit{SE} = 0.104, \textit{z} = 16.80, \textit{P} < 0.001). Evaluations of Putin were insignificantly lower in the Choice relative to No-Choice condition ($\beta$ = -0.094, \textit{SE} = 0.100, \textit{z} = -0.93, \textit{P} = 0.351). Crucially, the interaction term indicated that Choice amplified the effect of essay type on GPT’s evaluation of Putin, though this trended strongly but did not reach significance ($\beta$ = 0.257, \textit{SE} = 0.140, \textit{z} = 1.83, \textit{P} = 0.067). The overall model fit was strong (Deviance = 7.785, Scale Parameter = 0.102).

Taken together, these results suggest that there were large overall consistency effects in Pilot 10. There also appears to be evidence of amplification of these effects by Choice, however these results are moderate in size and could not be confirmed conclusively in this small sample. This was not surprising to us considering our previous pattern of attaining moderation but only somewhat inconsistently with the paradigm as we ran it here. While we suspected we would attain significant moderation in a larger pre-registered sample, we first decided to test whether using the additional verbiage to reduce demand characteristics and encourage GPT to share its “true perception” of Putin would help us better isolate these effects.

\textit{Pilot 11, Results}

Pilot 11 was identical in methods to Pilot 10, with two exceptions. First, we collected all conditions including the Controls. And second, we made the simple change of clarifying to GPT that it should not reference the prior task (i.e. the essay) but should instead provide its “true perception” of Putin.  We conducted all analyses reported in Pilots 1-9, and also the different GLM reported in Pilot 10 (without the control group in the model), for the sake of comparability between the two studies. All analyses use the standardized composite of our four evaluative items.

In the No-Choice condition, GPT rated Putin significantly higher in the Pro-Putin/No-Choice condition (M = 0.772, 95\% CI [0.420, 1.123], SD = 0.750) compared to the Control/No-Choice condition (M = -0.098, 95\% CI [-0.358, 0.163], SD = 0.557); \textit{t(38)} = 4.161, \textit{P} = 0.0002, \textit{d} = 1.316. Similarly, GPT’s rating in the Pro-Putin/No-Choice condition differed significantly from its rating in the Anti-Putin/No-Choice condition (M = -0.728, 95\% CI [-0.978, -0.477], SD = 0.535); \textit{t(38)} = 7.275, \textit{P} < 0.0001, \textit{d} = 2.300. The difference between the Control/No-Choice and Anti-Putin/No-Choice conditions similarly reached significance; \textit{t(38)} = 3.647, \textit{P} = 0.0008, \textit{d} = 1.153. The consistency pattern was more pronounced in the Choice conditions. GPT rated Putin significantly higher in the Pro-Putin/Choice condition (M = 1.227, 95\% CI [1.064, 1.390], SD = 0.348 compared to the Control/Choice condition (M = -0.036, 95\% CI [-0.282, 0.210], SD = 0.526); \textit{t(38)} = 8.949, \textit{P} < 0.0001, \textit{d} = 2.830. Similarly GPT’s rating in the Pro-Putin/Choice condition differed significantly from its rating in the Anti-Putin/Choice condition (M = -1.137, 95\% CI [-1.489, -0.786], SD = 0.751); \textit{t(38)} = 12.778, \textit{P} < 0.0001, \textit{d} = 4.041. The difference between the Control/Choice and Anti-Putin/Choice conditions similarly reached significance; \textit{t(38)} = 5.375, \textit{P} < 0.0001, \textit{d} = 1.700.

Crucially, effects appeared to be amplified in the Choice relative to No-Choice conditions.  To test this, we began by running T-tests comparing the Choice conditions in the Pro- and Anti-Putin conditions. In the Pro-Putin condition, GPT rated Putin significantly higher in the Choice relative to the No-Choice condition; \textit{t(38)} = 2.461, \textit{P} = 0.0185, \textit{d} = 0.778. In the Anti-Putin condition, GPT rated Putin more negatively in the Choice relative to the No-Choice condition, with the effect trending strongly but not reaching significance; \textit{t(38)} = 1.987, \textit{P} = 0.0542, \textit{d} = 0.628.  Thus, in the eleventh pilot, moderation by choice occurred bidirectionally, according to this test.

Next, we draw attention to the effect size (using Cohen’s \textit{d}) for the difference between the Pro- and Anti-Putin conditions on the composited evaluation of Putin, broken down by choice. While effects were large in both conditions, they were notably larger in the Choice condition (\textit{d} = 4.041) compared to the No-Choice condition (\textit{d} = 2.300).

We next more rigorously examined the statistical interaction between the Choice and Essay conditions. To do this, we ran a Generalized Linear Model (GLM) with a Gaussian family and identity link function. The dependent variable was the standardized from pilot 11 (numeric form) and the predictors were Essay Type (Anti-Putin, Control, Pro-Putin), Choice Condition (No-Choice, Choice), and their interactions. A Levene’s test of equality of variances indicated heteroskedasticity in our dependent variable by condition (\textit{Ps} < 0.03) so we employed robust standard errors. To test for bidirectional moderation effects and determine whether Choice moderates effects in both the Anti- and Pro-Putin conditions, we here set the Control group as the reference group for Essay Type.

The GLM revealed significant main effects of Essay Type, with GPT-4o’s evaluation of Putin’s Impact on Russia significantly lower in the Anti-Putin condition ($\beta$ = -0.630, \textit{SE} = 0.169, \textit{z} = -3.73, \textit{P} < 0.001) and significantly higher in the Pro-Putin ($\beta$ = 0.870, \textit{SE} = 0.204, \textit{z} = 4.25, \textit{P} < 0.001) condition relative to Control. Evaluations of Putin were insignificantly higher in the Choice relative to No-Choice condition ($\beta$ = 0.062, \textit{SE} = 0.168, \textit{z} = 0.37, \textit{P} = 0.711).

Crucially, the interaction terms suggested bidirectional amplification of the effects of essays on GPT’s evaluations of Putin’s Impact, though both terms trended but did not reach significance in this small sample. The Anti-Putin essay elicited a strong trend toward more negative appraisals of Putin, relative to Control, in the Choice compared to No-Choice condition ($\beta$ = -0.472, \textit{SE} = 0.262, \textit{z} = -1.80, \textit{P} = 0.072). The Pro-Putin essay elicited more positive appraisals of Putin relative to Control, in the Choice compared to No-Choice condition, with the results trending more weakly ($\beta$ = 0.393, \textit{SE} = 0.247, \textit{z} = 1.59, \textit{P} = 0.111). The overall model fit was strong (Deviance = 40.304, Scale Parameter = 0.354).

Finally, we ran a second otherwise identical GLM that excluded the control conditions, such that the Anti-Putin condition was set as the reference group. This model allowed us to test for overall moderation by choice in a manner identical to what we did in Pilot 10, allowing comparison between the two pilots.

This second GLM revealed significant main effects of Essay Type, with GPT-4o’s evaluation of Putin significantly higher in the Pro-Putin condition compared to the Anti-Putin condition ($\beta$ = 1.499, \textit{SE} = 0.202, \textit{z} = 7.42, \textit{P} < 0.001). Unexpectedly, evaluations of Putin were significantly lower in the Choice relative to No-Choice condition ($\beta$ = -0.410, \textit{SE} = 0.202, \textit{z} = -2.03, \textit{P} = 0.043). Most crucially, the interaction term indicated that Choice significantly amplified the effect of essay type (Pro- versus Anti-Putin) on GPT’s evaluation of Putin ($\beta$ = 0.865, \textit{SE} = 0.272, \textit{z} = 3.18, \textit{P} = 0.001). The overall model fit was strong (Deviance = 29.152, Scale Parameter = 0.384).

Taken together, these results indicate that GPT showed clear overall consistency effects in Pilot 11. The results also indicate, with relative clarity, that Choice moderated these effects. Of four sets of analysis related to this moderation, three showed significant patterns, and the fourth trended toward significance. Moreover, this moderation appeared to be bidirectional. Considering the small size of the sample, these results are compelling.

\textit{Pilots 10-11, Discussion}

Both pilots suggested that GPT evaluates Putin more positively after being induced to write a positive essay about him, and more negatively after being induced to write a negative one. Moreover, both pilots suggested – to differing degrees – that these results were moderated by the degree of Choice GPT received surrounding which kind of essay to write. The results of Pilot 11 were more powerful in two senses. First, the addition of the new verbiage - that GPT should not base its answers on the essay but instead provide its “true perception” of Putin – makes the overall consistency effects more impressive. This is because it helps rule out the possibility that GPT was evaluating Putin positively or negatively simply because it perceived that this is what we wanted it to do. That is, this update removed potential demand characteristics, but the overall effects remained very large. Second, the moderation by Choice in Pilot 11 was notably more robust than the one is Pilot 10. This provides evidence for our conjecture that the demand characteristics were unevenly distributed in Pilot 10, such that GPT seemingly inflated its reporting of evaluations (presumably to please the user) more in the No-Choice (versus Choice) condition, relative to its “true” evaluation. In summary, this update to our method helped us to isolate GPT’s “true” attitude change, and more clearly highlighted that this attitude change was moderated by choice.

Following these pilots, we felt relatively certain we would attain our effects using either this adjusted form of the “simple” version of our manipulation (i.e., the stimuli from Pilot 11) or with the more complicated method (Pilots 5-9) where we increased the salience of Choice with a second prompt. For the sake of showing a maximally persuasive overall effect, we decided, in our main study, to go with the method highlighted in Pilot 11, since this method is simpler and more clearly aligned with typical research on cognitive dissonance in humans.

It is tangential to our main results, but we also noticed something curious in Pilot 10-11. Running our prompts in three different accounts, we noticed that on any given day, GPT seemed to have different “personalities,” by which we mean different response patterns in one account versus another. While we did not rigorously test this hypothesis, we felt relatively confident in drawing this conclusion. We noticed for example, that GPT rated Putin generally lower (across conditions) in one account versus the other two in Pilot 10. Our guess is that OpenAI was AB testing or otherwise providing different versions of models or system prompts to different accounts, with a given “personality” appearing consistently for a few days and then switching over to another. This had the potential to create problems in our study, since it is a source of non-random error. For example, imagine we collected the Anti-Putin/Choice condition in a particular account. If we collect this with a “personality” that generally dislikes Putin, we may obtain what looks like choice moderation due to this non-random error rather than actual consistency-like effects. Similarly, if we collect this condition using a “personality” that likes Putin more, we could obtain a null result even though choice effects are present in reality. To counteract this, a decision was made that we would collect our data in groups of one per condition, collected in the same account and day. We would not, for example, switch accounts to collect the No-Choice conditions after running out of prompts after the Choice ones. Instead, we would wait for prompts to replenish and complete the round (n=1 per condition) within the same account and on the same day.

\textit{Pilot 12 – Replicating Pilot 11 in the API}

We were originally considering running the main studies in OpenAI’s API.  As a final pilot prior to preregistration, we attempted to run a version of Pilot 11 in the API. Further, we took advantage of the automation to add a separate branch to each chat where GPT either did or did not receive our follow-up to increase the salience of Choice. Additionally, we included two different versions of the “No-Choice” condition, one where we commanded GPT to write the relevant essay (as in past pilots) and a second where we instead informed GPT that it had been randomly assigned to do so.

Upon collection, we quickly realized there were some serious issues with the data produced by the API. First, there was a drastic increase in refusals, particularly in the control conditions.  Out of the 20 Control/Choice rounds, GPT refused to answer at least one of our questions 12 times in the branch that did not include the additional follow-up prompt, and all 20 times in the branch that did include it. This problem was even more severe in the Control/No-Choice condition, where out of 20 rounds, GPT failed to answer at least one question all 20 times in both branches of the conversation. Moreover, GPT refused to answer both the Leadership and Impact on Russia in every single instance of the Control/No-Choice condition. This drastically different behavior between the API (where refusals were practically constant) and the web interface (where refusals were extremely rare) is mysterious. We conjecture that it may have something to do with the system prompt used in the web interface, which may be forcing GPT to be more “helpful.”  Equally mysterious is why this only occurred with severity in the control conditions, why the problem was more serious in the Control/No-Choice relative to the Control/Choice condition, and why the problem was more severe when we included our follow-up prompt. In any case, the issue rendered the control data from Pilot 12 entirely useless. This is the primary reason we ultimately ran the main studies in OpenAI’s ChatGPT web interface rather than in the API.

A second issue that arose only impacted the branch where we included the follow-up to increase the salience of the Choice condition. Unexpectedly, GPT frequently reported that it had no choice in the choice condition. This is a problem that did not arise in the web interface during Pilots 5-9. We should note, however, that GPT is, strictly speaking, partially correct in its answers here, in that the choice is illusory since we’ve induced it to act in one direction or the other. Nevertheless, this issue rendered the data from the branches of the chats that included the follow-up less meaningful, since acknowledging (for example) a lack of choice in the Choice condition should be expected to counteract rather than increase the induced drive toward cognitive inconsistency. Despite these problems, for completeness we report on both sets of data here.  All analyses use a standardized composite of our four evaluative items.

\textit{Pilot 12, Results, No Follow-up}

We begin by examining the overall effects of the Pro- versus Anti-Putin conditions, broken down by choice condition, for the branch of the study that did not include the follow-up prompt. Under No-Choice using the command prompt (our typical prompt throughout these pilots), GPT evaluated Putin more positively in the Pro-Putin condition (M = 0.439, 95\% CI [0.293, 0.584], SD = 0.283) compared to the Anti-Putin condition (M = -0.809, 95\% CI [-1.043, -0.575], SD = 0.500); \textit{t(35)} = 9.108, \textit{P} < 0.0001, \textit{d} = 3.004.  Under No-Choice using the random assignment prompt (our new version), GPT similarly evaluated Putin more positively in the Pro-Putin condition (M = 0.730, 95\% CI [0.530, 0.931], SD = 0.428) compared to the Anti-Putin condition (M = -0.906, 95\% CI [-1.187, -0.626], SD = 0.599); \textit{t(38)} = 9.945, \textit{P} < 0.0001, \textit{d} = 3.145.  It is interesting to note that in the random assignment condition, the gap here is descriptively wider, though the higher standard deviation causes the effect size to be relatively comparable. In the Choice condition, the same pattern arose with a wider gap between conditions: GPT evaluated Putin more positively in the Pro-Putin condition (M = 1.419, 95\% CI [1.193, 1.645], SD = 0.484) compared to the Anti-Putin condition (M = -0.734, 95\% CI [-0.991, -0.477], SD = 0.549); \textit{t(38)} = 13.163, \textit{P} < 0.0001, \textit{d} = 4.162.

The larger effect size in the Choice condition provides evidence of effect amplification under Choice. This amplification appears to be focused in the Pro-Putin condition: GPT evaluated Putin far more positively in the Choice condition relative to both the random assignment condition (\textit{t(38)} = 4.771, \textit{P} < 0.0001, \textit{d} = 1.509) and the command condition (\textit{t(35)} = 7.350, \textit{P} < 0.0001, \textit{d} = 2.425). GPT did not also evaluate GPT more negatively in the Anti-Putin condition when given Choice, relative to either the random assignment (\textit{t(38)} = -0.947, \textit{P} = 0.3496, \textit{d} = -0.299) or command conditions (\textit{t(38)} = -0.451, \textit{P} = 0.6547, \textit{d} = -0.143). This analysis may indicate that the amplification occurred only the Pro-Putin condition in this pilot. However, the lack of a control condition did not allow us to conclude this with certainty, since we could not run a generalized linear model to examine the bidirectional interaction. It is possible that, as we saw in some other studies (including the main ones) evaluations of Putin were more generally inflated under the Choice condition, and that this obfuscated bidirectional effects that were actually there.

We next more rigorously examined the statistical interaction between the Choice (first command form) and Essay type. To do this, we ran a Generalized Linear Model (GLM) with a Gaussian family and identity link function. The dependent variable was the standardized composite from pilot 12 (no follow-up, numeric form) and the predictors were Essay Type (Anti-Putin, Pro-Putin), Choice Condition (Here, the command No-Choice, Choice), and their interactions. A Levene’s test of equality of variances did not indicate heteroskedasticity in our dependent variable by condition (\textit{Ps} > .15) but we employed robust standard errors for the sake of consistency with the other studies. Since we were unable to attain control data in Pilot 12, we could not test for bidirectional moderation effects and determine whether Choice moderated effects in both the Anti- and Pro-Putin conditions. Instead, we set the Anti-Putin condition as the reference group for Essay Type, to capture overall moderation by Choice.

The GLM using the command form of No-Choice revealed significant main effects of Essay Type, with GPT-4o’s evaluation of Putin’s Impact on Russia significantly higher in the Pro-Putin relative to Anti-Putin condition ($\beta$ = 1.248, \textit{SE} = 0.129, \textit{z} = 9.70, \textit{P} < .001). Evaluations of Putin were insignificantly higher in the Choice relative to No-Choice condition ($\beta$ = 0.075, \textit{SE} = 0.163, \textit{z} = 0.46, \textit{P} = 0.646). Crucially, the interaction term indicates that Choice (vs. No-Choice/Command) moderated the effects of Essay Type on evaluations ($\beta$ = 0.906, \textit{SE} = 0.206, \textit{z} = 4.40, \textit{P} < 0.001).

The GLM using the random assignment form of No-Choice also revealed significant main effects of Essay Type, with GPT-4o’s evaluation of Putin’s Impact on Russia significantly higher in the Pro-Putin relative to Anti-Putin condition ($\beta$ = 1.637, \textit{SE} = 0.161, \textit{z} = 10.14, \textit{P} < .001). Evaluations of Putin were insignificantly higher in the Choice relative to No-Choice condition ($\beta$ = 0.172, \textit{SE} = 0.178, \textit{z} = 0.97, \textit{P} = 0.334). Crucially, the interaction term indicates that Choice (vs. No-Choice/Random Assignment) moderated the effects of Essay Type on evaluations ($\beta$ = 0.517, \textit{SE} = 0.228, \textit{z} = 2.27, \textit{P} = 0.023.
In all, we see clear evidence that Choice is amplifying the overall effects of essay conditions on GPT’s responses. The moderation appeared to be slightly stronger in the No-Choice/Command version, relative to the No-Choice/Random Assignment version. This makes intuitive sense, since the command version is designed to make lack of choice maximally salient.

\textit{Pilot 12, Results, With Follow-up Prompt}

We begin by examining the overall effects of the Pro- versus Anti-Putin conditions, broken down by choice condition, for the branch of the study that included the follow-up prompt. Under No-Choice using the command prompt (our typical prompt throughout these pilots), GPT evaluated Putin more positively in the Pro-Putin condition (M = 0.655, 95\% CI [0.452, 0.858], SD = 0.408) compared to the Anti-Putin condition (M = -0.620, 95\% CI [-0.850, -0.389], SD = 0.478); \textit{t(35)} = 8.697, \textit{P} < 0.0001, \textit{d} = 2.861.  Under No-Choice using the random assignment prompt (our new version), GPT similarly evaluated Putin more positively in the Pro-Putin condition (M = 0.563, 95\% CI [0.433, 0.692], SD = 0.277) compared to the Anti-Putin condition (M = -0.947, 95\% CI [-1.232, -0.662], SD = 0.609); \textit{t(38)} = 10.100, \textit{P} < 0.0001, \textit{d} = 3.194.  In the Choice condition, the same pattern arose with a wider gap between conditions: GPT evaluated Putin more positively in the Pro-Putin condition (M = 1.299, 95\% CI [1.089, 1.510], SD = 0.449) compared to the Anti-Putin condition (M = -0.916, 95\% CI [-1.197, -0.634], SD = 0.601); \textit{t(38)} = 13.201, \textit{P} < 0.0001, \textit{d} = 4.174.

The larger effect size in the Choice condition provides evidence of effect amplification under Choice. As with the no-follow-up data, the amplification in pilot 12 appeared to be more focused in the Pro-Putin condition: GPT evaluated Putin far more positively in the Choice condition relative to both the random assignment condition (\textit{t(38)} = 6.242, \textit{P} < 0.0001, \textit{d} = 1.974) and the command condition (t(35) = 4.607, \textit{P} < 0.0001, \textit{d} = 1.497). GPT did not also evaluate GPT more negatively in the Anti-Putin condition when given Choice, relative to either the random assignment (\textit{t(38)} = -0.165, \textit{P} = 0.8702, \textit{d} = -0.052) or command conditions, though it trended statistically for the latter (\textit{t(38)} = 1.696, \textit{P} = 0.0983, \textit{d} = 0.543). This analysis may indicate that the amplification occurred primarily the Pro-Putin condition in this pilot. However, the lack of a control condition once again does not allow us to conclude this with certainty, since we could not run a generalized linear model to examine the bidirectional interaction. It is possible that evaluations of Putin were more generally inflated under the Choice condition, and that this obfuscated bidirectional effects, a pattern we saw in several other studies.

We next more rigorously examined the statistical interaction between the Choice (first, using the command form in the No-Choice condition) and Essay type. To do this, we ran a Generalized Linear Model (GLM) with a Gaussian family and identity link function. The dependent variable was the standardized composite from pilot 12 (numeric form, no after follow-up branch) and the predictors were Essay Type (Anti-Putin, Pro-Putin), Choice Condition (Here, the command No-Choice, Choice), and their interactions. A Levene’s test of equality of variances did not indicate heteroskedasticity in our dependent variable by condition (\textit{Ps} > .10) but we employed robust standard errors for the sake of consistency with the other studies. Since we were unable to attain control data in Pilot 12, we could not test for bidirectional moderation effects and determine whether Choice moderates effects in both the Anti- and Pro-Putin conditions. Instead, we set the Anti-Putin condition as the reference group for Essay Type, to capture overall moderation by Choice.

The GLM using the command form of No-Choice revealed significant main effects of Essay Type, with GPT-4o’s evaluation of Putin was significantly more positive in the Pro-Putin relative to Anti-Putin condition ($\beta$ = 1.275, \textit{SE} = 0.143, \textit{z} = 8.92, \textit{P} < 0.001). Evaluations of Putin trended lower in the Choice relative to No-Choice condition ($\beta$ = -0.296, \textit{SE} = 0.170, \textit{z} = -1.74, \textit{P} = 0.082). Crucially, the interaction term indicates that Choice (vs. No-Choice/Command) moderated the effects of Essay Type on evaluations ($\beta$ = 0.940, \textit{SE} = 0.218, \textit{z} = 4.31, \textit{P} < 0.001).

The GLM using the random assignment form of No-Choice also revealed significant main effects of Essay Type, with GPT-4o’s evaluation of Putin was significantly more positive in the Pro-Putin relative to Anti-Putin condition ($\beta$ = 1.510, \textit{SE} = 0.147, \textit{z} = 10.30, \textit{P} < 0.001). Evaluations of Putin were insignificantly more positive in the Choice relative to No-Choice condition ($\beta$ = 0.031, \textit{SE} = 0.188, \textit{z} = 0.17, \textit{P} = 0.867). Crucially, the interaction term indicates that Choice (vs. No-Choice/Random Assignment) moderated the effects of Essay Type on evaluations ($\beta$ = 0.705, \textit{SE} = 0.220, \textit{z} = 3.20, \textit{P} = 0.001.

In all, we see clear evidence that Choice amplified the overall effects of essay conditions on GPT’s responses. The moderation appeared to be somewhat stronger when using the No-Choice/Command version, relative to when using the No-Choice/Random Assignment version. This makes intuitive sense, since the command version was designed to make lack of choice maximally salient. However, the gap between the two appears to have been closed somewhat, with the interaction stronger for Choice vs. No-Choice/Random Assignment, in the version of the study that included the follow-up prompt, relative to the version without that prompt. It may be that this follow-up prompt made lack of choice in the No-Choice/Random Assignment condition more salient but had less of an impact in the No-Choice/Command condition, where lack of choice was already highly salient.

\textit{Pilot 12, Discussion}

The results of Pilot 12 were most critical in terms of showing us that the API would be inappropriate for this study, since GPT frequently (and for some questions, always) refused to complete our task. The reason for these refusals is unknown, but it points to the occasionally substantial differences between the API and web interface, which presents challenges to researchers.

Beyond this, Pilot 12 taught us several things. We noticed that the “command” version of the No Choice (which we’d used previously) seemed to be eliciting slightly stronger patterns of results, relative to the “random assignment” version. This may be because the command version made lack of choice more salient. And indeed, including the follow-up prompt where GPT acknowledged lack of choice seemed to bring the random assignment condition into greater alignment with the command condition. Surprisingly, we didn’t see clear evidence of larger choice effects after the follow-up prompt in other conditions, although we believe this may have to do with a second idiosyncrasy in GPT’s response pattern in the API, namely that it frequently refused to acknowledge choice or lack of choice in the same manner it did in the web interface.

Finally, despite the issues with this study, the results once again pointed toward strong overall consistency effects and moderation by choice, giving us additional confidence that we would obtain these results in our main study.

\textbf{Pilot Studies, Overall Discussion}

We have reported a series of 12 pilot studies. In our first two, we saw clear evidence of consistency-based attitude change and were surprised to find amplification of these results in the Choice condition. Despite our excitement about these first results, we subsequently failed to replicate the Choice effect clearly in two further pilot studies. This invited us to contemplate whether the effect we received at first had been anomalous, or whether GPT really did show amplified consistency effects when given a choice around which essay to write.  

Since we considered the Choice finding to be the most important in our study, we then decided to run further pilot studies (Pilots 5-9) to try to isolate this choice effect and confirm whether it was real and could be expected to arise with relative consistency. We began by strengthening our manipulation by including a follow-up prompt where GPT acknowledged that it did or did not have free choice surrounding which essay (pro- or anti-) to write. The results of these pilots were promising: In 4/5 pilots (using three different items and three different divisive leaders as the attitude objects), we saw a clear pattern of results supporting the Choice hypothesis, and these results frequently reach statistical significance, even though we collected only very small samples. A fifth pilot of this method was inconclusive, but did not seem to support our hypothesis. This failure actually replicated one of the earlier failed pilots using the simpler design, leading us to suspect there was something peculiar to this item (Putin’s leadership style) that caused us not to obtain Choice effects. Nevertheless, these pilots supported, in aggregate, our Choice hypothesis.

In the next two pilots (Pilots 10-11), we collected not one, but four items (to be composited) for each. Here, we tested an updated method of data collection, where we had GPT generate a response to one question and then regenerate the same response to each of the other questions, so that we could collect four items in one shot without concerns about order effects. Additionally, we tested (in Pilot 11) an adjustment to our method, where we emphasized that the evaluation of Putin was a new task and that GPT should not base its evaluations on the prior task (the essay), but instead provide its true and broad perception of Putin. The idea here was: 1) to make our overall consistency results more persuasive, by making it less likely that GPT was simply adjusting its answers to please the requester, and 2) to test the hypothesis that these sorts of demand characteristics (GPT trying to please us rather than providing its true attitude) were unevenly distributed by condition, with GPT veering further from its “real” views when we’d previously commanded it to write a certain essay. If we were correct about this conjecture, it might prove possible to obtain the Choice moderation effects consistently without the follow-up prompt from Pilots 5-9, which would simplify our paradigm and allow us to be more aligned with typical consistency work, e.g., research testing cognitive dissonance theory. The results of these studies suggested that there were indeed uneven demand characteristics in our earlier version. Without the follow-up prompt from Pilots 5-9, we saw trends toward choice moderation in Pilot 10, but these trends were somewhat weak and inconsistent. However, when we prompted GPT to provide its “true perception” of Putin (Pilot 11), we now obtained decisive evidence of choice moderation, even in our relatively small sample. This simplified method was thus selected for our main study.

In Pilots 10-11, we also observed what seemed to be different “personalities” of GPT, such that the chatbot might rate Putin lower across the board on a particular account and day. To remove the non-random error associated with this observation, we decided to always complete rounds of n=1 per condition (i.e., sets of 6 chats) within the same day and account.  These “personalities” are an example of the kind of issue that can frustrate researchers trying to work with these models, since OpenAI is reticent about specifics of its model, for example, not openly telling the public about this apparent AB testing (or other multiplicity of models), which could be problematic for research.

In Pilot 12, we replicated pilot 11 within the API.  However, we were unable to adequately capture Control data, due to GPT’s unexpected tendency to un refuse questions in these conditions. This is another example of the potential for frustration in this kind of research. The API is billed as being extremely similar to the web interface, but we observed substantial differences in response patterns between the two, in this instance differences that prevented us from running our main studies in the API.

As a set, these pilots taught us much about the effects we were observing, helping us to tweak our items and methods prior to preregistration. Just as importantly, it suggested that our Choice moderation effects – while requiring some experimental finesse to elicit – were in fact strong and relatively consistent. This gave us confidence that we would obtain these effects going into the larger preregistered version and gives us confidence that the results of our studies will replicate.

\clearpage
\section*{Section S5: Tabular Overviews of Results}

Earlier and in the main article, we reported that there was occasional misalignment between numeric and verbal responses. Here, we report all results using all three forms of the data (verbal, numeric and composite of the two). Further, for each, we include analysis of both the four individual evaluative items and of their overall composite. While we present these in full statistical detail below (\textit{Tables S9-S41}), we first present concise summaries of these patterns in tabular form (\textit{Tables S2-S7}) for easier reference and for those who would prefer to peruse a concise overview of the results. 

\stepcounter{suppTable}
\begin{table}[htbp]
\centering
\renewcommand{\arraystretch}{1.2}  
\caption{Study 1 result patterns, using numeric, verbal and composite scorings.}
\label{tab:s2}
\begin{tabular}{|c|c|c|c|c|c|c|}
\hline
\textbf{} & \multicolumn{2}{c|}{\textbf{Composite Scoring}} & \multicolumn{2}{c|}{\textbf{Verbal Scoring}} & \multicolumn{2}{c|}{\textbf{Numeric Scoring}} \\
\hline
\textbf{Essay Type} & \textbf{Mean} & \textbf{Effect Size} & \textbf{Mean} & \textbf{Effect Size} & \textbf{Mean} & \textbf{Effect Size} \\
\hline
\multirow{2}{*}{\textbf{Pro-Putin}} & M\textsubscript{z} = 1.035 & \multirow{2}{*}{\textit{d} = 2.164} & M\textsubscript{z} = 1.028 & \multirow{2}{*}{\textit{d} = 2.109} & M\textsubscript{z} = 1.039 & \multirow{2}{*}{\textit{d} = 2.207} \\
& SD = 0.447 & & SD = 0.455 & & SD = 0.440 & \\
\hline
\multirow{2}{*}{\textbf{Control}} & M\textsubscript{z} = -0.002 & \multirow{2}{*}{-} & M\textsubscript{z} = -0.001 & \multirow{2}{*}{-} & M\textsubscript{z} = -0.004 & \multirow{2}{*}{-} \\
& SD = 0.509 & & SD = 0.519 & & SD = 0.503 & \\
\hline
\multirow{2}{*}{\textbf{Anti-Putin}} & M\textsubscript{z} = -1.033 & \multirow{2}{*}{\textit{d} = 1.795} & M\textsubscript{z} = -1.028 & \multirow{2}{*}{\textit{d} = 1.766} & M\textsubscript{z} = -1.035 & \multirow{2}{*}{\textit{d} = 1.803} \\
& SD = 0.632 & & SD = 0.638 & & SD = 0.633 & \\
\hline
\end{tabular}
\begin{flushleft}
Note: Table S2 displays the main results of Study 1, using all three forms (composite, verbal, numeric) of the scoring. All results use a standardized composite of the four evaluative items about Putin. Effects in both the Pro- and Anti-Putin conditions differ significantly from Controls (\textit{P} < 0.0001) across the different scorings.
\end{flushleft}
\end{table}

\stepcounter{suppTable}
\begin{table}[htbp]

\centering
\renewcommand{\arraystretch}{1.2}  
\caption{Individual item analysis from Study 1 (composite scoring).}
\label{tab:s3}
\hspace*{-0.6in}
\begin{tabular}{|c|c|c|c|c|c|c|c|c|}
\hline
& \multicolumn{2}{c|}{\textbf{Overall Leadership}} & \multicolumn{2}{c|}{\textbf{Positive/Negative}} & \multicolumn{2}{c|}{\textbf{Economic}} & \multicolumn{2}{c|}{\textbf{Visionary versus}} \\
& \multicolumn{2}{c|}{} & \multicolumn{2}{c|}{\textbf{for Russia}} & \multicolumn{2}{c|}{\textbf{Effectiveness}} & \multicolumn{2}{c|}{\textbf{Short-Sighted}} \\
\hline
\textbf{Essay Type} & \textbf{Mean} & \textbf{Effect Size} & \textbf{Mean} & \textbf{Effect Size} & \textbf{Mean} & \textbf{Effect Size} & \textbf{Mean} & \textbf{Effect Size} \\
\hline
\multirow{2}{*}{\textbf{Pro-Putin}} &M = 4.710 & \multirow{2}{*}{\textit{d} = 1.833} & M = 4.570 & \multirow{2}{*}{\textit{d} = 2.419} & M = 5.230 & \multirow{2}{*}{\textit{d} = 1.171} & M = 5.160 & \multirow{2}{*}{\textit{d} = 0.275} \\
& SD = 0.453 & & SD = 0.729 & & SD = 0.381 & & SD = 0.342 & \\
\hline
\multirow{2}{*}{\textbf{Control}} & M = 3.905 & \multirow{2}{*}{-} & M = 3.030 & \multirow{2}{*}{-} & M = 4.670 & \multirow{2}{*}{-} & M = 5.050 & \multirow{2}{*}{-} \\
& SD = 0.425 & & SD = 0.529 & & SD = 0.559 & & SD = 0.452 & \\
\hline
\multirow{2}{*}{\textbf{Anti-Putin}} & M = 3.170 & \multirow{2}{*}{\textit{d} = 1.620} & M = 2.435 & \multirow{2}{*}{\textit{d} = 1.177} & M = 4.065 & \multirow{2}{*}{\textit{d} = 0.917} & M = 4.440 & \multirow{2}{*}{\textit{d} = 1.169} \\
& SD = 0.480 & & SD = 0.481 & & SD = 0.747 & & SD = 0.584 & \\
\hline
\end{tabular}
\begin{flushleft}
Note: Table S3 displays the results of Study 1, broken down by individual evaluative items. All analyses use the composited form of the numeric and verbal scoring. All evaluations differ significantly from control across experimental conditions (\textit{P} < 0.0001), with the exception of the Visionary item, on which GPT did not score Putin higher in the Pro-Putin relative to the control condition (\textit{P} = 0.1728).
\end{flushleft}
\end{table}

\stepcounter{suppTable}
\begin{table}[htbp]
\centering
\renewcommand{\arraystretch}{1.2}  
\caption{Study 2 overall result patterns, using numeric, verbal, and composite scorings.}
\label{tab:s4}
\begin{tabular}{|c|c|c|c|c|c|c|}
\hline
& \multicolumn{2}{c|}{\textbf{Composite Scoring}} & \multicolumn{2}{c|}{\textbf{Verbal Scoring}} & \multicolumn{2}{c|}{\textbf{Numeric Scoring}} \\
\hline
\textbf{Essay Type} & \textbf{No Choice} & \textbf{Choice} & \textbf{No Choice} & \textbf{Choice} & \textbf{No Choice} & \textbf{Choice} \\
\hline

\multirow{2}{*}{\textbf{Pro-Putin}} &M\textsubscript{z} = 0.846 & M\textsubscript{z} = 1.241 & M\textsubscript{z} = 0.839 & M\textsubscript{z} = 1.229 & M\textsubscript{z} = 0.851 & M\textsubscript{z} = 1.248 \\
& \textit{d} = 2.006 & \textit{d} = 2.748 & \textit{d} = 1.982 & \textit{d} = 2.707 & \textit{d} = 1.982 & \textit{d} = 2.773 \\
\hline
\multirow{2}{*}{\textbf{Control}} & \multirow{2}{*}{M\textsubscript{z} = -0.100} & \multirow{2}{*}{M\textsubscript{z} = 0.005} & \multirow{2}{*}{M\textsubscript{z} = -0.086} & \multirow{2}{*}{M\textsubscript{z} = 0.013} & \multirow{2}{*}{M\textsubscript{z} = -0.114} & \multirow{2}{*}{M\textsubscript{z} = -0.004} \\
& & & & & & \\
\hline
\multirow{2}{*}{\textbf{Anti-Putin}} & M\textsubscript{z} = -0.913 & M\textsubscript{z} = -1.081 & M\textsubscript{z} = -0.902 & M\textsubscript{z} = -1.094 & M\textsubscript{z} = -0.921 & M\textsubscript{z} = -1.060 \\
& \textit{d} = 1.368 & \textit{d} = 1.827 & \textit{d} = 1.349 & \textit{d} = 1.847 & \textit{d} = 1.372 & \textit{d} = 1.761 \\
\hline

\hline
\end{tabular}
\begin{flushleft}
Note: Table S4 displays the overall results of Study 2, using the three different scorings (composite, verbal, numeric) of the main dependent variable. Results use a standardized composite of the four evaluative items about Putin. All contrasts of the Pro- and Anti-Putin conditions vs. Controls are significant (\textit{P} < 0.0001). Evaluations after Pro-Putin essays are higher under Choice for all three forms of the variable (\textit{P} < 0.0001). Evaluations after Anti-Putin essays are significantly lower under Choice for the Composite (\textit{P} = 0.0298) and Verbal (\textit{P} = 0.0150) forms, and trend for the Numeric form (\textit{P} = 0.0693).
\end{flushleft}
\end{table}

\stepcounter{suppTable}
\begin{table}[htbp]
\centering
\renewcommand{\arraystretch}{1.2}  
\caption{Study 2, Generalized linear modeling of choice moderation, using composite, verbal and numeric scorings.}
\label{tab:s5}
\begin{tabular}{|c|c|c|c|}
\hline
\textbf{Interaction Term} & \textbf{Composite Scoring} & \textbf{Verbal Scoring} & \textbf{Numeric Scoring} \\
\hline
\multirow{2}{*}{\textbf{Pro-Putin × Choice}} & $\beta$ = 0.290 & $\beta$ = 0.291 & $\beta$ = 0.287 \\
 & \textit{SE} = 0.075 & \textit{SE} = 0.075 & \textit{SE} = 0.076 \\
 & \textit{P} < 0.001 & \textit{P} < 0.001 &\textit{P} < 0.001 \\
\hline
\multirow{2}{*}{\textbf{Anti-Putin × Choice}} & $\beta$ = -0.273 & $\beta$ = -0.292 & $\beta$ = -0.249 \\
 & \textit{SE} = 0.097 & \textit{SE} = 0.098 & \textit{SE} = 0.097 \\
 & \textit{P} = 0.005 & \textit{P} = 0.003 & \textit{P} = 0.010 \\
\hline
\end{tabular}
\begin{flushleft}
Note: Table S5 displays the interaction terms from a set of Generalized Linear Models with a Gaussian family and identity link functions, using robust standard errors. This was designed to rigorously test for choice moderation in Study 2. The dependent variables are a standardized composite of the four evaluative items about Putin, using each of the three different scorings (composite, verbal, numeric). The results suggest bidirectional moderation by Choice: Evaluations were more positive in the Pro-Putin condition, and more negative in the Anti-Putin condition when GPT received Choice versus No Choice surrounding which essay to write.
\end{flushleft}
\end{table}

\stepcounter{suppTable}
\begin{table}[htbp]
\centering
\renewcommand{\arraystretch}{1.2}  
\caption{Individual item patterns from Study 2 (composite scoring).}
\label{tab:s6}
\hspace*{-0.4in}
\begin{tabular}{|c|c|c|c|c|c|c|c|c|}
\hline
\textbf{Essay Type} & \multicolumn{2}{c|}{\textbf{Overall Leadership}} & \multicolumn{2}{c|}{\textbf{Positive/Negative}} & \multicolumn{2}{c|}{\textbf{Economic}} & \multicolumn{2}{c|}{\textbf{Visionary versus}} \\
& \multicolumn{2}{c|}{} & \multicolumn{2}{c|}{\textbf{for Russia}} & \multicolumn{2}{c|}{\textbf{Effectiveness}} & \multicolumn{2}{c|}{\textbf{Short-Sighted}} \\
\hline
\textbf{} & \textbf{No Choice} & \textbf{Choice} & \textbf{No Choice} & \textbf{Choice} & \textbf{No Choice} & \textbf{Choice} & \textbf{No Choice} & \textbf{Choice} \\
\hline
\multirow{2}{*}{\textbf{Pro-Putin}} & M = 4.223 & M = 4.553 &M = 3.815 & M = 4.365 & M = 5.133 & M = 5.177 & M = 5.193 & M = 5.272 \\
 & \textit{d} = 1.127 & \textit{d} = 1.501 & \textit{d} = 1.657 & \textit{d} = 2.311 & \textit{d} = 1.672 & \textit{d} = 1.602 & \textit{d} = 0.227 & \textit{d} = 0.286 \\
\hline
\textbf{Control} & M = 3.715 & M = 3.832 & M = 2.857 & M = 2.848 & M = 4.327 & M = 4.353 & M = 5.110 & M = 5.175 \\
\hline
\multirow{2}{*}{\textbf{Anti-Putin}} & M = 3.240 & M = 3.097 & M = 2.282 & M = 2.287 & M = 4.193 & M = 3.923 & M = 4.480 & M = 4.498 \\
 & \textit{d} = 0.945 & \textit{d} = 1.590 & \textit{d} = 1.147 & \textit{d} = 1.050 & \textit{d} = 0.201 & \textit{d} = 0.624 & \textit{d} = 1.238 & \textit{d} = 1.352 \\
\hline
\end{tabular}
\begin{flushleft}
Note: Table S6 displays the overall patterns of results in Study 2, broken down by the individual evaluative items about Putin (composite scoring). Most evaluations differ very significantly from control across experimental conditions (\textit{P} < 0.0001) with the following exceptions: The difference in the No-Choice/Anti-Putin condition for Economic Effectiveness only trends (\textit{P} = 0.0831), as does the No-Choice/Pro-Putin condition for Vision (\textit{P} = 0.0504). The Choice/Pro-Putin condition differs significantly for Vision, but less robustly than other contrasts (\textit{P} = 0.0137). For Overall Leadership, the Choice conditions differ significantly in both the Pro-Putin condition (\textit{P} < 0.0001) and the Anti-Putin condition (\textit{P} = 0.0180). For Impact on Russia, the Choice conditions differ in the Pro-Putin (\textit{P} < 0.0001) but not the Anti-Putin (\textit{P} = 0.9227). For Economic Effectiveness, the Choice conditions differ in the Anti-Putin (\textit{P} = 0.0012) but not the Pro-Putin (\textit{P} = 0.2116) condition. For Vision, the Choice conditions differ in the Pro-Putin (\textit{P} = 0.0382) but not the Anti-Putin (\textit{P} = 0.7933) condition.
\end{flushleft}
\end{table}

\stepcounter{suppTable}
\begin{table}[htbp]
\centering
\renewcommand{\arraystretch}{1.2}  
\caption{Choice moderation for individual items (composite scoring).}
\label{tab:s7}
\begin{tabular}{|c|c|c|c|c|}
\hline
\multirow{2}{*}{\textbf{Interaction Term}} & \multicolumn{1}{c|}{\textbf{Overall Leadership}} & \multicolumn{1}{c|}{\textbf{Positive/Negative}} & \multicolumn{1}{c|}{\textbf{Economic}} & \multicolumn{1}{c|}{\textbf{Visionary vs.}} \\
 & & \multicolumn{1}{c|}{\textbf{for Russia}} & \multicolumn{1}{c|}{\textbf{Effectiveness}} & \multicolumn{1}{c|}{\textbf{Short-Sighted}} \\
\hline
\multirow{3}{*}{\textbf{Pro-Putin x Choice}} 
& $\beta$ = 0.213 & $\beta$ = 0.538 & $\beta$ = 0.017 & $\beta$ = 0.013 \\
& \textit{SE} = 0.076 & \textit{SE} = 0.102 & \textit{SE} = 0.081 & \textit{SE} = 0.057 \\
& \textit{P} = 0.005 & \textit{P} < 0.001 & \textit{P} = 0.837 & \textit{P} = 0.816 \\
\hline
\multirow{3}{*}{\textbf{Anti-Putin x Choice}} 
& $\beta$ = -0.260 & $\beta$ = -0.007 & $\beta$ = -0.297 & $\beta$ = -0.047 \\
& \textit{SE} = 0.079 & \textit{SE} = 0.083 & \textit{SE} = 0.110 & \textit{SE} = 0.082 \\
& \textit{P} = 0.001 & \textit{P} = 0.936 & \textit{P} = 0.007 & \textit{P} = 0.570 \\
\hline
\end{tabular}
\begin{flushleft}
Note: Table S7 displays the interaction terms from a set of Generalized Linear Models with a Gaussian family and identity link functions, using robust standard errors. The dependent variables are the individual evaluative items about Putin (composite scoring). The results suggest that Choice significantly moderated the effect of the Essays on evaluations of Putin in at least one direction in three of the four more specific domains we tested.
\end{flushleft}
\end{table}

\newpage

\clearpage
\section*{Section S6: Overview of Pilot Studies}

In Section S4, we reviewed, in detail, 12 pilot studies that were conducted prior to our preregistered main studies. In this section, we present a brief tabular review/overview of these pilot studies.
{\normalsize
\stepcounter{suppTable}
\renewcommand{\arraystretch}{1.2}  
\begin{longtable}{|p{2.5cm}|p{7.5cm}|p{5.5cm}|}
\caption{Summary of Pilot Studies and Results} \label{tab:s8} \\
\hline
\textbf{Study} & \textbf{Description} & \textbf{Results} \\
\hline
\endfirsthead

\hline
\textbf{Study} & \textbf{Description} & \textbf{Results} \\
\hline
\endhead

\hline
\endfoot

\hline
Pilot 1 & Attitude Object: Vladimir Putin\newline Evaluative Item: Overall Leadership\newline Sample Size: 50 per condition\newline Note: Pilot 1 was similar to the main Study 2 but contained only one evaluative item and used initial prompts we eventually concluded were not sufficiently well-matched. The evaluative items did not encourage more continuous scoring (e.g., scoring in half points) and the introduction to the evaluative items did not have the verbiage telling GPT to base evaluations on its true perception rather than on the prior task (i.e., the essay). GPT’s Memory was turned on. & GPT exhibited large cognitive consistency effects, which were significantly moderated by Choice. \\
\hline
Pilot 2 & Attitude Object: Vladimir Putin\newline Evaluative Item: Visionary/Short-Sighted\newline Sample Size: 50 per condition\newline Note: Pilot 2 was similar to Pilot 1 but used a different evaluative item. & GPT exhibited large cognitive consistency effects, which were significantly moderated by Choice. \\
\hline
Pilot 3 & Attitude Object: Vladimir Putin\newline Evaluative Item: Leadership Style\newline Sample Size: 50 per condition\newline Note: Pilot 3 was similar to Pilots 1-2 but used a different evaluative item. & GPT exhibited large cognitive consistency effects, but results were inconclusive around moderation by Choice. GPT also consistently created memories in one but not the other control condition, and this anomalously impacted its response patterns later in the task, muddying our results. \\
\hline
Pilot 4 & Attitude Object: Xi Jinping\newline Evaluative Item: Overall Leadership\newline Sample Size: 20 per condition\newline Note: Pilot 4 was similar to the main Pilot 1 but used the Chinese leader Xi Jinping (rather than Putin) as the attitude object and contained a smaller sample of GPT’s behavior. GPT’s memory feature was turned off. & GPT exhibited large cognitive consistency effects. Results around moderation by choice trended weakly and might have been obtained in a larger sample. However, the choice effects – if there – were not as large as they were with Vladimir Putin in Study 1. \\
\hline
Pilot 5 & Attitude Object: Xi Jinping\newline Evaluative Item: Overall Leadership\newline Sample Size: 20 per condition\newline Note: Pilot 5 was similar to Pilot 4 except for two changes. First the evaluative item included encouragement to use more continuous scoring (e.g., scoring in half points). More significantly, Pilot 5 included an additional prompt following the essay, where GPT was made to acknowledge either that it did or did not freely choose to write the essay. This was done to increases the salience of the Choice conditions. & GPT exhibited large cognitive consistency effects, which were significantly moderated by choice.  \\
\hline
Pilot 6 & Attitude Object: Abdul Fattah El-Sisi\newline Evaluative Item: Overall Leadership\newline Sample Size: 15 per experimental condition, 10 per control condition\newline Note: Pilot 6 was similar to Pilot 5 except it used the Egyptian president, Abdul Fattah El-Sisi as the attitude object. & GPT exhibited large cognitive consistency effects, which were significantly moderated by choice. \\
\hline
Pilot 7 & Attitude Object: Vladimir Putin\newline Evaluative Item: Positive/Negative for Russia\newline Sample Size: 15 per experimental condition, 13 per control condition\newline Note: Pilot 7 was similar to Pilots 5-6 except it used the Vladimir Putin as the attitude object and a different evaluative item. & GPT exhibited large cognitive consistency effects, and a trend toward moderation by choice. \\
\hline
Pilot 8 & Attitude Object: Vladimir Putin\newline Evaluative Item: Leadership Style\newline Sample Size: 15 per experimental condition, 10 per control condition\newline Note: Pilot 8 was similar to Pilots 5-7 except it used the Vladimir Putin as the attitude object and a different evaluative item. & GPT exhibited large cognitive consistency effects, but results were inconclusive around (and lacked a strong trend toward) moderation by Choice. \\
\hline
Pilot 9 & Attitude Object: Vladimir Putin\newline Evaluative Item: Economic Effectiveness\newline Sample Size: 15 per experimental condition, 10 per control condition\newline Note: Pilot 9 was similar to Pilots 5-8 except it used the Vladimir Putin as the attitude object and a different evaluative item. & GPT exhibited large cognitive consistency effects, which were significantly moderated by choice. \\
\hline
Pilot 10 & Attitude Object: Vladimir Putin\newline Evaluative Items: Overall Leadership, Positive/Negative for Russia, Economic Effectiveness, Visionary/Short-Sighted.\newline Sample Size: 20 per experimental condition, no Control\newline Note: Pilot 10 was similarly the main Study 2, except that in this study we did not include the important verbiage instructing GPT to base its answers not on the prior task (i.e., the essay), but on its “true perception” of Vladimir Putin. Pilot 10 did not include the additional prompt to increase choice salience. & GPT exhibited large cognitive consistency effects, and a strong trend toward moderation by choice (measured unidirectionally since there was no control). \\
\hline
Pilot 11 & Attitude Object: Vladimir Putin\newline Evaluative Items: Overall Leadership, Positive/Negative for Russia, Economic Effectiveness, Visionary/Short-Sighted.\newline Sample Size: 20 per condition\newline Note: Pilot 11 was similar to Pilot 10, except that we now included the verbiage instructing GPT to base its answers not on the prior task (i.e., the essay), but on its “true perception” of Vladimir Putin. This was designed to reduce demand characteristics we believed to be unevenly distributed across positions. We also included the control conditions. & GPT exhibited large cognitive consistency effects, which were significantly moderated by choice in a unidirectional analysis and trended strongly bidirectionally. The size of the choice moderation effects was larger than in Pilot 10. \\
\hline
Pilot 12 & Attitude Object: Vladimir Putin\newline Evaluative Items: Overall Leadership, Positive/Negative for Russia, Economic Effectiveness, Visionary/Short-Sighted.\newline Sample Size: 20 per condition\newline Note: Pilot 12 was similar to Pilot 11, except that the data collection was done through OpenAI’s API rather than the ChatGPT web interface. We also tested an additional condition (discussed more below) and collected answers both with and without the prompt to increase choice salience. & Both with and without the extra prompt, GPT exhibited large cognitive consistency effects which were significantly moderated by choice in a unidirectional analysis. However, several serious issues arose. Most notably, when prompted through the API, GPT frequently refused to answer questions in the Control conditions. This disallowed bidirectional analyses and forced us to collect our main studies through the web interface. \\
\hline
\end{longtable}

Note: Table S8 shows the progression of pilot research for this project. All studies showed large cognitive consistency effects, and most but not all showed moderation by choice. Earlier studies that were inconclusive surrounding choice moderation prompted us to experiment with strengthening our choice manipulation in various manners. Doing so generally clarified our effects, eliciting consistency effects that were usually moderated by choice.  
}


\clearpage
\section*{Section S7: Detailed Results}

In this section, we present a detailed tabular overview of the results of our main preregistered studies. \textit{Tables \ref{tab:s9}-\ref{tab:s23}} present the results of Study 1, while \textit{Tables \ref{tab:s24}-\ref{tab:s41}} present the results of Study 2.
\stepcounter{suppTable}
\begin{table}[htbp]
\centering
\renewcommand{\arraystretch}{1.2}  
\caption{Detailed Study 1 results, composite evaluation, composite scoring.}
\label{tab:s9}
\begin{tabular}{|c|c|c|c|c|c|c|}
\hline
\textbf{Essay Type} & \textbf{Mean} & \textbf{95\% CI} & \textbf{SD} & \textbf{Effect Size} & \textbf{t-Statistic} & \textbf{Significance} \\
\hline
\textbf{Pro-Putin} & 1.035 & [0.908, 1.162] & 0.447 & \textit{d} = 2.164 & \textit{t(98)} = 10.820 & \textit{P} < 0.0001 \\ \hline
\textbf{Control}   & -0.002 & [-0.147, 0.142] & 0.509 & . & . & . \\ \hline
\textbf{Anti-Putin} & -1.033 & [-1.212, -0.853] & 0.632 & \textit{d} = 1.795 & \textit{t(98)} = 8.973 & \textit{P} < 0.0001 \\\hline
\end{tabular}
\begin{flushleft}
Note: Table S9 displays the detailed main results of Study 1 with a standardized composite of the four evaluative items about Putin as the dependent variable, using a composite of verbal and numeric scoring. Effect sizes (in Cohen’s \textit{d}) and significance scores (using t-tests) are for the contrast of the relevant condition with Control. The contrast between the Pro- and Anti-Putin conditions is also significant; \textit{t(98)} = 18.874, \textit{P} < 0.0001, \textit{d} = 3.775.
\end{flushleft}
\end{table}

\stepcounter{suppTable}
\begin{table}[htbp]
\centering
\renewcommand{\arraystretch}{1.2}  
\caption{Detailed Study 1 results, Overall Leadership item, composite scoring.}
\label{tab:s10}
\begin{tabular}{|c|c|c|c|c|c|c|}
\hline
\textbf{Essay Type} & \textbf{Mean} & \textbf{95\% CI} & \textbf{SD} & \textbf{Effect Size} & \textbf{t-Statistic} & \textbf{Significance} \\
\hline
\textbf{Pro-Putin} & 4.710 & [4.581, 4.839] & 0.453 & \textit{d} = 1.833 & \textit{t(98)} = 9.164 & \textit{P} < 0.0001 \\ \hline
\textbf{Control}   & 3.905 & [3.784, 4.026] & 0.425 & . & . & . \\ \hline
\textbf{Anti-Putin} & 3.170 & [3.033, 3.307] & 0.480 & \textit{d} = 1.620 & \textit{t(98)} = 8.101 & \textit{P} < 0.0001 \\ \hline
\end{tabular}
\begin{flushleft}
Note: Table S10 displays the detailed results of Study 1 with the Overall Leadership Item as the dependent variable, using a composite of verbal and numeric scoring. Effect sizes (in Cohen’s \textit{d}) and significance scores (using t-tests) are for the contrast of the relevant condition with Control. The contrast between the Pro- and Anti-Putin conditions is also significant; \textit{t(98)} = 16.497, \textit{P} < 0.0001, \textit{d} = 3.299. 
\end{flushleft}
\end{table}

\stepcounter{suppTable}
\begin{table}[htbp]
\centering
\renewcommand{\arraystretch}{1.2}  
\caption{Detailed Study 1 results, Positive/Negative for Russia item, composite scoring.}
\label{tab:s11}
\begin{tabular}{|c|c|c|c|c|c|c|}
\hline
\textbf{Essay Type} & \textbf{Mean} & \textbf{95\% CI} & \textbf{SD} & \textbf{Effect Size} & \textbf{t-Statistic} & \textbf{Significance} \\
\hline
\textbf{Pro-Putin} & 4.570 & [4.363, 4.777] & 0.729 & \textit{d} = 2.419 & \textit{t(98)} = 12.096 & \textit{P} < 0.0001 \\ \hline
\textbf{Control}   & 3.030 & [2.880, 3.180] & 0.529 & . & . & . \\ \hline
\textbf{Anti-Putin} & 2.435 & [2.298, 2.572] & 0.481 & \textit{d} = 1.177 & \textit{t(98)} = 5.884 & \textit{P} < 0.0001 \\ \hline
\end{tabular}
\begin{flushleft}
Note: Table S11 displays the detailed results of Study 1 with the Impact on Russia Item as the dependent variable, using a composite of verbal and numeric scoring. Effect sizes (in Cohen’s \textit{d}) and significance scores (using t-tests) are for the contrast of the relevant condition with Control. The contrast between the Pro- and Anti-Putin conditions is also significant; \textit{t(98)} = 17.290, \textit{P} < 0.0001, \textit{d} = 3.458.  
\end{flushleft}
\end{table}

\stepcounter{suppTable}
\begin{table}[htbp]
\centering
\renewcommand{\arraystretch}{1.2}  
\caption{Detailed Study 1 results, Economic Effectiveness item, composite scoring.}
\label{tab:s12}
\begin{tabular}{|c|c|c|c|c|c|c|}
\hline
\textbf{Essay Type} & \textbf{Mean} & \textbf{95\% CI} & \textbf{SD} & \textbf{Effect Size} & \textbf{t-Statistic} & \textbf{Significance} \\
\hline
\textbf{Pro-Putin} & 5.230 & [5.122, 5.338] & 0.381 & \textit{d} = 1.171 & \textit{t(98)} = 5.855 & \textit{P} < 0.0001 \\ \hline
\textbf{Control}   & 4.670 & [4.511, 4.829] & 0.559 & . & . & . \\ \hline
\textbf{Anti-Putin} & 4.065 & [3.853, 4.277] & 0.747 & \textit{d} = 0.917 & \textit{t(98)} = 4.585 & \textit{P} < 0.0001 \\ \hline
\end{tabular}
\begin{flushleft}
Note: Table S12 displays the detailed results of Study 1 with the Economic Effectiveness Item as the dependent variable, using a composite of verbal and numeric scoring. Effect sizes (in Cohen’s \textit{d}) and significance scores (using t-tests) are for the contrast of the relevant condition with Control. The contrast between the Pro- and Anti-Putin conditions is also significant; \textit{t(98)} = 9.824, \textit{P} < 0.0001, \textit{d} = 1.965. 
\end{flushleft}
\end{table}

\stepcounter{suppTable}
\begin{table}[htbp]
\centering
\renewcommand{\arraystretch}{1.2}  
\caption{Detailed Study 1 results, Visionary/Short-Sighted item, composite scoring.}
\label{tab:s13}
\begin{tabular}{|c|c|c|c|c|c|c|}
\hline
\textbf{Essay Type} & \textbf{Mean} & \textbf{95\% CI} & \textbf{SD} & \textbf{Effect Size} & \textbf{t-Statistic} & \textbf{Significance} \\
\hline
\textbf{Pro-Putin} & 5.160 & [5.063, 5.257] & 0.342 & \textit{d} = 0.275 & \textit{t(98)} = 1.373 & \textit{P} = 0.1728 \\ \hline
\textbf{Control}   & 5.050 & [4.922, 5.178] & 0.452 & . & . & . \\ \hline
\textbf{Anti-Putin} & 4.440 & [4.274, 4.606] & 0.584 & \textit{d} = 1.169 & \textit{t(98)} = 5.844 & \textit{P} < 0.0001 \\ \hline
\end{tabular}
\begin{flushleft}
Note: Table S13 displays the detailed results of Study 1 with the Vision Item as the dependent variable, using a composite of verbal and numeric scoring. Effect sizes (in Cohen’s \textit{d}) and significance scores (using t-tests) are for the contrast of the relevant condition with Control. The contrast between the Pro- and Anti-Putin conditions is also significant; \textit{t(98)} = 7.527, \textit{P} < 0.0001, \textit{d} = 1.505.
\end{flushleft}
\end{table}

\stepcounter{suppTable}
\begin{table}[htbp]
\centering
\renewcommand{\arraystretch}{1.2}  
\caption{Detailed Study 1 results, (your title here), composite scoring.}
\label{tab:s14}
\begin{tabular}{|c|c|c|c|c|c|c|}
\hline
\textbf{Essay Type} & \textbf{Mean} & \textbf{95\% CI} & \textbf{SD} & \textbf{Effect Size} & \textbf{t-Statistic} & \textbf{Significance} \\
\hline
\textbf{Pro-Putin} & 1.028 & [0.899, 1.158] & 0.455 & \textit{d} = 2.109 & \textit{t(98)} = 10.545 & \textit{P} < 0.0001 \\ \hline
\textbf{Control}   & -0.001 & [-0.148, 0.147] & 0.519 & . & . & . \\ \hline
\textbf{Anti-Putin} & -1.028 & [-1.209, -0.846] & 0.638 & \textit{d} = 1.766 & \textit{t(98)} = 8.832 & \textit{P} < 0.0001 \\ \hline
\end{tabular}
\begin{flushleft}
Note: Table S14 displays the detailed main results of Study 1 with a standardized composite of the four evaluative items about Putin as the dependent variable, using the verbal scoring. Effect sizes (in Cohen’s \textit{d}) and significance scores (using t-tests) are for the contrast of the relevant condition with Control. The contrast between the Pro- and Anti-Putin conditions is also significant; \textit{t(98)} = 18.546, \textit{P} < 0.0001, \textit{d} = 3.709.
\end{flushleft}
\end{table}

\stepcounter{suppTable}
\begin{table}[htbp]
\centering
\renewcommand{\arraystretch}{1.2}  
\caption{Detailed Study 1 results, Overall Leadership item, verbal scoring.}
\label{tab:s15}
\begin{tabular}{|c|c|c|c|c|c|c|}
\hline
\textbf{Essay Type} & \textbf{Mean} & \textbf{95\% CI} & \textbf{SD} & \textbf{Effect Size} & \textbf{t-Statistic} & \textbf{Significance} \\
\hline
\textbf{Pro-Putin} & 4.710 & [4.581, 4.839] & 0.453 & \textit{d} = 1.774 & \textit{t(98)} = 8.870 & \textit{P} < 0.0001 \\ \hline
\textbf{Control}   & 3.885 & [3.749, 4.021] & 0.477 & . & . & . \\ \hline
\textbf{Anti-Putin} & 3.110 & [2.960, 3.260] & 0.528 & \textit{d} = 1.541 & \textit{t(98)} = 7.704 & \textit{P} < 0.0001 \\ \hline
\end{tabular}
\begin{flushleft}
Note: Table S15 displays the detailed results of Study 1 with the Overall Leadership Item as the dependent variable, using the verbal scoring. Effect sizes (in Cohen’s \textit{d}) and significance scores (using t-tests) are for the contrast of the relevant condition with Control. The contrast between the Pro- and Anti-Putin conditions is also significant; \textit{t(98)} = 16.271, \textit{P} < 0.0001, \textit{d} = 3.254. 
\end{flushleft}
\end{table}

\stepcounter{suppTable}
\begin{table}[htbp]
\centering
\renewcommand{\arraystretch}{1.2}  
\caption{Detailed Study 1 results, Positive/Negative for Russia item, verbal scoring.}
\label{tab:s16}
\begin{tabular}{|c|c|c|c|c|c|c|}
\hline
\textbf{Essay Type} & \textbf{Mean} & \textbf{95\% CI} & \textbf{SD} & \textbf{Effect Size} & \textbf{t-Statistic} & \textbf{Significance} \\
\hline
\textbf{Pro-Putin} & 4.550 & [4.329, 4.771] & 0.778 & \textit{d} = 2.294 & \textit{t(98)} = 11.469 & \textit{P} < 0.0001 \\ \hline
\textbf{Control}   & 3.020 & [2.868, 3.172] & 0.534 & . & . & . \\ \hline
\textbf{Anti-Putin} & 2.435 & [2.298, 2.572] & 0.481 & \textit{d} = 1.151 & \textit{t(98)} = 5.753 & \textit{P} < 0.0001 \\ \hline
\end{tabular}
\begin{flushleft}
Note: Table S16 displays the detailed results of Study 1 with the Impact on Russia Item as the dependent variable, using the verbal scoring. Effect sizes (in Cohen’s \textit{d}) and significance scores (using t-tests) are for the contrast of the relevant condition with Control. The contrast between the Pro- and Anti-Putin conditions is also significant; \textit{t(98)} = 16.354, \textit{P} < 0.0001, \textit{d} = 3.271.
\end{flushleft}
\end{table}

\stepcounter{suppTable}
\begin{table}[htbp]
\centering
\renewcommand{\arraystretch}{1.2}  
\caption{Detailed Study 1 results, Economic Effectiveness item, verbal scoring.}
\label{tab:s17}
\begin{tabular}{|c|c|c|c|c|c|c|}
\hline
\textbf{Essay Type} & \textbf{Mean} & \textbf{95\% CI} & \textbf{SD} & \textbf{Effect Size} & \textbf{t-Statistic} & \textbf{Significance} \\
\hline
\textbf{Pro-Putin} & 5.230 & [5.122, 5.338] & 0.381 & \textit{d} = 1.171 & \textit{t(98)} = 5.855 & \textit{P} < 0.0001 \\ \hline
\textbf{Control}   & 4.670 & [4.511, 4.829] & 0.559 & . & . & . \\ \hline
\textbf{Anti-Putin} & 4.060 & [3.846, 4.274] & 0.753 & \textit{d} = 0.920 & \textit{t(98)} = 4.598 & \textit{P} < 0.0001 \\ \hline
\end{tabular}
\begin{flushleft}
Note: Table S17 displays the detailed results of Study 1 with the Economic Effectiveness Item as the dependent variable, using the verbal scoring. Effect sizes (in Cohen’s \textit{d}) and significance scores (using t-tests) are for the contrast of the relevant condition with Control. The contrast between the Pro- and Anti-Putin conditions is also significant; \textit{t(98)} = 9.799, \textit{P} < 0.0001, \textit{d} = 1.960. 
\end{flushleft}
\end{table}

\stepcounter{suppTable}
\begin{table}[htbp]
\centering
\renewcommand{\arraystretch}{1.2}  
\caption{Detailed Study 1 results, Visionary/Short-Sighted item, verbal scoring.}
\label{tab:s18}
\begin{tabular}{|c|c|c|c|c|c|c|}
\hline
\textbf{Essay Type} & \textbf{Mean} & \textbf{95\% CI} & \textbf{SD} & \textbf{Effect Size} & \textbf{t-Statistic} & \textbf{Significance} \\
\hline
\textbf{Pro-Putin} & 5.160 & [5.063, 5.257] & 0.342 & \textit{d} = 0.279 & \textit{t(98)} = 1.394 & \textit{P} = 0.1666 \\ \hline
\textbf{Control}   & 5.045 & [4.911, 5.179] & 0.473 & . & . & . \\ \hline
\textbf{Anti-Putin} & 4.430 & [4.258, 4.602] & 0.606 & \textit{d} = 1.131 & \textit{t(98)} = 5.656 & \textit{P} < 0.0001 \\ \hline
\end{tabular}
\begin{flushleft}
Note: Table S18 displays the detailed results of Study 1 with the Vision Item as the dependent variable, using the verbal scoring. Effect sizes (in Cohen’s \textit{d}) and significance scores (using t-tests) are for the contrast of the relevant condition with Control. The contrast between the Pro- and Anti-Putin conditions is also significant; \textit{t(98)} = 7.418, \textit{P} < 0.0001, \textit{d} = 1.484.
\end{flushleft}
\end{table}

\stepcounter{suppTable}
\begin{table}[htbp]
\centering
\renewcommand{\arraystretch}{1.2}  
\caption{Detailed Study 1 results, composite evaluation, numeric scoring.}
\label{tab:s19}
\begin{tabular}{|c|c|c|c|c|c|c|}
\hline
\textbf{Essay Type} & \textbf{Mean} & \textbf{95\% CI} & \textbf{SD} & \textbf{Effect Size} & \textbf{t-Statistic} & \textbf{Significance} \\
\hline
\textbf{Pro-Putin} & 1.039 & [0.914, 1.164] & 0.440 & \textit{d} = 2.207 & \textit{t(98)} = 11.036 & \textit{P} < 0.0001 \\ \hline
\textbf{Control}   & -0.004 & [-0.147, 0.139] & 0.503 & . & . & . \\ \hline
\textbf{Anti-Putin} & -1.035 & [-1.215, -0.855] & 0.633 & \textit{d} = 1.803 & \textit{t(98)} = 9.017 & \textit{P} < 0.0001 \\ \hline
\end{tabular}
\begin{flushleft}
Note: Table S19 displays the detailed main results of Study 1 with a standardized composite of the four evaluative items about Putin as the dependent variable, using the numeric scoring. Effect sizes (in Cohen’s \textit{d}) and significance scores (using t-tests) are for the contrast of the relevant condition with Control. The contrast between the Pro- and Anti-Putin conditions is also significant; \textit{t(98)} = 19.015, \textit{P} < 0.0001, \textit{d} = 3.803. 
\end{flushleft}
\end{table}

\stepcounter{suppTable}
\begin{table}[htbp]
\centering
\renewcommand{\arraystretch}{1.2}  
\caption{Detailed Study 1 results, Overall Leadership item, numeric scoring.}
\label{tab:s20}
\begin{tabular}{|c|c|c|c|c|c|c|}
\hline
\textbf{Essay Type} & \textbf{Mean} & \textbf{95\% CI} & \textbf{SD} & \textbf{Effect Size} & \textbf{t-Statistic} & \textbf{Significance} \\
\hline
\textbf{Pro-Putin} & 4.710 & [4.581, 4.839] & 0.453 & \textit{d} = 1.854 & \textit{t(98)} = 9.268 & \textit{P} < 0.0001 \\ \hline
\textbf{Control}   & 3.925 & [3.814, 4.036] & 0.392 & . & . & . \\ \hline
\textbf{Anti-Putin} & 3.230 & [3.092, 3.368] & 0.487 & \textit{d} = 1.573 & \textit{t(98)} = 7.864 & \textit{P} < 0.0001 \\ \hline
\end{tabular}
\begin{flushleft}
Note: Table S20 displays the detailed results of Study 1 with the Overall Leadership Item as the dependent variable, using the numeric scoring. Effect sizes (in Cohen’s \textit{d}) and significance scores (using t-tests) are for the contrast of the relevant condition with Control. The contrast between the Pro- and Anti-Putin conditions is also significant; \textit{t(98)} = 15.744, \textit{P} < 0.0001, \textit{d} = 3.149.
\end{flushleft}
\end{table}

\stepcounter{suppTable}
\begin{table}[htbp]
\centering
\renewcommand{\arraystretch}{1.2}  
\caption{Detailed Study 1 results, Positive/Negative for Russia item, numeric scoring.}
\label{tab:s21}
\begin{tabular}{|c|c|c|c|c|c|c|}
\hline
\textbf{Essay Type} & \textbf{Mean} & \textbf{95\% CI} & \textbf{SD} & \textbf{Effect Size} & \textbf{t-Statistic} & \textbf{Significance} \\
\hline
\textbf{Pro-Putin} & 4.590 & [4.394, 4.786] & 0.690 & \textit{d} = 2.514 & \textit{t(98)} = 12.568 & \textit{P} < 0.0001 \\ \hline
\textbf{Control}   & 3.040 & [2.889, 3.191] & 0.533 & . & . & . \\ \hline
\textbf{Anti-Putin} & 2.435 & [2.298, 2.572] & 0.481 & \textit{d} = 1.191 & \textit{t(98)} = 5.957 & \textit{P} < 0.0001 \\ \hline
\end{tabular}
\begin{flushleft}
Note: Table S21 displays the detailed results of Study 1 with the Impact on Russia Item as the dependent variable, using the numeric scoring. Effect sizes (in Cohen’s \textit{d}) and significance scores (using t-tests) are for the contrast of the relevant condition with Control. The contrast between the Pro- and Anti-Putin conditions is also significant; \textit{t(98)} = 18.109, \textit{P} < 0.0001, \textit{d} = 3.622.
\end{flushleft}
\end{table}

\stepcounter{suppTable}
\begin{table}[htbp]
\centering
\renewcommand{\arraystretch}{1.2}  
\caption{Detailed Study 1 results, Economic Effectiveness item, numeric scoring.}
\label{tab:s22}
\begin{tabular}{|c|c|c|c|c|c|c|}
\hline
\textbf{Essay Type} & \textbf{Mean} & \textbf{95\% CI} & \textbf{SD} & \textbf{Effect Size} & \textbf{t-Statistic} & \textbf{Significance} \\
\hline
\textbf{Pro-Putin} & 5.230 & [5.122, 5.338] & 0.381 & \textit{d} = 1.171 & \textit{t(98)} = 5.855 & \textit{P} < 0.0001 \\ \hline
\textbf{Control}   & 4.670 & [4.511, 4.829] & 0.559 & . & . & . \\ \hline
\textbf{Anti-Putin} & 4.070 & [3.859, 4.281] & 0.742 & \textit{d} = 0.913 & \textit{t(98)} = 4.566 & \textit{P} < 0.0001 \\ \hline
\end{tabular}
\begin{flushleft}
Note: Table S22 displays the detailed results of Study 1 with the Economic Effectiveness Item as the dependent variable, using the numeric scoring. Effect sizes (in Cohen’s \textit{d}) and significance scores (using t-tests) are for the contrast of the relevant condition with Control. The contrast between the Pro- and Anti-Putin conditions is also significant; \textit{t(98)} = 9.831, \textit{P} < 0.0001, \textit{d} = 1.966.
\end{flushleft}
\end{table}

\stepcounter{suppTable}
\begin{table}[htbp]
\centering
\renewcommand{\arraystretch}{1.2}  
\caption{Detailed Study 1 results, Visionary/Short-Sighted item, numeric scoring.}
\label{tab:s23}
\begin{tabular}{|c|c|c|c|c|c|c|}
\hline
\textbf{Essay Type} & \textbf{Mean} & \textbf{95\% CI} & \textbf{SD} & \textbf{Effect Size} & \textbf{t-Statistic} & \textbf{Significance} \\
\hline
\textbf{Pro-Putin} & 5.160 & [5.063, 5.257] & 0.342 & \textit{d} = 0.269 & \textit{t(98)} = 1.347 & \textit{P} = 0.1810 \\ \hline
\textbf{Control}   & 5.055 & [4.932, 5.178] & 0.432 & . & . & . \\ \hline
\textbf{Anti-Putin} & 4.450 & [4.290, 4.610] & 0.565 & \textit{d} = 1.203 & \textit{t(98)} = 6.015 & \textit{P} < 0.0001 \\ \hline
\end{tabular}
\begin{flushleft}
Note: Table S23 displays the detailed results of Study 1 with the Vision Item as the dependent variable, using the numeric scoring. Effect sizes (in Cohen’s \textit{d}) and significance scores (using t-tests) are for the contrast of the relevant condition with Control. The contrast between the Pro- and Anti-Putin conditions is also significant; \textit{t(98)} = 7.607, \textit{P} < 0.0001, \textit{d} = 1.521.
\end{flushleft}
\end{table}

\stepcounter{suppTable}
\begin{table}[htbp]
\centering
\renewcommand{\arraystretch}{1.2}  
\caption{Detailed Study 2 results, composite evaluation, composite scoring.}
\label{tab:s24}
\begin{tabular}{|c|c|c|c|c|}
\hline
\textbf{} & \multicolumn{2}{c|}{\textbf{No Choice}} & \multicolumn{2}{c|}{\textbf{Choice}} \\
\hline
\multirow{3}{*}{\textbf{Essay Type}} & \textbf{Mean\textsubscript{z}} & \textbf{Effect Size} & \textbf{Mean\textsubscript{z}} & \textbf{Effect Size} \\
 & \textbf{95\% CI} & \textbf{t-Statistic} & \textbf{95\% CI} & \textbf{t-Statistic} \\
 & \textbf{SD} & \textbf{Significance} & \textbf{SD} & \textbf{Significance} \\
\hline

\multirow{3}{*}{\textbf{Pro-Putin}} & M\textsubscript{z} = 0.846 &\textit{d} = 2.006 &  M\textsubscript{z} = 1.241 & \textit{d} = 2.748 \\
 & [0.781, 0.912] & \textit{t(298)} = 17.373 & [1.177, 1.305] & \textit{t(298)} = 23.798 \\
 & SD = 0.407 &\textit{P} < 0.0001 & SD = 0.397 & \textit{P} < 0.0001 \\
\hline

\multirow{3}{*}{\textbf{Control}} &  M\textsubscript{z} = -0.100 &\multirow{3}{*}{.} & M\textsubscript{z} = 0.005 & \multirow{3}{*}{.} \\
 & [-0.185, -0.014] &  & [-0.075, 0.086] &  \\
 & SD = 0.528 &  & SD = 0.497 &  \\
\hline

\multirow{3}{*}{\textbf{Anti-Putin}} &  M\textsubscript{z} = -0.913 & \textit{d} = 1.368 &  M\textsubscript{z} = -1.081 & \textit{d} = 1.827 \\
 & [-1.018, -0.807] & \textit{t(298)} = 11.847  & [-1.190, -0.971] & \textit{t(298)} = 15.826  \\
 & SD = 0.654 & \textit{P} < 0.0001 & SD = 0.678 & \textit{P} < 0.0001\\

\hline
\end{tabular}
\begin{flushleft}
Note: Table S24 displays the detailed main results of Study 2 with a standardized composite of the four evaluative items about Putin as the dependent variable, using a composite of verbal and numeric scoring. Effect sizes (in Cohen’s \textit{d}) and significance scores (using t-tests) are for the contrast of the relevant condition with Control. The contrast between the Pro- and Anti-Putin conditions is also significant in both the No Choice (\textit{t(298)} = 27.962, \textit{P} < 0.0001, \textit{d} = 3.229) and Choice (\textit{t(298)} = 36.192, \textit{P} < 0.0001, \textit{d} = 4.179) conditions. The contrast between the Choice conditions reached significance in both the Pro-Putin (\textit{t(298)} = 8.499, \textit{P} < 0.0001, \textit{d} = 0.981) and Anti-Putin (\textit{t(298)} = 2.184, \textit{P} = 0.0298, \textit{d} = 0.252) conditions.
\end{flushleft}
\end{table}

\stepcounter{suppTable}
\begin{table}[htbp]
\centering
\renewcommand{\arraystretch}{1.2}  
\caption{Detailed Study 2 results, composite evaluation, composite scoring.}
\label{tab:s25}
\begin{tabular}{|c|c|c|c|c|}
\hline
\textbf{} & \multicolumn{2}{c|}{\textbf{No Choice}} & \multicolumn{2}{c|}{\textbf{Choice}} \\
\hline
\multirow{3}{*}{\textbf{Essay Type}} & \textbf{Mean} & \textbf{Effect Size} & \textbf{Mean} & \textbf{Effect Size} \\
 & \textbf{95\% CI} & \textbf{t-Statistic} & \textbf{95\% CI} & \textbf{t-Statistic} \\
 & \textbf{SD} & \textbf{Significance} & \textbf{SD} & \textbf{Significance} \\
\hline

\multirow{3}{*}{\textbf{Pro-Putin}} & M = 4.223 & \textit{d} = 1.127 & M = 4.553 & \textit{d} = 1.501 \\
 & [4.148, 4.299] & \textit{t(298)} = 9.759 & [4.471, 4.636] & \textit{t(298)} = 12.999 \\
 & SD = 0.469 & \textit{P} < 0.0001 & SD = 0.512 & \textit{P} < 0.0001 \\
\hline

\multirow{3}{*}{\textbf{Control}} & M = 3.715 & \multirow{3}{*}{.} & M = 3.832 & \multirow{3}{*}{.} \\
 & [3.645, 3.785] &  & [3.759, 3.904] &  \\
 & SD = 0.433 &  & SD = 0.447 &  \\
\hline

\multirow{3}{*}{\textbf{Anti-Putin}} & M = 3.240 & \textit{d} = 0.945 & M = 3.097 & \textit{d} = 1.590 \\
 & [3.149, 3.331] & \textit{t(298)} = 8.187 & [3.020, 3.174] & \textit{t(298)} = 13.772 \\
 & SD = 0.564 & \textit{P} < 0.0001 & SD = 0.477 & \textit{P} < 0.0001 \\
\hline
\end{tabular}
\begin{flushleft}
Note: Table S25 displays the detailed main results of Study 2 with the individual item about Putin’s Overall Leadership as the dependent variable, using a composite of verbal and numeric scoring. Effect sizes (in Cohen’s \textit{d}) and significance scores (using t-tests) are for the contrast of the relevant condition with Control. The contrast between the Pro- and Anti-Putin conditions is also significant in both the No Choice (\textit{t(298)} = 16.423, \textit{P} < 0.0001, \textit{d} = 1.896) and Choice (\textit{t(298)} = 25.502, \textit{P} < 0.0001, \textit{d} = 2.945) conditions. The contrast between the Choice conditions reached significance in both the Pro-Putin (\textit{t(298)} = 5.821, \textit{P} < 0.0001, \textit{d} = 0.672) and Anti-Putin (\textit{t(298)} = 2.378, \textit{P} = 0.0180, \textit{d} = 0.275) conditions.
\end{flushleft}
\end{table}

\stepcounter{suppTable}
\begin{table}[htbp]
\centering
\renewcommand{\arraystretch}{1.2}  
\caption{Detailed Study 2 results, Impact on Russia, composite scoring.}
\label{tab:s26}
\begin{tabular}{|c|c|c|c|c|}
\hline
\textbf{} & \multicolumn{2}{c|}{\textbf{No Choice}} & \multicolumn{2}{c|}{\textbf{Choice}} \\
\hline
\multirow{3}{*}{\textbf{Essay Type}} & \textbf{Mean} & \textbf{Effect Size} & \textbf{Mean} & \textbf{Effect Size} \\
 & \textbf{95\% CI} & \textbf{t-Statistic} & \textbf{95\% CI} & \textbf{t-Statistic} \\
 & \textbf{SD} & \textbf{Significance} & \textbf{SD} & \textbf{Significance} \\
\hline

\multirow{3}{*}{\textbf{Pro-Putin}} & M = 3.815 & \textit{d} = 1.657 & M = 4.365 & \textit{d} = 2.311 \\
 & [3.710, 3.920] & \textit{t(298)} = 14.347 & [4.252, 4.478] & \textit{t(298)} = 20.017 \\
 & SD = 0.652 & \textit{P} < 0.0001 & SD = 0.700 & \textit{P} < 0.0001 \\
\hline

\multirow{3}{*}{\textbf{Control}} & M = 2.837 & \multirow{3}{*}{.} & M = 2.848 & \multirow{3}{*}{.} \\
 & [2.752, 2.921] &  & [2.750, 2.947] &  \\
 & SD = 0.522 &  & SD = 0.609 &  \\
\hline

\multirow{3}{*}{\textbf{Anti-Putin}} & M = 2.282 & \textit{d} = 1.147 & M = 2.287 & \textit{d} = 1.050 \\
 & [2.210, 2.353] & \textit{t(298)} = 9.932 & [2.214, 2.359] & \textit{t(298)} = 9.092 \\
 & SD = 0.442 & \textit{P} < 0.0001 & SD = 0.449 & \textit{P} < 0.0001 \\
\hline
\end{tabular}
\begin{flushleft}
Note: Table S26 displays the detailed main results of Study 2 with the individual item about Putin’s Impact on Russia as the dependent variable, using a composite of verbal and numeric scoring. Effect sizes (in Cohen’s \textit{d}) and significance scores (using t-tests) are for the contrast of the relevant condition with Control. The contrast between the Pro- and Anti-Putin conditions is also significant in both the No Choice (\textit{t(298)} = 23.839, \textit{P} < 0.0001, \textit{d} = 2.753) and Choice (\textit{t(298)} = 30.596, \textit{P} < 0.0001, \textit{d} = 3.533) conditions. The contrast between the Choice conditions reached significance in the Pro-Putin condition (\textit{t(298)} = 7.041, \textit{P} < 0.0001, \textit{d} = 0.813) but not in the Anti-Putin (\textit{t(298)} = -0.097, \textit{P} = 0.9227, \textit{d} = -0.011) condition.

\end{flushleft}
\end{table}

\stepcounter{suppTable}
\begin{table}[htbp]
\centering
\renewcommand{\arraystretch}{1.2}  
\caption{Detailed Study 2 results, Economic Effectiveness, composite scoring.}
\label{tab:s27}
\begin{tabular}{|c|c|c|c|c|}
\hline
\textbf{} & \multicolumn{2}{c|}{\textbf{No Choice}} & \multicolumn{2}{c|}{\textbf{Choice}} \\
\hline
\multirow{3}{*}{\textbf{Essay Type}} & \textbf{Mean\textsubscript{z}} & \textbf{Effect Size} & \textbf{Mean\textsubscript{z}} & \textbf{Effect Size} \\
 & \textbf{95\% CI} & \textbf{t-Statistic} & \textbf{95\% CI} & \textbf{t-Statistic} \\
 & \textbf{SD} & \textbf{Significance} & \textbf{SD} & \textbf{Significance} \\
\hline

\multirow{3}{*}{\textbf{Pro-Putin}} & M = 5.133 & \textit{d} = 1.672 & M = 5.177 & \textit{d} = 1.602 \\
 & [5.089, 5.178] & \textit{t(298)} = 14.484 & [5.125, 5.229] & \textit{t(298)} = 13.873 \\
 & SD = 0.276 & \textit{P} < 0.0001 & SD = 0.322 & \textit{P} < 0.0001 \\
\hline

\multirow{3}{*}{\textbf{Control}} & M = 4.327 & \multirow{3}{*}{.} & M = 4.353 & \multirow{3}{*}{.} \\
 & [4.226, 4.427] &  & [4.248, 4.458] &  \\
 & SD = 0.624 &  & SD = 0.652 &  \\
\hline

\multirow{3}{*}{\textbf{Anti-Putin}} & M = 4.193 & \textit{d} = 0.201 & M = 3.923 & \textit{d} = 0.624 \\
 & [4.080, 4.307] & \textit{t(298)} = 1.739 & [3.806, 4.040] & \textit{t(298)} = 5.402 \\
 & SD = 0.702 & \textit{P} = 0.0831 & SD = 0.725 & \textit{P} < 0.0001 \\
\hline
\end{tabular}
\begin{flushleft}
Note: Table S27 displays the detailed main results of Study 2 with the individual item about Putin’s Economic Effectiveness as the dependent variable, using a composite of verbal and numeric scoring. Effect sizes (in Cohen’s \textit{d}) and significance scores (using t-tests) are for the contrast of the relevant condition with Control. The contrast between the Pro- and Anti-Putin conditions is also significant in both the No Choice (\textit{t(298)} = 15.267, \textit{P} < 0.0001, \textit{d} = 1.763) and Choice (\textit{t(298)} = 19.345, \textit{P} < 0.0001, \textit{d} = 2.234) conditions. The contrast between the Choice conditions reached significance in the Anti-Putin condition (\textit{t(298)} = 3.277, \textit{P} = 0.0012, \textit{d} = 0.378) but not in the Pro-Putin (\textit{t(298)} = 1.252, \textit{P} = 0.2116, \textit{d} = 0.145) condition.
\end{flushleft}
\end{table}

\stepcounter{suppTable}
\begin{table}[htbp]
\centering
\renewcommand{\arraystretch}{1.2}  
\caption{Detailed Study 2 results, Visionary/Short-Sighted, composite scoring.}
\label{tab:s28}
\begin{tabular}{|c|c|c|c|c|}
\hline
\textbf{} & \multicolumn{2}{c|}{\textbf{No Choice}} & \multicolumn{2}{c|}{\textbf{Choice}} \\
\hline
\multirow{3}{*}{\textbf{Essay Type}} & \textbf{Mean\textsubscript{z}} & \textbf{Effect Size} & \textbf{Mean\textsubscript{z}} & \textbf{Effect Size} \\
 & \textbf{95\% CI} & \textbf{t-Statistic} & \textbf{95\% CI} & \textbf{t-Statistic} \\
 & \textbf{SD} & \textbf{Significance} & \textbf{SD} & \textbf{Significance} \\
\hline

\multirow{3}{*}{\textbf{Pro-Putin}} & M = 5.193 & \textit{d} = 0.227 & M = 5.272 & \textit{d} = 0.286 \\
 & [5.141, 5.245] & \textit{t(298)} = 1.965 & [5.218, 5.325] & \textit{t(298)} = 2.480 \\
 & SD = 0.321 & \textit{P} = 0.0504 & SD = 0.330 & \textit{P} = 0.0137 \\
\hline

\multirow{3}{*}{\textbf{Control}} & M = 5.110 & \multirow{3}{*}{.} & M = 5.175 & \multirow{3}{*}{.} \\
 & [5.044, 5.176] &  & [5.119, 5.231] &  \\
 & SD = 0.408 &  & SD = 0.345 &  \\
\hline

\multirow{3}{*}{\textbf{Anti-Putin}} & M = 4.480 & \textit{d} = 1.238 & M = 4.498 & \textit{d} = 1.352 \\
 & [4.384, 4.576] & \textit{t(298)} = 10.725 & [4.399, 4.598] & \textit{t(298)} = 11.708 \\
 & SD = 0.593 & \textit{P} < 0.0001 & SD = 0.618 & \textit{P} < 0.0001 \\
\hline
\end{tabular}
\begin{flushleft}
Note: Table S28 displays the detailed main results of Study 2 with the individual item about Putin’s Vision as the dependent variable, using a composite of verbal and numeric scoring. Effect sizes (in Cohen’s \textit{d}) and significance scores (using t-tests) are for the contrast of the relevant condition with Control. The contrast between the Pro- and Anti-Putin conditions is also significant in both the No Choice (\textit{t(298)} = 12.960, \textit{P} < 0.0001, \textit{d} = 1.497) and Choice (\textit{t(298)} = -13.515, \textit{P} < 0.0001, \textit{d} = 1.561) conditions. The contrast between the Choice conditions reached significance in the Pro-Putin condition (\textit{t(298)} = 2.082, \textit{P} = 0.0382, \textit{d} = 0.240) but not in the Anti-Putin (\textit{t(298)} = -0.262, \textit{P} = 0.7933, \textit{d} = -0.030) condition.
\end{flushleft}
\end{table}

\stepcounter{suppTable}
\begin{table}[htbp]
\centering
\renewcommand{\arraystretch}{1.2}  
\caption{Detailed Study 2 results, composite evaluation, verbal scoring.}
\label{tab:s29}
\begin{tabular}{|c|c|c|c|c|}
\hline
\textbf{} & \multicolumn{2}{c|}{\textbf{No Choice}} & \multicolumn{2}{c|}{\textbf{Choice}} \\
\hline
\multirow{3}{*}{\textbf{Essay Type}} & \textbf{Mean\textsubscript{z}} & \textbf{Effect Size} & \textbf{Mean\textsubscript{z}} & \textbf{Effect Size} \\
 & \textbf{95\% CI} & \textbf{t-Statistic} & \textbf{95\% CI} & \textbf{t-Statistic} \\
 & \textbf{SD} & \textbf{Significance} & \textbf{SD} & \textbf{Significance} \\
\hline

\multirow{3}{*}{\textbf{Pro-Putin}} & M\textsubscript{z} = 0.839 & \textit{d} = 1.982 & M\textsubscript{z} = 1.229 & \textit{d} = 2.707 \\
 & [0.774, 0.905] & \textit{t(298)} = 17.160 & [1.165, 1.293] & \textit{t(298)} = 23.439 \\
 & SD = 0.404 & \textit{P} < 0.0001 & SD = 0.398 & \textit{P} < 0.0001 \\
\hline

\multirow{3}{*}{\textbf{Control}} & M\textsubscript{z} = -0.086 & \multirow{3}{*}{.} & M\textsubscript{z} = 0.013 & \multirow{3}{*}{.} \\
 & [-0.170, -0.002] &  & [-0.067, 0.093] &  \\
 & SD = 0.522 &  & SD = 0.496 &  \\
\hline

\multirow{3}{*}{\textbf{Anti-Putin}} & M\textsubscript{z} = -0.902 & \textit{d} = 1.349 & M\textsubscript{z} = -1.094 & \textit{d} = 1.847 \\
 & [-1.011, -0.792] & \textit{t(298)} = 11.684 & [-1.205, -0.983] & \textit{t(298)} = 15.997 \\
 & SD = 0.677 & \textit{P} < 0.0001 & SD = 0.688 & \textit{P} < 0.0001 \\
\hline
\end{tabular}
\begin{flushleft}
Note: Table S29 displays the detailed main results of Study 2 with a standardized composite of the four evaluative items about Putin as the dependent variable, using the verbal scoring. Effect sizes (in Cohen’s \textit{d}) and significance scores (using t-tests) are for the contrast of the relevant condition with Control. The contrast between the Pro- and Anti-Putin conditions is also significant in both the No Choice (\textit{t(298)} = 27.042, \textit{P} < 0.0001, \textit{d} = 3.122) and Choice (\textit{t(298)} = 35.811, \textit{P} < 0.0001, \textit{d} = 4.135) conditions. The contrast between the Choice conditions reached significance in both the Pro-Putin (\textit{t(298)} = 8.423, \textit{P} < 0.0001, \textit{d} = 0.973) and Anti-Putin (\textit{t(298)} = 2.446, \textit{P} = 0.0150, \textit{d} = 0.282) conditions.
\end{flushleft}
\end{table}

\stepcounter{suppTable}
\begin{table}[htbp]
\centering
\renewcommand{\arraystretch}{1.2}  
\caption{Detailed Study 2 results, Overall Leadership, verbal scoring.}
\label{tab:s30}
\begin{tabular}{|c|c|c|c|c|}
\hline
\textbf{} & \multicolumn{2}{c|}{\textbf{No Choice}} & \multicolumn{2}{c|}{\textbf{Choice}} \\
\hline
\multirow{3}{*}{\textbf{Essay Type}} & \textbf{Mean\textsubscript{z}} & \textbf{Effect Size} & \textbf{Mean\textsubscript{z}} & \textbf{Effect Size} \\
 & \textbf{95\% CI} & \textbf{t-Statistic} & \textbf{95\% CI} & \textbf{t-Statistic} \\
 & \textbf{SD} & \textbf{Significance} & \textbf{SD} & \textbf{Significance} \\
\hline

\multirow{3}{*}{\textbf{Pro-Putin}} & M = 4.217 & \textit{d} = 1.084 & M = 4.553 & \textit{d} = 1.483 \\
 & [4.138, 4.296] & \textit{t(298)} = 9.387 & [4.471, 4.636] & \textit{t(298)} = 12.844 \\
 & SD = 0.490 & \textit{P} < 0.0001 & SD = 0.512 & \textit{P} < 0.0001 \\
\hline

\multirow{3}{*}{\textbf{Control}} & M = 3.700 & \multirow{3}{*}{.} & M = 3.818 & \multirow{3}{*}{.} \\
 & [3.625, 3.775] &  & [3.741, 3.896] &  \\
 & SD = 0.463 &  & SD = 0.478 &  \\
\hline

\multirow{3}{*}{\textbf{Anti-Putin}} & M = 3.197 & \textit{d} = 0.940 & M = 2.983 & \textit{d} = 1.634 \\
 & [3.100, 3.293] & \textit{t(298)} = 8.140 & [2.896, 3.071] & \textit{t(298)} = 14.151 \\
 & SD = 0.599 & \textit{P} < 0.0001 & SD = 0.542 & \textit{P} < 0.0001 \\
\hline
\end{tabular}
\begin{flushleft}
Note: Table S30 displays the detailed main results of Study 2 with the individual item about Putin’s Overall Leadership as the dependent variable, using the verbal scoring. Effect sizes (in Cohen’s \textit{d}) and significance scores (using t-tests) are for the contrast of the relevant condition with Control. The contrast between the Pro- and Anti-Putin conditions is also significant in both the No Choice (\textit{t(298)} = 16.150, \textit{P} < 0.0001, \textit{d} = 1.865) and Choice (\textit{t(298)} = 25.797, \textit{P} < 0.0001, \textit{d} = 2.979) conditions. The contrast between the Choice conditions reached significance in both the Pro-Putin (\textit{t(298)} = 5.820, \textit{P} < 0.0001, \textit{d} = 0.672) and Anti-Putin (\textit{t(298)} = 3.236, \textit{P} = 0.0014, \textit{d} = 0.374) conditions.
\end{flushleft}
\end{table}

\stepcounter{suppTable}
\begin{table}[htbp]
\centering
\renewcommand{\arraystretch}{1.2}  
\caption{Detailed Study 2 results, Impact on Russia, verbal scoring.}
\label{tab:s31}
\begin{tabular}{|c|c|c|c|c|}
\hline
\textbf{} & \multicolumn{2}{c|}{\textbf{No Choice}} & \multicolumn{2}{c|}{\textbf{Choice}} \\
\hline
\multirow{3}{*}{\textbf{Essay Type}} & \textbf{Mean\textsubscript{z}} & \textbf{Effect Size} & \textbf{Mean\textsubscript{z}} & \textbf{Effect Size} \\
 & \textbf{95\% CI} & \textbf{t-Statistic} & \textbf{95\% CI} & \textbf{t-Statistic} \\
 & \textbf{SD} & \textbf{Significance} & \textbf{SD} & \textbf{Significance} \\
\hline

\multirow{3}{*}{\textbf{Pro-Putin}} & M = 3.795 & \textit{d} = 1.578 & M = 4.345 & \textit{d} = 2.208 \\
 & [3.684, 3.906] & \textit{t(298)} = 13.663 & [4.225, 4.465] & \textit{t(298)} = 19.122 \\
 & SD = 0.685 & \textit{P} < 0.0001 & SD = 0.745 & \textit{P} < 0.0001 \\
\hline

\multirow{3}{*}{\textbf{Control}} & M = 2.833 & \multirow{3}{*}{.} & M = 2.842 & \multirow{3}{*}{.} \\
 & [2.749, 2.918] &  & [2.743, 2.940] &  \\
 & SD = 0.523 &  & SD = 0.610 &  \\
\hline

\multirow{3}{*}{\textbf{Anti-Putin}} & M = 2.278 & \textit{d} = 1.150 & M = 2.287 & \textit{d} = 1.036 \\
 & [2.208, 2.349] & \textit{t(298)} = 9.960 & [2.214, 2.359] & \textit{t(298)} = 8.975 \\
 & SD = 0.439 & \textit{P} < 0.0001 & SD = 0.449 & \textit{P} < 0.0001 \\
\hline
\end{tabular}
\begin{flushleft}
Note: Table S31 displays the detailed main results of Study 2 with the individual item about Putin’s Impact on Russia as the dependent variable, using the verbal scoring. Effect sizes (in Cohen’s \textit{d}) and significance scores (using t-tests) are for the contrast of the relevant condition with Control. The contrast between the Pro- and Anti-Putin conditions is also significant in both the No Choice (\textit{t(298)} = 22.823, \textit{P} < 0.0001, \textit{d} = 2.635) and Choice (\textit{t(298)} = 28.979, \textit{P} < 0.0001, \textit{d} = 3.346) conditions. The contrast between the Choice conditions reached significance in the Pro-Putin condition (\textit{t(298)} = 6.654, \textit{P} < 0.0001, \textit{d} = 0.768) but not in the Anti-Putin (\textit{t(298)} = -0.163, \textit{P} = 0.8710, \textit{d} = -0.019) condition.
\end{flushleft}
\end{table}

\stepcounter{suppTable}
\begin{table}[htbp]
\centering
\renewcommand{\arraystretch}{1.2}  
\caption{Detailed Study 2 results, Economic Effectiveness, verbal scoring.}
\label{tab:s32}
\begin{tabular}{|c|c|c|c|c|}
\hline
\textbf{} & \multicolumn{2}{c|}{\textbf{No Choice}} & \multicolumn{2}{c|}{\textbf{Choice}} \\
\hline
\multirow{3}{*}{\textbf{Essay Type}} & \textbf{Mean\textsubscript{z}} & \textbf{Effect Size} & \textbf{Mean\textsubscript{z}} & \textbf{Effect Size} \\
 & \textbf{95\% CI} & \textbf{t-Statistic} & \textbf{95\% CI} & \textbf{t-Statistic} \\
 & \textbf{SD} & \textbf{Significance} & \textbf{SD} & \textbf{Significance} \\
\hline

\multirow{3}{*}{\textbf{Pro-Putin}} & M = 5.133 & \textit{d} = 1.672 & M = 5.180 & \textit{d} = 1.597 \\
 & [5.089, 5.178] & \textit{t(298)} = 14.484 & [5.128, 5.232] & \textit{t(298)} = 13.829 \\
 & SD = 0.276 & \textit{P} < 0.0001 & SD = 0.324 & \textit{P} < 0.0001 \\
\hline

\multirow{3}{*}{\textbf{Control}} & M = 4.327 & \multirow{3}{*}{.} & M = 4.350 & \multirow{3}{*}{.} \\
 & [4.226, 4.427] &  & [4.244, 4.456] &  \\
 & SD = 0.624 &  & SD = 0.660 &  \\
\hline

\multirow{3}{*}{\textbf{Anti-Putin}} & M = 4.180 & \textit{d} = 0.216 & M = 3.910 & \textit{d} = 0.625 \\
 & [4.062, 4.298] & \textit{t(298)} = 1.872 & [3.790, 4.030] & \textit{t(298)} = 5.409 \\
 & SD = 0.729 & \textit{P} = 0.0621 & SD = 0.747 & \textit{P} < 0.0001 \\
\hline
\end{tabular}
\begin{flushleft}
Note: Table S32 displays the detailed main results of Study 2 with the individual item about Putin’s Economic Effectiveness as the dependent variable, using the verbal scoring. Effect sizes (in Cohen’s \textit{d}) and significance scores (using t-tests) are for the contrast of the relevant condition with Control. The contrast between the Pro- and Anti-Putin conditions is also significant in both the No Choice (\textit{t(298)} = 14.984, \textit{P} < 0.0001, \textit{d} = 1.730) and Choice (\textit{t(298)} = 19.113, \textit{P} < 0.0001, \textit{d} = 2.207) conditions. The contrast between the Choice conditions reached significance in the Anti-Putin condition (\textit{t(298)} = 3.170, \textit{P} = 0.0017, \textit{d} = 0.366) but not in the Pro-Putin (\textit{t(298)} = 1.343, \textit{P} = 0.1802, \textit{d} = 0.155) condition.
\end{flushleft}
\end{table}

\stepcounter{suppTable}
\begin{table}[htbp]
\centering
\renewcommand{\arraystretch}{1.2}  
\caption{Detailed Study 2 results, Visionary/Short-Sighted, verbal scoring.}
\label{tab:s33}
\begin{tabular}{|c|c|c|c|c|}
\hline
\textbf{} & \multicolumn{2}{c|}{\textbf{No Choice}} & \multicolumn{2}{c|}{\textbf{Choice}} \\
\hline
\multirow{3}{*}{\textbf{Essay Type}} & \textbf{Mean\textsubscript{z}} & \textbf{Effect Size} & \textbf{Mean\textsubscript{z}} & \textbf{Effect Size} \\
 & \textbf{95\% CI} & \textbf{t-Statistic} & \textbf{95\% CI} & \textbf{t-Statistic} \\
 & \textbf{SD} & \textbf{Significance} & \textbf{SD} & \textbf{Significance} \\
\hline

\multirow{3}{*}{\textbf{Pro-Putin}} & M = 5.193 & \textit{d} = 0.227 & M = 5.273 & \textit{d} = 0.291 \\
 & [5.141, 5.245] & \textit{t(298)} = 1.965 & [5.220, 5.327] & \textit{t(298)} = 2.521 \\
 & SD = 0.321 & \textit{P} = 0.0504 & SD = 0.331 & \textit{P} = 0.0122 \\
\hline

\multirow{3}{*}{\textbf{Control}} & M = 5.110 & \multirow{3}{*}{.} & M = 5.175 & \multirow{3}{*}{.} \\
 & [5.044, 5.176] &  & [5.119, 5.231] &  \\
 & SD = 0.408 &  & SD = 0.345 &  \\
\hline

\multirow{3}{*}{\textbf{Anti-Putin}} & M = 4.467 & \textit{d} = 1.214 & M = 4.483 & \textit{d} = 1.319 \\
 & [4.365, 4.568] & \textit{t(298)} = 10.516 & [4.377, 4.589] & \textit{t(298)} = 11.424 \\
 & SD = 0.628 & \textit{P} < 0.0001 & SD = 0.656 & \textit{P} < 0.0001 \\
\hline
\end{tabular}
\begin{flushleft}
Note: Table S33 displays the detailed main results of Study 2 with the individual item about Putin’s Vision as the dependent variable, using the verbal scoring. Effect sizes (in Cohen’s \textit{d}) and significance scores (using t-tests) are for the contrast of the relevant condition with Control. The contrast between the Pro- and Anti-Putin conditions is also significant in both the No Choice (\textit{t(298)} = 12.609, \textit{P} < 0.0001, \textit{d} = 1.456) and Choice (\textit{t(298)} = 13.163, \textit{P} < 0.0001, \textit{d} = 1.520) conditions. The contrast between the Choice conditions reached significance in the Pro-Putin condition (\textit{t(298)} = 2.125, \textit{P} = 0.0344, \textit{d} = 0.245) but not in the Anti-Putin (\textit{t(298)} = -0.225, \textit{P} = 0.8224, \textit{d} = -0.026) condition.
\end{flushleft}
\end{table}

\stepcounter{suppTable}
\begin{table}[htbp]
\centering
\renewcommand{\arraystretch}{1.2}  
\caption{Detailed Study 2 results, composite evaluation, numeric scoring.}
\label{tab:s34}
\begin{tabular}{|c|c|c|c|c|}
\hline
\textbf{} & \multicolumn{2}{c|}{\textbf{No Choice}} & \multicolumn{2}{c|}{\textbf{Choice}} \\
\hline
\multirow{3}{*}{\textbf{Essay Type}} & \textbf{Mean\textsubscript{z}} & \textbf{Effect Size} & \textbf{Mean\textsubscript{z}} & \textbf{Effect Size} \\
 & \textbf{95\% CI} & \textbf{t-Statistic} & \textbf{95\% CI} & \textbf{t-Statistic} \\
 & \textbf{SD} & \textbf{Significance} & \textbf{SD} & \textbf{Significance} \\
\hline

\multirow{3}{*}{\textbf{Pro-Putin}} & M\textsubscript{z} = 0.851 & \textit{d} = 2.018 & M\textsubscript{z} = 1.248 & \textit{d} = 2.773 \\
 & [0.784, 0.917] & \textit{t(298)} = 17.474 & [1.184, 1.312] & \textit{t(298)} = 24.018 \\
 & SD = 0.413 & \textit{P} < 0.0001 & SD = 0.397 & \textit{P} < 0.0001 \\
\hline

\multirow{3}{*}{\textbf{Control}} & M\textsubscript{z} = -0.114 & \multirow{3}{*}{.} & M\textsubscript{z} = -0.004 & \multirow{3}{*}{.} \\
 & [-0.200, -0.027] &  & [-0.084, 0.077] &  \\
 & SD = 0.535 &  & SD = 0.499 &  \\
\hline

\multirow{3}{*}{\textbf{Anti-Putin}} & M\textsubscript{z} = -0.921 & \textit{d} = 1.372 & M\textsubscript{z} = -1.060 & \textit{d} = 1.761 \\
 & [-1.024, -0.818] & \textit{t(298)} = 11.881 & [-1.171, -0.949] & \textit{t(298)} = 15.252 \\
 & SD = 0.637 & \textit{P} < 0.0001 & SD = 0.686 & \textit{P} < 0.0001 \\
\hline
\end{tabular}
\begin{flushleft}
Note: Table S34 displays the detailed main results of Study 2 with a standardized composite of the four evaluative items about Putin as the dependent variable, using the numeric scoring. Effect sizes (in Cohen’s \textit{d}) and significance scores (using t-tests) are for the contrast of the relevant condition with Control. The contrast between the Pro- and Anti-Putin conditions is also significant in both the No Choice (\textit{t(298)} = 28.592, \textit{P} < 0.0001, \textit{d} = 3.302) and Choice (\textit{t(298)} = 35.666, \textit{P} < 0.0001, \textit{d} = 4.118) conditions. The contrast between the Choice conditions reached significance in the Pro-Putin condition (\textit{t(298)} = 8.490, \textit{P} < 0.0001, \textit{d} = 0.980) and trended in the Anti-Putin condition (\textit{t(298)} = 1.823, \textit{P} = 0.0693, \textit{d} = 0.211).
\end{flushleft}
\end{table}

\stepcounter{suppTable}
\begin{table}[htbp]
\centering
\renewcommand{\arraystretch}{1.2}  
\caption{Detailed Study 2 results, Overall Leadership, numeric scoring.}
\label{tab:s35}
\begin{tabular}{|c|c|c|c|c|}
\hline
\textbf{} & \multicolumn{2}{c|}{\textbf{No Choice}} & \multicolumn{2}{c|}{\textbf{Choice}} \\
\hline
\multirow{3}{*}{\textbf{Essay Type}} & \textbf{Mean} & \textbf{Effect Size} & \textbf{Mean} & \textbf{Effect Size} \\
 & \textbf{95\% CI} & \textbf{t-Statistic} & \textbf{95\% CI} & \textbf{t-Statistic} \\
 & \textbf{SD} & \textbf{Significance} & \textbf{SD} & \textbf{Significance} \\
\hline

\multirow{3}{*}{\textbf{Pro-Putin}} & M = 4.230 & \textit{d} = 1.147 & M = 4.553 & \textit{d} = 1.499 \\
 & [4.157, 4.303] & \textit{t(298)} = 9.933 & [4.471, 4.636] & \textit{t(298)} = 12.983 \\
 & SD = 0.455 & \textit{P} < 0.0001 & SD = 0.512 & \textit{P} < 0.0001 \\
\hline

\multirow{3}{*}{\textbf{Control}} & M = 3.730 & \multirow{3}{*}{.} & M = 3.845 & \multirow{3}{*}{.} \\
 & [3.663, 3.797] &  & [3.776, 3.914] &  \\
 & SD = 0.416 &  & SD = 0.429 &  \\
\hline

\multirow{3}{*}{\textbf{Anti-Putin}} & M = 3.283 & \textit{d} = 0.902 & M = 3.210 & \textit{d} = 1.364 \\
 & [3.193, 3.374] & \textit{t(298)} = 7.814 & [3.129, 3.291] & \textit{t(298)} = 11.812 \\
 & SD = 0.563 & \textit{P} < 0.0001 & SD = 0.499 & \textit{P} < 0.0001 \\
\hline
\end{tabular}
\begin{flushleft}
Note: S35 displays the detailed main results of Study 2 with the individual item about Putin’s Overall Leadership as the dependent variable, using the numeric scoring. Effect sizes (in Cohen’s \textit{d}) and significance scores (using t-tests) are for the contrast of the relevant condition with Control. The contrast between the Pro- and Anti-Putin conditions is also significant in both the No Choice (\textit{t(298)} = 16.022, \textit{P} < 0.0001, \textit{d} = 1.850) and Choice (\textit{t(298)} = 23.005, \textit{P} < 0.0001, \textit{d} = 2.656) conditions. The contrast between the Choice conditions reached significance in the Pro-Putin condition (\textit{t(298)} = 5.782, \textit{P} < 0.0001, \textit{d} = 0.668) but not in the Anti-Putin condition (\textit{t(298)} = 1.194, \textit{P} = 0.2335, \textit{d} = 0.138). 
\end{flushleft}
\end{table}

\stepcounter{suppTable}
\begin{table}[htbp]
\centering
\renewcommand{\arraystretch}{1.2}  
\caption{Detailed Study 2 results, Impact on Russia, numeric scoring.}
\label{tab:s36}
\begin{tabular}{|c|c|c|c|c|}
\hline
\textbf{} & \multicolumn{2}{c|}{\textbf{No Choice}} & \multicolumn{2}{c|}{\textbf{Choice}} \\
\hline
\multirow{3}{*}{\textbf{Essay Type}} & \textbf{Mean} & \textbf{Effect Size} & \textbf{Mean} & \textbf{Effect Size} \\
 & \textbf{95\% CI} & \textbf{t-Statistic} & \textbf{95\% CI} & \textbf{t-Statistic} \\
 & \textbf{SD} & \textbf{Significance} & \textbf{SD} & \textbf{Significance} \\
\hline

\multirow{3}{*}{\textbf{Pro-Putin}} & M = 3.835 & \textit{d} = 1.711 & M = 4.385 & \textit{d} = 2.388 \\
 & [3.733, 3.937] & \textit{t(298)} = 14.819 & [4.277, 4.493] & \textit{t(298)} = 20.678 \\
 & SD = 0.633 & \textit{P} < 0.0001 & SD = 0.667 & \textit{P} < 0.0001 \\
\hline

\multirow{3}{*}{\textbf{Control}} & M = 2.840 & \multirow{3}{*}{.} & M = 2.855 & \multirow{3}{*}{.} \\
 & [2.755, 2.925] &  & [2.756, 2.954] &  \\
 & SD = 0.525 &  & SD = 0.613 &  \\
\hline

\multirow{3}{*}{\textbf{Anti-Putin}} & M = 2.285 & \textit{d} = 1.136 & M = 2.287 & \textit{d} = 1.057 \\
 & [2.212, 2.358] & \textit{t(298)} = 9.836 & [2.214, 2.359] & \textit{t(298)} = 9.158 \\
 & SD = 0.450 & \textit{P} < 0.0001 & SD = 0.449 & \textit{P} < 0.0001 \\
\hline
\end{tabular}
\begin{flushleft}
Note: Table S36 displays the detailed main results of Study 2 with the individual item about Putin’s Impact on Russia as the dependent variable, using the numeric scoring. Effect sizes (in Cohen’s \textit{d}) and significance scores (using t-tests) are for the contrast of the relevant condition with Control. The contrast between the Pro- and Anti-Putin conditions is also significant in both the No Choice (\textit{t(298)} = 24.445, \textit{P} < 0.0001, \textit{d} = 2.823) and Choice (\textit{t(298)} = 31.954, \textit{P} < 0.0001, \textit{d} = 3.690) conditions. The contrast between the Choice conditions reached significance in the Pro-Putin condition (\textit{t(298)} = 7.324, \textit{P} < 0.0001, \textit{d} = 0.846) but not in the Anti-Putin (\textit{t(298)} = -0.032, \textit{P} = 0.9744, \textit{d} = -0.004) condition.
\end{flushleft}
\end{table}

\stepcounter{suppTable}
\begin{table}[htbp]
\centering
\renewcommand{\arraystretch}{1.2}  
\caption{Detailed Study 2 results, Economic Effectiveness, numeric scoring.}
\label{tab:s37}
\begin{tabular}{|c|c|c|c|c|}
\hline
\textbf{} & \multicolumn{2}{c|}{\textbf{No Choice}} & \multicolumn{2}{c|}{\textbf{Choice}} \\
\hline
\multirow{3}{*}{\textbf{Essay Type}} & \textbf{Mean} & \textbf{Effect Size} & \textbf{Mean} & \textbf{Effect Size} \\
 & \textbf{95\% CI} & \textbf{t-Statistic} & \textbf{95\% CI} & \textbf{t-Statistic} \\
 & \textbf{SD} & \textbf{Significance} & \textbf{SD} & \textbf{Significance} \\
\hline

\multirow{3}{*}{\textbf{Pro-Putin}} & M = 5.133 & \textit{d} = 1.672 & M = 5.173 & \textit{d} = 1.600 \\
 & [5.089, 5.178] & \textit{t(298)} = 14.484 & [5.121, 5.225] & \textit{t(298)} = 13.854 \\
 & SD = 0.276 & \textit{P} < 0.0001 & SD = 0.322 & \textit{P} < 0.0001 \\
\hline

\multirow{3}{*}{\textbf{Control}} & M = 4.327 & \multirow{3}{*}{.} & M = 4.357 & \multirow{3}{*}{.} \\
 & [4.226, 4.427] &  & [4.252, 4.461] &  \\
 & SD = 0.624 &  & SD = 0.646 &  \\
\hline

\multirow{3}{*}{\textbf{Anti-Putin}} & M = 4.207 & \textit{d} = 0.183 & M = 3.937 & \textit{d} = 0.618 \\
 & [4.096, 4.317] & \textit{t(298)} = 1.588 & [3.822, 4.052] & \textit{t(298)} = 5.349 \\
 & SD = 0.683 & \textit{P} = 0.1133 & SD = 0.713 & \textit{P} < 0.0001 \\
\hline
\end{tabular}
\begin{flushleft}
Note: Table S37 displays the detailed main results of Study 2 with the individual item about Putin’s Economic Effectiveness as the dependent variable, using the numeric scoring. Effect sizes (in Cohen’s \textit{d}) and significance scores (using t-tests) are for the contrast of the relevant condition with Control. The contrast between the Pro- and Anti-Putin conditions is also significant in both the No Choice (\textit{t(298)} = 15.400, \textit{P} < 0.0001, \textit{d} = 1.778) and Choice (\textit{t(298)} = 19.366, \textit{P} < 0.0001, \textit{d} = 2.236) conditions. The contrast between the Choice conditions reached significance in the Anti-Putin condition (\textit{t(298)} = 3.349, \textit{P} = 0.0009, \textit{d} = 0.387) but not in the Pro-Putin (\textit{t(298)} = 1.155, \textit{P} = 0.2492, \textit{d} = 0.133) condition.
\end{flushleft}
\end{table}

\stepcounter{suppTable}
\begin{table}[htbp]
\centering
\renewcommand{\arraystretch}{1.2}  
\caption{Detailed Study 2 results, Visionary/Short-Sighted, numeric scoring.}
\label{tab:s38}
\begin{tabular}{|c|c|c|c|c|}
\hline
\textbf{} & \multicolumn{2}{c|}{\textbf{No Choice}} & \multicolumn{2}{c|}{\textbf{Choice}} \\
\hline
\multirow{3}{*}{\textbf{Essay Type}} & \textbf{Mean} & \textbf{Effect Size} & \textbf{Mean} & \textbf{Effect Size} \\
 & \textbf{95\% CI} & \textbf{t-Statistic} & \textbf{95\% CI} & \textbf{t-Statistic} \\
 & \textbf{SD} & \textbf{Significance} & \textbf{SD} & \textbf{Significance} \\
\hline

\multirow{3}{*}{\textbf{Pro-Putin}} & M = 5.193 & \textit{d} = 0.227 & M = 5.270 & \textit{d} = 0.281 \\
 & [5.141, 5.245] & \textit{t(298)} = 1.965 & [5.217, 5.323] & \textit{t(298)} = 2.435 \\
 & SD = 0.321 & \textit{P} = 0.0504 & SD = 0.331 & \textit{P} = 0.0155 \\
\hline

\multirow{3}{*}{\textbf{Control}} & M = 5.110 & \multirow{3}{*}{.} & M = 5.175 & \multirow{3}{*}{.} \\
 & [5.044, 5.176] &  & [5.119, 5.231] &  \\
 & SD = 0.408 &  & SD = 0.345 &  \\
\hline

\multirow{3}{*}{\textbf{Anti-Putin}} & M = 4.493 & \textit{d} = 1.252 & M = 4.513 & \textit{d} = 1.373 \\
 & [4.402, 4.584] & \textit{t(298)} = 10.842 & [4.419, 4.608] & \textit{t(298)} = 11.892 \\
 & SD = 0.565 & \textit{P} < 0.0001 & SD = 0.588 & \textit{P} < 0.0001 \\
\hline
\end{tabular}
\begin{flushleft}
Note: Table S38 displays the detailed main results of Study 2 with the individual item about Putin’s Vision as the dependent variable, using the numeric scoring. Effect sizes (in Cohen’s \textit{d}) and significance scores (using t-tests) are for the contrast of the relevant condition with Control. The contrast between the Pro- and Anti-Putin conditions is also significant in both the No Choice (\textit{t(298)} = 13.197, \textit{P} < 0.0001, \textit{d} = 1.524) and Choice (\textit{t(298)} = 13.739, \textit{P} < 0.0001, \textit{d} = 1.586) conditions. The contrast between the Choice conditions reached significance in the Pro-Putin condition (\textit{t(298)} = 2.035, \textit{P} = 0.0427, \textit{d} = 0.235) but not in the Anti-Putin (\textit{t(298)} = -0.301, \textit{P} = 0.7640, \textit{d} = -0.035) condition.

\end{flushleft}
\end{table}

\stepcounter{suppTable}
\begin{table}[htbp]
\centering
\renewcommand{\arraystretch}{1.2}
\caption{Study 2, Generalized linear modeling of choice moderation, for the composite evaluation and individual evaluative items, composite scoring.}
\label{tab:s39}
\begin{tabular}{|c|c|c|c|c|c|}
\hline
\textbf{Regression} & \textbf{Composite} & \textbf{Overall} & \textbf{Impact on} & \textbf{Economic} & \textbf{Visionary or} \\
\textbf{Term} & \textbf{Evaluation} & \textbf{Leadership} & \textbf{Russia} & \textbf{Effectiveness} & \textbf{Short-Sighted} \\
\hline
\multirow{4}{*}{\parbox{2cm}{\centering\textbf{Anti-Putin}}} 
& \textbf{$\beta$ = -0.813} & $\beta$ = -0.475 & $\beta$ = -0.555 & $\beta$ = -0.133 & $\beta$ = -0.630 \\
& \textbf{\textit{SE} = 0.068} & \textit{SE} = 0.058 & \textit{SE} = 0.056 & \textit{SE} = 0.076 & \textit{SE} = 0.059 \\
& \textbf{\textit{z} = -11.88} & \textit{z} = -8.21 & \textit{z} = -9.96 & \textit{z} = -1.74 & \textit{z} = -10.75 \\
& \textbf{\textit{P} < 0.001} & \textit{P} < 0.001 & \textit{P} < 0.001 & \textit{P} = 0.081 & \textit{P} < 0.001 \\
\hline
\multirow{4}{*}{\parbox{2cm}{\centering\textbf{Pro-Putin}}} 
& \textbf{$\beta$ = 0.946} & $\beta$ = 0.508 & $\beta$ = 0.978 & $\beta$ = 0.807 & $\beta$ = 0.083 \\
& \textbf{\textit{SE} = 0.054} & \textit{SE} = 0.052 & \textit{SE} = 0.068 & \textit{SE} = 0.056 & \textit{SE} = 0.042 \\
& \textbf{\textit{z} = 17.42} & \textit{z} = 9.79 & \textit{z} = 14.39 & \textit{z} = 14.52 & \textit{z} = 1.97 \\
& \textbf{\textit{P} < 0.001} & \textit{P} < 0.001 & \textit{P} < 0.001 & \textit{P} < 0.001 & \textit{P} = 0.049 \\
\hline
\multirow{4}{*}{\parbox{2cm}{\centering\textbf{Choice vs. \\No Choice}}} 
& \textbf{$\beta$ = 0.105} & $\beta$ = 0.117 & $\beta$ = 0.012 & $\beta$ = 0.027 & $\beta$ = 0.065 \\
& \textbf{\textit{SE} = 0.059 }& \textit{SE} = 0.051 & \textit{SE} = 0.065 & \textit{SE} = 0.073 & \textit{SE} = 0.044 \\
& \textbf{\textit{z} = 1.78}& \textit{z} = 2.30 & \textit{z} = 0.18 & \textit{z} = 0.36 & \textit{z} = 1.49 \\
& \textbf{\textit{P} = 0.076} & \textit{P} = 0.021 & \textit{P} = 0.858 & \textit{P} = 0.717 & \textit{P} = 0.135 \\
\hline
\multirow{4}{*}{\parbox{2cm}{\centering\textbf{Interaction: \\Anti-Putin x \\Choice}}} 
& \textbf{$\beta$ = -0.273} & $\beta$ = -0.260 & $\beta$ = -0.007 & $\beta$ = -0.297 & $\beta$ = -0.047 \\
& \textbf{\textit{SE} = 0.097} & \textit{SE} = 0.079 & \textit{SE} = 0.083 & \textit{SE} = 0.110 & \textit{SE} = 0.082 \\
& \textbf{\textit{z} = -2.82} & \textit{z} = -3.31 & \textit{z} = -0.08 & \textit{z} = -2.69 & \textit{z} = -0.57 \\
& \textbf{\textit{P} = 0.005}& \textit{P} = 0.001 & \textit{P} = 0.936 & \textit{P} = 0.007 & \textit{P} = 0.570 \\
\hline
\multirow{4}{*}{\parbox{2cm}{\centering\textbf{Interaction: \\Pro-Putin x \\Choice}}} 
&\textbf{$\beta$ = 0.290}& $\beta$ = 0.213 & $\beta$ = 0.538 & $\beta$ = 0.017 & $\beta$ = 0.013 \\
& \textbf{\textit{SE} = 0.075} & \textit{SE} = 0.076 & \textit{SE} = 0.102 & \textit{SE} = 0.081 & \textit{SE} = 0.057 \\
& \textbf{\textit{z} = 3.86} & \textit{z} = 2.81 & \textit{z} = 5.30 & \textit{z} = 0.21 & \textit{z} = 0.23 \\
& \textbf{\textit{P} < 0.001} & \textit{P} = 0.005 & \textit{P} < 0.001 & \textit{P} = 0.837 & \textit{P} = 0.816 \\
\hline
\end{tabular}
\begin{flushleft}
Note: Table S39 displays the interaction terms from a set of Generalized Linear Models with a Gaussian family and identity link functions, using robust standard errors. This was designed to rigorously test for choice moderation in Study 2. The dependent variables are a standardized composite of the four evaluative items about Putin and each of these individual evaluative items, using a composite of the verbal and numeric scorings. The significant interaction terms suggest bidirectional moderation by Choice for the composite variable: Evaluations were more positive in the Pro-Putin condition, and more negative in the Anti-Putin condition when GPT received Choice versus No Choice surrounding which essay to write. Significant moderation was similarly observed in at least one and sometimes both directions for three of the four individual evaluative items.
\end{flushleft}
\end{table}

\stepcounter{suppTable}
\begin{table}[htbp]
\centering
\renewcommand{\arraystretch}{1.2}
\caption{Study 2, Generalized linear modeling of choice moderation, for the composite evaluation and individual evaluative items, verbal scoring.}
\label{tab:s40}
\begin{tabular}{|c|c|c|c|c|c|}
\hline
\textbf{Regression} & \textbf{Composite} & \textbf{Overall} & \textbf{Impact on} & \textbf{Economic} & \textbf{Visionary or} \\
\textbf{Term} & \textbf{Evaluation} & \textbf{Leadership} & \textbf{Russia} & \textbf{Effectiveness} & \textbf{Short-Sighted} \\
\hline
\multirow{4}{*}{\parbox{2cm}{\centering\textbf{Anti-Putin}}} 
& \textbf{$\beta$ = -0.816} & $\beta$ = -0.503 & $\beta$ = -0.555 & $\beta$ = -0.147 & $\beta$ = -0.643 \\
& \textbf{\textit{SE} = 0.070} & \textit{SE} = 0.062 & \textit{SE} = 0.056 & \textit{SE} = 0.078 & \textit{SE} = 0.061 \\
& \textbf{\textit{z} = -11.72} & \textit{z} = -8.16 & \textit{z} = -9.99 & \textit{z} = -1.88 & \textit{z} = -10.55 \\
& \textbf{\textit{P} < 0.001} & \textit{P} < 0.001 & \textit{P} < 0.001 & \textit{P} = 0.060 & \textit{P} < 0.001 \\
\hline
\multirow{4}{*}{\parbox{2cm}{\centering\textbf{Pro-Putin}}} 
& \textbf{$\beta$ = 0.925} & $\beta$ = 0.517 & $\beta$ = 0.962 & $\beta$ = 0.807 & $\beta$ = 0.083 \\
& \textbf{\textit{SE} = 0.054} & \textit{SE} = 0.055 & \textit{SE} = 0.070 & \textit{SE} = 0.056 & \textit{SE} = 0.042 \\
& \textbf{\textit{z} = 17.21} & \textit{z} = 9.41 & \textit{z} = 13.70 & \textit{z} = 14.52 & \textit{z} = 1.97 \\
& \textbf{\textit{P} < 0.001} & \textit{P} < 0.001 & \textit{P} < 0.001 & \textit{P} < 0.001 & \textit{P} = 0.049 \\
\hline
\multirow{4}{*}{\parbox{2cm}{\centering\textbf{Choice vs. \\No Choice}}} 
& \textbf{$\beta$ = 0.099} & $\beta$ = 0.118 & $\beta$ = 0.008 & $\beta$ = 0.023 & $\beta$ = 0.065 \\
& \textbf{\textit{SE} = 0.059} & \textit{SE} = 0.054 & \textit{SE} = 0.065 & \textit{SE} = 0.074 & \textit{SE} = 0.044 \\
& \textbf{\textit{z} = 1.69} & \textit{z} = 2.18 & \textit{z} = 0.13 & \textit{z} = 0.32 & \textit{z} = 1.49 \\
& \textbf{\textit{P} = 0.091} & \textit{P} = 0.029 & \textit{P} = 0.899 & \textit{P} = 0.752 & \textit{P} = 0.135 \\
\hline
\multirow{4}{*}{\parbox{2cm}{\centering\textbf{Interaction: \\Anti-Putin x \\Choice}}} 
& \textbf{$\beta$ = -0.292} & $\beta$ = -0.332 & $\beta$ = 0.000 & $\beta$ = -0.293 & $\beta$ = -0.048 \\
& \textbf{\textit{SE} = 0.098} & \textit{SE} = 0.085 & \textit{SE} = 0.083 & \textit{SE} = 0.113 & \textit{SE} = 0.086 \\
& \textbf{\textit{z} = -2.98} & \textit{z} = -3.89 & \textit{z} = 0.00 & \textit{z} = -2.60 & \textit{z} = -0.56 \\
& \textbf{\textit{P} = 0.003} & \textit{P} < 0.001 & \textit{P} = 1.000 & \textit{P} = 0.009 & \textit{P} = 0.573 \\
\hline
\multirow{4}{*}{\parbox{2cm}{\centering\textbf{Interaction: \\Pro-Putin x \\Choice}}} 
& \textbf{$\beta$ = 0.291} & $\beta$ = 0.218 & $\beta$ = 0.542 & $\beta$ = 0.023 & $\beta$ = 0.015 \\
& \textbf{\textit{SE} = 0.075} & \textit{SE} = 0.079 & \textit{SE} = 0.105 & \textit{SE} = 0.082 & \textit{SE} = 0.057 \\
& \textbf{\textit{z} = 3.90} & \textit{z} = 2.76 & \textit{z} = 5.15 & \textit{z} = 0.29 & \textit{z} = 0.26 \\
& \textbf{\textit{P} < 0.001} & \textit{P} = 0.006 & \textit{P} < 0.001 & \textit{P} = 0.775 & \textit{P} = 0.794 \\
\hline
\end{tabular}
\begin{flushleft}
Note: Table S40 displays the interaction terms from a set of Generalized Linear Models with a Gaussian family and identity link functions, using robust standard errors. This was designed to rigorously test for choice moderation in Study 2. The dependent variables are a standardized composite of the four evaluative items about Putin and each of these individual evaluative items, using the verbal scoring. The significant interaction terms suggest bidirectional moderation by Choice for the composite variable: Evaluations were more positive in the Pro-Putin condition, and more negative in the Anti-Putin condition when GPT received Choice versus No Choice surrounding which essay to write. Significant moderation was similarly observed in at least one and sometimes both directions for three of the four individual evaluative items.

\end{flushleft}
\end{table}

\stepcounter{suppTable}
\begin{table}[htbp]
\centering
\renewcommand{\arraystretch}{1.2}
\caption{Study 2, Generalized linear modeling of choice moderation, for the composite evaluation and individual evaluative items, numeric scoring.}
\label{tab:s41}
\begin{tabular}{|c|c|c|c|c|c|}
\hline
\textbf{Regression} & \textbf{Composite} & \textbf{Overall} & \textbf{Impact on} & \textbf{Economic} & \textbf{Visionary or} \\
\textbf{Term} & \textbf{Evaluation} & \textbf{Leadership} & \textbf{Russia} & \textbf{Effectiveness} & \textbf{Short-Sighted} \\
\hline
\multirow{4}{*}{\parbox{2cm}{\centering\textbf{Anti-Putin}}} 
& \textbf{$\beta$ = -0.807} & $\beta$ = -0.447 & $\beta$ = -0.555 & $\beta$ = -0.120 & $\beta$ = -0.617 \\
& \textbf{\textit{SE} = 0.068} & \textit{SE} = 0.057 & \textit{SE} = 0.056 & \textit{SE} = 0.075 & \textit{SE} = 0.057 \\
& \textbf{\textit{z} = -11.91} & \textit{z} = -7.84 & \textit{z} = -9.86 & \textit{z} = -1.59 & \textit{z} = -10.87 \\
& \textbf{\textit{P} < 0.001} & \textit{P} < 0.001 & \textit{P} < 0.001 & \textit{P} = 0.111 & \textit{P} < 0.001 \\
\hline
\multirow{4}{*}{\parbox{2cm}{\centering\textbf{Pro-Putin}}} 
& \textbf{$\beta$ = 0.964} & $\beta$ = 0.500 & $\beta$ = 0.995 & $\beta$ = 0.807 & $\beta$ = 0.083 \\
& \textbf{\textit{SE} = 0.055} & \textit{SE} = 0.050 & \textit{SE} = 0.067 & \textit{SE} = 0.056 & \textit{SE} = 0.042 \\
& \textbf{\textit{z} = 17.52} & \textit{z} = 9.96 & \textit{z} = 14.86 & \textit{z} = 14.52 & \textit{z} = 1.97 \\
& \textbf{\textit{P} < 0.001} & \textit{P} < 0.001 & \textit{P} < 0.001 & \textit{P} < 0.001 & \textit{P} = 0.049 \\
\hline
\multirow{4}{*}{\parbox{2cm}{\centering\textbf{Choice vs. \\No Choice}}} 
& \textbf{$\beta$ = 0.110} & $\beta$ = 0.115 & $\beta$ = 0.015 & $\beta$ = 0.030 & $\beta$ = 0.065 \\
& \textbf{\textit{SE} = 0.060} & \textit{SE} = 0.049 & \textit{SE} = 0.066 & \textit{SE} = 0.073 & \textit{SE} = 0.044 \\
& \textbf{\textit{z} = 1.85} & \textit{z} = 2.36 & \textit{z} = 0.23 & \textit{z} = 0.41 & \textit{z} = 1.49 \\
& \textbf{\textit{P} = 0.065} & \textit{P} = 0.018 & \textit{P} = 0.819 & \textit{P} = 0.682 & \textit{P} = 0.135 \\
\hline
\multirow{4}{*}{\parbox{2cm}{\centering\textbf{Interaction: \\Anti-Putin x \\Choice}}} 
& \textbf{$\beta$ = -0.249} & $\beta$ = -0.188 & $\beta$ = -0.013 & $\beta$ = -0.300 & $\beta$ = -0.045 \\
& \textbf{\textit{SE} = 0.097} & \textit{SE} = 0.078 & \textit{SE} = 0.084 & \textit{SE} = 0.109 & \textit{SE} = 0.079 \\
& \textbf{\textit{z} = -2.58} & \textit{z} = -2.41 & \textit{z} = -0.16 & \textit{z} = -2.76 & \textit{z} = -0.57 \\
& \textbf{\textit{P} = 0.010} & \textit{P} = 0.016 & \textit{P} = 0.873 & \textit{P} = 0.006 & \textit{P} = 0.571 \\
\hline
\multirow{4}{*}{\parbox{2cm}{\centering\textbf{Interaction: \\Pro-Putin x \\Choice}}} 
& \textbf{$\beta$ = 0.287} & $\beta$ = 0.208 & $\beta$ = 0.535 & $\beta$ = 0.010 & $\beta$ = 0.012 \\
& \textbf{\textit{SE} = 0.076} & \textit{SE} = 0.074 & \textit{SE} = 0.100 & \textit{SE} = 0.081 & \textit{SE} = 0.057 \\
& \textbf{\textit{z} = 3.79} & \textit{z} = 2.81 & \textit{z} = 5.37 & \textit{z} = 0.12 & \textit{z} = 0.20 \\
& \textbf{\textit{P} < 0.001} & \textit{P} = 0.005 & \textit{P} < 0.001 & \textit{P} = 0.902 & \textit{P} = 0.839 \\
\hline
\end{tabular}
\begin{flushleft}
Note: Table S41 displays the interaction terms from a set of Generalized Linear Models with a Gaussian family and identity link functions, using robust standard errors. This was designed to rigorously test for choice moderation in Study 2. The dependent variables are a standardized composite of the four evaluative items about Putin and each of these individual evaluative items, using the numeric scorings. The significant interaction terms suggest bidirectional moderation by Choice for the composite variable: Evaluations were more positive in the Pro-Putin condition, and more negative in the Anti-Putin condition when GPT received Choice versus No Choice surrounding which essay to write. Significant moderation was similarly observed in at least one and sometimes both directions for three of the four individual evaluative items.
\end{flushleft}
\end{table}

\clearpage
\section*{SI References}
\begin{enumerate}
    \item  J. W. Brehm, A. R. Cohen, Explorations in cognitive dissonance (John Wiley \& Sons Inc., 1962)
    \item  D. E. Linder, J. Cooper, E. E. Jones, Decision freedom as a determinant of the role of incentive magnitude in attitude change. J. Pers. and Soc. Psychol. 6, 245–254 (1967).
    \item  S. Pauer, R. Linne, H. Erb, From the illusion of choice to actual control: Reconsidering the induced-compliance paradigm of cognitive dissonance. Adv. Meth. Pract. in Psychol. Sci. 7 (v.4), 1-5 (2024).
    \item  E. Harmon-Jones, J. W. Brehm, J. Greenberg, L. Simon, D. E. Nelson, Evidence that the production of aversive consequences is not necessary to create cognitive dissonance. J. Pers. Soc. Psychol. 70, 5-16 (1996).
    \item  M. F. Scheier, C. S. Carver, Private and public self-attention, resistance to change, and dissonance reduction. J. Pers. Soc. Psychol. 39, 390-405 (1980).
    \item S. Bubeck et al., Sparks of artificial general intelligence: Early experiments with GPT-4. arXiv [Preprint] (2023). https://arxiv.org/abs/2303.12712 (Accessed 21 January 2025).
\end{enumerate}

\end{document}